\documentclass{article}

\usepackage{arxiv}

\usepackage[utf8]{inputenc} 
\usepackage[T1]{fontenc}    
\usepackage{hyperref}       
\usepackage{url}            
\usepackage{booktabs}       
\usepackage{tabularx}       
\usepackage{array}          
\newcolumntype{L}[1]{>{\raggedright\arraybackslash}p{#1}}
\newcolumntype{Y}{>{\raggedright\arraybackslash}X}
\usepackage{amsfonts}       
\usepackage{amsmath}
\usepackage{amsthm}
\usepackage{bm}             
\usepackage{nicefrac}       
\usepackage{float}          
\usepackage[section]{placeins} 
\usepackage{microtype}      
\usepackage{lipsum}
\usepackage{graphicx}
\usepackage{subcaption}
\usepackage{tikz}
\usetikzlibrary{arrows.meta,positioning,fit,backgrounds,shapes.multipart, calc}
\usepackage[ruled]{algorithm2e} 
\usepackage{listings}
\definecolor{codegreen}{rgb}{0.0,0.45,0.12}
\definecolor{codegray}{rgb}{0.5,0.5,0.5}
\definecolor{codeblue}{rgb}{0.10,0.10,0.60}
\definecolor{codebg}{rgb}{0.97,0.97,0.97}
\lstdefinestyle{pythonstyle}{
  language=Python,
  backgroundcolor=\color{codebg},
  basicstyle=\ttfamily\footnotesize,
  keywordstyle=\color{codeblue}\bfseries,
  commentstyle=\color{codegreen}\itshape,
  stringstyle=\color{codegreen},
  numberstyle=\ttfamily\scriptsize\color{codegray},
  numbers=left,
  numbersep=8pt,
  showstringspaces=false,
  breaklines=true,
  frame=single,
  rulecolor=\color{codegray},
  captionpos=b,
  columns=fullflexible,
  keepspaces=true,
}
\graphicspath{ {./images/} }

\definecolor{domainblue}{RGB}{226,238,249}
\definecolor{gridgray}{RGB}{135,145,155}
\definecolor{rayblue}{RGB}{20,91,150}
\definecolor{cornergreen}{RGB}{30,130,76}
\definecolor{cellorange}{RGB}{238,153,48}
\definecolor{cutred}{RGB}{190,55,55}

\newtheorem{defin}{Definition}[section]
\newtheorem{remark}{Remark}[section]

\title{PoliVEM: a Python-driven virtual element framework for computational solid mechanics}

\author{
 Paulo Akira F. Enabe \\
    Escola Politécnica\\
    University of São Paulo\\
    Department of Structural and Geotechnical Engineering\\
  \texttt{paulo.enabe@usp.br} \\
  \And
 Rodrigo Provasi \\
    Escola Politécnica\\
    University of São Paulo\\
    Department of Structural and Geotechnical Engineering\\
  \texttt{provasi@usp.br} \\
}

\begin{document}
\raggedbottom
\maketitle
\begin{abstract}
This work presents \texttt{PoliVEM}, a software framework for the Virtual Element Method (VEM) in computational solid and structural mechanics. A C++17 computational core and a Python interface place one-dimensional beams, two- and three-dimensional elasticity, axisymmetric elasticity, transient diffusion, and finite-strain hyperelasticity in a common implementation. The framework stores vertex, edge, face, and cell degrees of freedom in one hierarchy, constructs the energy, strain, and $L^2$ projections from common polynomial data, and retains the consistency--stabilization split at the element level. The core separates the mesh, material, element, assembler, and solver responsibilities and combines them by composition. A new formulation supplies its projection, discrete form, and stabilization while reusing the mesh representation, degree-of-freedom numbering, boundary-condition treatment, sparse assembly, algebraic solvers, and Python binding pattern. The numerical infrastructure includes polygonal and polyhedral mesh input, higher-order entity numbering, cached projection operators, static condensation, coloured sparse assembly, linear solver selection based on the matrix structure, an incremental Newton method with line search and regularization, and explicit and implicit time integration. Three numerical experiments examine the shared implementation through polynomial reproduction, recovery after static condensation, higher-order degree-of-freedom handling, polygonal and polyhedral meshes, patch tests, stiffness spectra, and comparisons with independent finite element calculations. The results verify the common core for the beam and two- and three-dimensional elasticity formulations. They also identify the scaling of high-order polynomial bases on anisotropic cells and the memory required by sparse direct factorizations as the main limitations of the current implementation.
\end{abstract}


\section{Introduction}
\label{sec:introduction}

\paragraph{} This work presents \texttt{PoliVEM}, a software framework for the Virtual Element Method (VEM) in computational solid and structural mechanics. The framework combines a C++17 computational core with a Python interface and implements one-dimensional beams, two- and three-dimensional elasticity at different polynomial orders, axisymmetric elasticity, transient diffusion, and finite-strain hyperelasticity. These formulations use the same mesh representation, projection machinery, assembly infrastructure, and algebraic solution layer. This paper describes that common structure, its implementation, and three numerical studies of the shared infrastructure.

\paragraph{} The VEM was introduced in \cite{beirao2013vem} as a conforming Galerkin discretization of a two-dimensional elliptic problem on polygonal meshes. A detailed account of its construction, including the ingredients needed for polyhedral elements, is given in \cite{beirao2014hitchhiker}. A local virtual element space contains a prescribed polynomial subspace, although its nonpolynomial basis functions are not evaluated explicitly in the element interior. Projections computable from the degrees of freedom provide the polynomially consistent part of the discrete bilinear form, while a stabilization term supplies the required stability on the kernel of the projection. This construction accommodates cells with variable numbers of vertices and faces, including nonconvex cells, without mapping each cell to a fixed reference element. Hanging nodes can be treated as additional polygon vertices. VEM thereby extends the finite element framework from standard simplex and tensor-product element families to general polygonal and polyhedral meshes.

\paragraph{} The range of VEM formulations now extends well beyond the scalar elliptic model problem considered in the original construction. Applications include elastic and inelastic solids on polytopal meshes \cite{beirao2015inelastic}, arbitrary-order two-dimensional elasticity \cite{artioli2017}, parabolic and hyperbolic time-dependent problems \cite{vacca2015parabolic,vacca2017hyperbolic}, Euler--Bernoulli and Timoshenko beams \cite{wriggers2022,wriggers2023}, and isotropic and transversely isotropic hyperelasticity at finite strain \cite{vanHuyssteen2020isotropic,vanHuyssteen2021transverselyisotropic}. More recent developments include higher-order three-dimensional elasticity \cite{xu2024highorder3d}, curvilinear elements \cite{artioli2020curvilinear}, studies of polygonal and polyhedral mesh quality \cite{sorgente2021meshquality,sorgente2022polyhedral}, and stabilization-free formulations based on enriched projections \cite{berrone2024divergencefree,xu20243dSFVEM}. The term VEM now covers a family of discretizations.

\paragraph{} This breadth also complicates software design. A VEM implementation must manage variable-length connectivity; degrees of freedom attached to vertices, edges, faces, and cell interiors; polynomial moments on general polytopes; projection operators that do not require pointwise evaluation of the virtual basis functions; and stabilization terms whose form and scaling depend on the governing problem. Higher-order and three-dimensional formulations add orientation and entity-numbering requirements, while nonlinear and transient problems require repeated assembly, tangent updates, mass operators, and time integration. Implementations developed for one model problem seldom expose these components for reuse.

\paragraph{} Section~\ref{sec:comparison_vem_softwares} compares PoliVEM with existing VEM software. Among the released packages surveyed there, those with high-level Python interfaces focus mainly on canonical partial differential equations, while codes aimed specifically at structural analysis largely stop at two-dimensional linear elastostatics. Advanced structural formulations, including beams, axisymmetric elasticity, and finite-strain hyperelasticity, are still commonly distributed as code attached to individual method papers. Moving from one structural VEM formulation to another therefore often requires a separate implementation even when the formulations use the same projections, degree-of-freedom patterns, assembly operations, and solvers.

\paragraph{} PoliVEM was developed during the first author's doctoral research to collect these common operations in one code base. The implemented formulations draw on established VEM literature and on the authors' previous studies of Euler--Bernoulli beams \cite{enabe2025hybrid}, axisymmetric elasticity \cite{enabe2025axisymmetric}, mass-lumped transient problems \cite{enabe2026masslumped}, and stabilization at finite strain \cite{enabe2026stabilization}. This paper treats those formulations as cited inputs and examines how their common mathematical structure is represented in software and how much formulation-specific code is needed when the mesh, projection, assembly, solver, and binding layers are reused.

\paragraph{} The main contributions of this work are:
\begin{itemize}
  \item \textbf{A common computational representation of VEM formulations.} Energy, strain, and $L^2$ projections are constructed from common polynomial and degree-of-freedom data, and the local operators retain the consistency--stabilization split. This representation applies across spatial dimensions, approximation orders, and the static, transient, and nonlinear problems included in the framework.
  \item \textbf{A two-tier implementation for engineering analysis.} The C++ core contains the element calculations, sparse assembly, and solution procedures. Python drives mesh preparation, model definition, execution, and post-processing through NumPy-compatible data structures. A new formulation follows the same binding and execution pattern as the existing solvers.
  \item \textbf{Numerical infrastructure shared by the structural formulations.} The software includes polygonal and polyhedral mesh input, higher-order entity numbering, static condensation, a coloured sparse-assembly procedure, linear solver selection based on matrix structure, an incremental Newton method with line search, and explicit and implicit time integration. Section~\ref{sec:numerical_implementations} states the algorithms used by the implementation and the criteria used to select them.
  \item \textbf{Targeted verification of the shared infrastructure.} Section~\ref{sec:numerical_experiments} reports new experiments for the general-order beam formulation and higher-order two- and three-dimensional elasticity. These experiments test polynomial reproduction, static condensation, higher-order degree-of-freedom handling, general polytopal meshes, and the assembled linear systems. The numerical assessment of the axisymmetric, transient, and finite-strain formulations is not repeated; the corresponding method papers are cited instead.
\end{itemize}

\paragraph{} The mathematical formulations retain their original attribution. The three numerical experiments use new meshes and test configurations, with no figures or data carried over from the preceding method papers.

\paragraph{} The remainder of the paper is organized as follows. Section~\ref{sec:comparison_vem_softwares} compares PoliVEM with existing VEM software and defines the scope of the software claim. Section~\ref{sec:unified_vem_core} presents the common virtual spaces, degrees of freedom, projections, and consistency--stabilization split used by the code. Section~\ref{sec:software_arch} describes the layered architecture, mesh facilities, core abstractions, extension procedure, and Python workflow. Section~\ref{sec:numerical_implementations} gives the projection, sparse-assembly, linear-solution, nonlinear-solution, and time-integration algorithms. Section~\ref{sec:numerical_experiments} reports the three new numerical studies and cites the earlier validation of the remaining formulations. Section~\ref{sec:conclusion} presents the conclusions.

\section{Comparison with existing VEM software}
\label{sec:comparison_vem_softwares}

\paragraph{} The implementation of the VEM differs from that of the FEM in several concrete respects. There is no reference element, the basis functions are not known in explicit form, and numerical integration must be carried out on arbitrary polytopes. These requirements have motivated a dedicated body of VEM software over the past decade, ranging from short didactic scripts to general-purpose libraries. Table~\ref{tab:vem-software} summarizes the packages most relevant to the present work, together with PoliVEM, and characterizes them by spatial dimension, polynomial order, the physics actually covered, the programming interface, and the availability model.

\begin{table}[!htbp]
  \centering
  \renewcommand{\arraystretch}{1.3}
  \setlength{\tabcolsep}{4pt}
  \footnotesize
  \caption{Selected VEM software packages compared with PoliVEM. The two
  positioning axes are the \emph{high-level Python interface} and the
  \emph{structural-mechanics coverage beyond linear elastostatics}. Only
  \emph{dune-vem}, PolyDiM and PoliVEM expose the former, and among those only
  PoliVEM targets structural mechanics up to finite-strain hyperelasticity.}
  \label{tab:vem-software}
  \begin{tabularx}{\textwidth}{@{}L{1.9cm}L{1.35cm}L{1.35cm}L{0.85cm}YL{0.85cm}L{1.0cm}L{1.9cm}@{}}
  \toprule
  \textbf{Software} & \textbf{Lang.} & \textbf{Dim.} & \textbf{Order} &
  \textbf{Physics / PDE coverage} & \textbf{Aud.$^{b}$} & \textbf{Py$^{a}$} &
  \textbf{Availability} \\
  \midrule
  Vem++ \cite{dassi2025vempp} & C++ & 2D/3D & arb. &
  Poisson, elasticity (incl.\ mixed H--R), Stokes/N--S, Maxwell, viscoelasticity, plasticity &
  M/E & No & Freeware on request (closed) \\
  
  Veamy \cite{veamy2019} & C++ & 2D & $k{=}1$ &
  Linear elastostatics, Poisson & E & No & GPL (dormant, 2019) \\
  
  VEMLab \cite{vemlab} & MATLAB & 2D & $k{=}1$ &
  Linear elastostatics (plane strain/stress), Poisson & E & No & GPL-3.0 (2024) \\
  
  \emph{dune-vem} \cite{dednerhodson2024} & C++/Py & 2D & arb. &
  Elliptic (2nd/4th order), nonlinear elliptic, biharmonic, Stokes/N--S &
  M & Yes (UFL) & GPL \\
  
  PolyDiM \cite{polydim2025} & C++/Py/M & 2D/3D & arb. &
  Elliptic, parabolic, Stokes/N--S, linear elasticity, fracture networks & M & Yes (pybind11) & GPL-3.0 \\
  
  VEM in 50 lines \cite{sutton2017} & MATLAB & 2D & $k{=}1$ &
  Poisson (didactic) & M & No & Published in article \\
  
  mVEM \cite{yu2022mvem} & MATLAB & 2D(+3D) & $\le 3$ &
  Poisson, linear elasticity, plate bending, mixed (Darcy/Stokes), variational inequalities, adaptive &
  M/E & No & GPL-3.0 \\
  
  VEMcomp \cite{vemcomp2024} & MATLAB & 2D/3D & $k{=}1$ &
  Elliptic \& parabolic, bulk--surface PDEs & M & No & Open source \\
  
  High-order 2D VEM \cite{herrera2023} & MATLAB & 2D & arb. &
  Poisson / elliptic (high order) & M & No & Published in article \\
  \midrule
  \textbf{PoliVEM (this work)} & C++/Py & 1D/2D/3D & $k\ge1$ &
  Beams, linear elasticity (2D/3D, higher order), axisymmetric elasticity,
  transient diffusion, finite-strain hyperelasticity & E & Yes (pybind11) &
  Not publicly available at publication \\
  \bottomrule
  \end{tabularx}
  
  \vspace{2pt}
  \begin{flushleft}\footnotesize
  $^{a}$ High-level Python / scientific-computing interface (UFL or NumPy/SciPy),
  not merely C++ that can be called from Python.\\
  $^{b}$ Primary audience: M = applied mathematics, E = engineering.
  \end{flushleft}
\end{table}

\paragraph{} Several general-purpose VEM codes are written in C++. \texttt{Vem++}, presented in \cite{dassi2025vempp}, handles two- and three-dimensional problems at arbitrary order, including the Laplacian, the Stokes and Navier--Stokes equations, magnetostatics in a Maxwell formulation, linear elasticity through a mixed Hellinger--Reissner principle, and inelastic solid mechanics. The reported examples include a viscoelastic cylinder and an elasto-plastic perforated plate. The library supports interchangeable polytope quadrature rules, linear solvers, and virtual element spaces, and it includes a set of tutorials. Its C++ interface has no high-level scripting layer, and the corresponding publication provides neither a public repository nor a software license.

\paragraph{} \texttt{PolyDiM} \cite{polydim2025} is a C++ library for polytopal discretizations with a Python binding and a MATLAB interface for prototyping and post-processing. It supports the VEM and FEM in two and three dimensions at high order. Its applications include elliptic and parabolic problems, the Stokes and Navier--Stokes equations, and linear elasticity, with emphasis on mixed-dimensional and discrete-fracture-network problems. Its solid-mechanics coverage is limited to small-strain linear elasticity and does not extend to beams, axisymmetric formulations, finite deformations, plasticity, or contact.

\paragraph{} The VEM can also be embedded within an established finite element framework. The \texttt{dune-vem} module \cite{dednerhodson2024}, built on \texttt{DUNE-FEM}, provides a high-level interface in which the weak form is specified symbolically in the Unified Form Language (UFL). Assembly and solution are driven from Python, with finite and virtual element discretizations used interchangeably. The module supports arbitrary order together with conforming, nonconforming, $C^1$, and divergence- or curl-free spaces. Its reported applications are the Laplacian, incompressible Navier--Stokes flow, and a nonlinear fourth-order Willmore flow. According to the authors, the implementation is restricted to two spatial dimensions, and the extension to three dimensions is described only at a conceptual level. Structural applications such as elasticity are not among the demonstrated problems.

\paragraph{} Several VEM packages written in MATLAB are intended for teaching and methodological prototyping. The fifty-line implementation in \cite{sutton2017}, reported as the first publicly available VEM code, solves the lowest-order Poisson problem in two dimensions and is explicitly didactic. It gives a concrete implementation of the mathematical formulation in \cite{beirao2013vem}. The implementation in \cite{herrera2023} extends this approach to arbitrary order for nodal elliptic problems. The \texttt{mVEM} package \cite{yu2022mvem} collects conforming and nonconforming VEM implementations for the Poisson problem up to order three, linear elasticity, plate bending, mixed formulations for Darcy and Stokes flow, variational inequalities, and adaptive refinement, together with polygonal mesh-refinement routines. Its three-dimensional support is limited to the lowest-order Poisson problem, and nonlinear solid mechanics is listed among the intended future extensions. \texttt{VEMcomp} \cite{vemcomp2024} addresses time-dependent and surface problems. It solves elliptic and parabolic, linear and semilinear equations on bulk, surface, and coupled bulk--surface domains in two and three dimensions, using the lowest-order VEM in space and a first-order implicit-explicit Euler scheme in time. Apart from the linear-elasticity and plate elements of \texttt{mVEM}, these MATLAB packages address canonical partial differential equations rather than structural mechanics.

\paragraph{} A smaller group of codes is written for an engineering audience. The authors in \cite{veamy2019} present \texttt{Veamy}, an object-oriented C++ library that adopts the notation of finite element analysis and provides its own polygonal mesh generator. It solves the two-dimensional Poisson and linear elastostatic problems at lowest order and includes a companion finite element module for direct comparison. It has no higher-order, three-dimensional, or scripting capability. Its MATLAB counterpart \texttt{VEMLab} in \cite{vemlab}, developed by the same research group as \texttt{Veamy}, pursues comparable objectives in MATLAB. It implements lowest-order VEM on polygonal elements together with three- and four-node finite elements for comparison. The library provides several stabilization strategies for elastostatics, supports multiple materials in the Poisson problem, runs in Octave, and exports results to GiD, VTK, and ParaView. Like \texttt{Veamy}, it is restricted to two dimensions and lowest order and covers linear elastostatics and the scalar Poisson problem.

\paragraph{} High-level Python interfaces to the VEM already exist in \texttt{dune-vem} and \texttt{PolyDiM}, so Python access alone does not distinguish PoliVEM. No released and maintained VEM library, however, combines such an interface with structural-mechanics coverage beyond linear elastostatics. \texttt{dune-vem} targets canonical problems of applied mathematics, while \texttt{PolyDiM} stops at small-strain linear elasticity. The engineering-oriented packages \texttt{Veamy}, \texttt{VEMLab}, and \texttt{mVEM} provide no high-level Python layer and, apart from the plate elements of \texttt{mVEM}, do not advance past linear elastostatics. Structural formulations such as beams, axisymmetric elasticity, and finite-strain hyperelasticity remain available mainly as research code accompanying individual method papers.

\paragraph{} PoliVEM collects previously published formulations for one-dimensional beams, higher-order two- and three-dimensional elasticity, axisymmetric elasticity, transient diffusion, and finite-strain hyperelasticity under one projection--stabilization core. A performance-oriented C++ implementation supplies the numerical kernels, while a high-level Python interface drives the simulations. The mathematical formulations were published separately; this work contributes their unification, accessibility, and packaging for structural mechanics.

\section{The unified VEM core}
\label{sec:unified_vem_core}

\paragraph{} The formulations implemented in this work address one-dimensional beams, two- and three-dimensional elasticity, axisymmetric elasticity, transient diffusion, and finite-strain hyperelasticity through the same set of virtual element ingredients. This section writes that common core in matrix form and shows how every solver uses the same projection and stabilization machinery. The construction follows the foundational virtual element literature \cite{beirao2013vem,beirao2014hitchhiker}, the polynomial reconstruction of \cite{ahmad2013}, the higher-order elastic projection of \cite{artioli2017}, the vector-valued treatment of \cite{beirao2015inelastic}, the one-dimensional formulations of \cite{wriggers2022,wriggers2023}, and the time-dependent VEM of \cite{vacca2015parabolic}. The matrix form mirrors the implementation and provides the common notation used in the sections that follow.

\subsection{Local spaces and degrees of freedom}
\label{subsec:local_spaces}
\paragraph{} Let $\mathcal{T}_h$ be a decomposition of the geometric domain $\Omega$ into non-overlapping simple polytopes. The approximation and stability properties of the Virtual Element Method hold under geometric regularity requirements on the partition. The two-dimensional condition is stated explicitly below. In three dimensions, analogous star-shapedness assumptions are imposed on each polyhedron and its polygonal faces, together with lower bounds on face and edge sizes.

\begin{defin}[Admissible mesh family]\label{def:admissible_mesh_family}
A family $\{\mathcal{T}_h\}_h$ of decompositions of $\Omega$ into non-overlapping simple polygons $E$ of diameter $h_E$, with $h = \max_E h_E$, is admissible with parameter $\gamma > 0$ if every element $E$ of every $\mathcal{T}_h$ is star-shaped with respect to a ball of radius at least $\gamma h_E$, and every edge $e \in E$ has length at least $\gamma h_E$.
\end{defin}

\paragraph{} The admissibility condition of Definition \ref{def:admissible_mesh_family}, or its three-dimensional analogue, is assumed throughout. Under these conditions, the stability constants of Section \ref{subsec:projection_and_stab_split} remain independent of the mesh. Each element $E$ has diameter $h_E$, measure $|E|$, centroid $\mathbf{x}_E$, and outward edge or face normals $\mathbf{n}$. For a method of order $k$, the local polynomial space $\mathbb{P}_k(E)$ has dimension
\begin{equation}
  \label{eq:polynomial_space_dim}
  N_k = \begin{cases}
    \frac{(k+1)(k+2)}{2}, \; &\text{in two dimensions}, \\
    \frac{(k+1)(k+2)(k+3)}{6}, \; &\text{in three dimensions},
  \end{cases}
\end{equation}
and is represented in the scaled monomial basis
\begin{equation}
  \label{eq:scaled_monomials}
  m_{\alpha}(\mathbf{x}) = \left( \frac{\mathbf{x} - \mathbf{x}_E}{h_E} \right)^{\bm{\alpha}},
\end{equation}
with $|\bm{\alpha}| \leq k$. Define the complementary monomial space
\begin{equation}
  \mathbb{P}^{\star}_{k-1,k}(E)
  =
  \operatorname{span}\{m_{\alpha}: k-1 \leq |\bm{\alpha}| \leq k\}.
\end{equation}
For a polygon, the local enhanced virtual element space $V_{h,k,E}$ collects functions that are polynomials of degree at most $k$ on each edge, whose Laplacian is a polynomial of degree at most $k$ in the interior, and that satisfy the enhancement constraint of \cite{ahmad2013},
\begin{equation}
  \int \limits_E (\Pi^\nabla_{k,E}v)q\,dE = \int \limits_E vq\,dE,
  \qquad \forall q\in \mathbb{P}^{\star}_{k-1,k}(E),
\end{equation}
where $\Pi^\nabla_{k,E}$ is the energy projection introduced in Section \ref{subsec:projection_operators}. On a polyhedron, the boundary trace is assembled from the corresponding two-dimensional virtual spaces on its faces. The space $\mathbb{P}^{\star}_{k-1,k}(E)$ is a chosen algebraic complement of $\mathbb{P}_{k-2}(E)$ in $\mathbb{P}_k(E)$; it is not assumed to be an $L^2$-orthogonal complement. The interior moments of degree at most $k-2$ are already degrees of freedom through (\ref{eq:degrees_of_freedom}), and the constraint ties the remaining moments, of degrees $k-1$ and $k$, to the energy projection. That projection is computable from the boundary degrees of freedom together with the interior moments of degree at most $k-2$. These data make the full $L^2$ projection onto $\mathbb{P}_k(E)$ computable from the degrees of freedom alone, which is the property exploited by the mass operator and the load term. Functions in $V_{h,k,E}$ are never evaluated in closed form inside the element. They are handled only through a set of degrees of freedom and through polynomial projections built from those degrees of freedom.

\paragraph{} The scalar degrees of freedom are linear functionals
\begin{equation}
  \label{eq:degrees_of_freedom}
  \mathcal{X}_i^V(v) = v(V_i), \quad \mathcal{X}^e_j(v) = \frac{1}{|e|}\int \limits_e v \hat{L}_j ds, \quad \mathcal{X}^E_\alpha (v) = \frac{1}{|E|} \int \limits_E vm_\alpha dE,
\end{equation}
with $V_i$ the vertices of $E$, $\hat{L}_j$ the Legendre polynomial of degree $j=0,\ldots,k-2$ on the edge $e \subset \partial E$, and $|\bm{\alpha}| \leq k - 2$. The framework stores these functionals in a fixed hierarchy: vertex values, edge moments, face moments in three dimensions, and interior moments. This ordering is used in one, two, and three dimensions. It gives every formulation the same local-to-global map structure and allows the assembly stage to be written once. For a vector field, the functionals of (\ref{eq:degrees_of_freedom}) act componentwise. The local scalar count on a polygon with $N_v$ vertices is
\begin{equation}
  \label{eq:number_of_dofs}
  N^{\mathrm{scal}}_E = N_v + N_v(k-1) + \frac{1}{2}k(k-1),
\end{equation}
namely vertex values, edge moments, and interior moments, and the count is $N^{\mathrm{dof}}_E = d N^{\mathrm{scal}}_E$ with $d$ the spatial dimension. In three dimensions a polyhedron with $N_v$ vertices, $N_e$ edges, and $N_f$ faces has the analogous count
\begin{equation}
  \label{eq:number_of_dofs_3d}
  N_E^{\mathrm{scal},\mathrm{3D}} = N_v + N_e(k-1) + N_f \dim \mathbb{P}^{(2)}_{k-2} + \dim \mathbb{P}^{(3)}_{k-2},
\end{equation}
with
\begin{equation}
  \dim \mathbb{P}^{(2)}_{m} = \frac{1}{2} (m+1)(m+2)
\end{equation}
the face-moment count, and
\begin{equation}
  \dim \mathbb{P}^{(3)}_m = \frac{1}{6}(m+1)(m+2)(m+3)
\end{equation}
the interior-moment count, both taken at degree $k-2$ in agreement with the interior moments of (\ref{eq:degrees_of_freedom}). Equations (\ref{eq:number_of_dofs}) and (\ref{eq:number_of_dofs_3d}) size every matrix that follows.

\subsection{Projection operators}
\label{subsec:projection_operators}

\paragraph{} Three projections turn the inaccessible virtual functions into computable polynomials, and each one feeds a different operator of the discrete problem. 

\paragraph{} The energy projection $\Pi^\nabla_{k,E}: V_{h,k,E} \longrightarrow \mathbb{P}_k(E)$ reproduces the gradient and is defined by the orthogonality condition together with a fixing of the constant,
\begin{equation}
  \label{eq:energy_projection}
  \int \limits_E \nabla (\Pi^\nabla_{k,E} v) \cdot \nabla q dE = \int \limits_E \nabla v \cdot \nabla q dE, \quad P_0(\Pi^\nabla_{k,E} v) = P_0(v),
\end{equation}
for all $q \in \mathbb{P}_k(E)$, where $P_0$ is the vertex average for $k=1$ and the element average for $k>1$. The energy projection supplies the lowest-order elastic stiffness and, through its gradient, the deformation gradient of the finite-strain element.

\paragraph{} The strain projection $\Pi^\varepsilon_{k-1,E}:[V_{h,k,E}]^d \longrightarrow [\mathbb{P}_{k-1}(E)]^{d \times d}_{\mathrm{sym}}$ reproduces the symmetric gradient and is the operator that drives the general-order elasticity of \cite{artioli2017}. It is the $L^2$ projection of the strain onto the symmetric polynomial tensors of degree $k-1$,
\begin{equation}
  \label{eq:strain_projection}
  \int \limits_E (\Pi^\varepsilon_{k-1,E} \mathbf{u}) : \mathbf{q} dE = \int \limits_E \bm{\varepsilon} (\mathbf{u}) : \mathbf{q} dE, \; \forall \mathbf{q} \in [\mathbb{P}_{k-1}(E)]^{d \times d}_{\mathrm{sym}},
\end{equation}
and it is computable from the degrees of freedom because integration by parts expresses its right-hand side through boundary traces and the interior moments of (\ref{eq:degrees_of_freedom}). The implementation uses the energy projection (\ref{eq:energy_projection}) for the lowest order $k=1$ and the strain projection (\ref{eq:strain_projection}) for the higher orders $k\geq2$. At $k=1$, the strain projection reduces to the same constant-strain consistency term produced by the energy projection.

\paragraph{} The $L^2$-projection $\Pi^0_{k,E}: V_{h,k,E} \longrightarrow \mathbb{P}_k(E)$ reproduces the field itself and is defined by
\begin{equation}
  \label{eq:l2_projection}
  \int \limits_E (\Pi^0_{k,E} v)q dE = \int \limits_E vqdE, \; \forall q \in \mathbb{P}_k(E).
\end{equation}
It supplies the mass operator and the load, and is essential in the transient setting. The energy and $L^2$ projections reproduce scalar polynomials of degree at most $k$. For a vector polynomial displacement of degree at most $k$, the strain projection reproduces its symmetric gradient, which has degree at most $k-1$. All three projections are computable from the degrees of freedom alone.

\subsection{Matrix realization}
\label{subsec:matrix_realization}

\paragraph{} Each projection is assembled from a few small matrices written in the monomial basis. For the scalar energy projection, define
\begin{equation}
  \mathbf{G}_{ab} = \int \limits_E \nabla m_a \cdot \nabla m_b dE, \quad \mathbf{B}_{ai} = \int \limits_E \nabla m_a \cdot \nabla \psi_i dE, \quad \mathbf{D}_{ia} = \mathcal{X}_i (m_a),
\end{equation}
where $\{ \psi_i \}$ is the canonical basis of $V_{h,k, E}$. Following \cite{beirao2013vem,beirao2014hitchhiker}, the first rows of $\mathbf{G}$ and $\mathbf{B}$ are replaced by the constant-fixing condition $P_0(\cdot)$, which makes the system nonsingular. Integration by parts expresses the entries of $\mathbf{B}$ through boundary traces and, for $k\geq2$, the interior moments. The projection then has two representations,
\begin{equation}
  \label{eq:energy_star_projection_matrix}
  \bm{\Pi}^* = \mathbf{G}^{-1}\mathbf{B},
\end{equation}
with size $N_k \times N^{\mathrm{scal}}_E$, and
\begin{equation}
  \label{eq:energy_projection_matrix}
  \bm{\Pi} = \mathbf{D} \bm{\Pi}^*,
\end{equation}
with size $N^{\mathrm{scal}}_E \times N^{\mathrm{scal}}_E$. Both matrices take the local scalar degree-of-freedom vector that identifies a virtual function. The matrix $\bm{\Pi}^*$ returns the projected field as $N_k$ monomial coefficients, while $\bm{\Pi}$ returns the same field as $N^{\mathrm{scal}}_E$ degree-of-freedom values. They are connected by $\mathbf{D}$, whose entry $\mathbf{D}_{ia} = \mathcal{X}_i(m_a)$ evaluates the $i$-th degree of freedom on the $a$-th monomial and maps monomial coefficients to degrees of freedom. Read from right to left, $\bm{\Pi} = \mathbf{D}\bm{\Pi}^*$ first computes the polynomial coefficients and then maps them back to the degree-of-freedom basis. For a vector field, the scalar construction is applied componentwise. Its block form has $dN_k$ coefficient rows and $N_E^{\mathrm{dof}}=dN_E^{\mathrm{scal}}$ columns.

\paragraph{} The consistency term presented in Section \ref{subsec:projection_and_stab_split} contracts the polynomial operator with $\bm{\Pi}^*$, because the polynomial operator lives in the coefficient space, while the stabilization term acts in the degree of freedom space through $\mathbf{D}$ or $\bm{\Pi}$. The strain projection of (\ref{eq:strain_projection}) is built by the same recipe, with $\mathbf{G}$ and $\mathbf{B}$ replaced by the strain-tensor mass matrix and the strain traction term, giving the coefficient operator $\bm{\Pi}^{\varepsilon *}$. The $L^2$ projection of (\ref{eq:l2_projection}) uses the monomial mass matrix and the enhanced moment matrix,
\begin{equation}
  \label{eq:l2_projection_enhanced_matrices}
  \mathbf{H}_{ab} = \int \limits_E m_a m_b dE, \quad \mathbf{C}^0_{ai} = \int \limits_E m_a \psi_i dE, \quad \bm{\Pi}^{0*} = \mathbf{H}^{-1}\mathbf{C}^0,
\end{equation} 
where the enhancement constraint of \cite{ahmad2013} makes $\mathbf{C}^0$ computable from the degrees of freedom.

\subsection{The projection and stability split}
\label{subsec:projection_and_stab_split}

\paragraph{} The structural feature shared by every formulation is the way the local operator is built from a projection. The projection captures only the polynomial content of a virtual field, so the operator obtained from the projected part alone is rank-deficient on the non-polynomial complement. A stabilization term restores the missing rank without disturbing the polynomial response. The local operator is the sum of a consistency term and a stabilization term,
\begin{equation}
  \label{eq:local_general_stiffness_matrix}
  \mathbf{K}_E = \underbrace{(\bm{\Pi}^{c*})^{\mathrm{T}} \mathbf{A}^{\mathbb{P}}_E\bm{\Pi}^{c*}}_{\text{consistency}} + \underbrace{\mathbf{S}_E}_{\text{stabilization}},
\end{equation}
where $\bm{\Pi}^{c*}$ is the coefficient projection appropriate to the formulation, namely the energy coefficient projection $\bm{\Pi}^*$ at lowest order and the strain coefficient projection $\bm{\Pi}^{\varepsilon *}$ for higher-order elasticity. The matrix $\mathbf{A}^{\mathbb{P}}_E$ is the constitutive operator contracted with the kinematic map and integrated over the polynomial space, so that the consistency term has the same algebraic form in every formulation. It contains the constitutive data of the consistency term; the stabilization may also use material-dependent scaling. For the higher-order elastic formulation, the projection $\bm{\Pi}^{\varepsilon *}$ already targets the strain. Let $\mathbf{Q}_{\varepsilon}$ collect the symmetric strain-polynomial basis functions in matrix form. Then
\begin{equation}
  \mathbf{A}^{\mathbb{P}}_E = \int \limits_E \mathbf{Q}_{\varepsilon}^{\mathrm{T}} \mathbf{C}\mathbf{Q}_{\varepsilon}\,dE
\end{equation}
is the constitutive-weighted mass matrix of that basis, with $\mathbf{C}$ the elasticity matrix in plane stress, plane strain, axisymmetric, or three-dimensional form. For the finite-strain hyperelastic formulation, the projection $\bm{\Pi}^*$ targets the displacement gradient, and the kinematic map instead enters $\mathbf{A}^{\mathbb{P}}_E$ through $\mathbf{B}_{F} = \partial \mathbf{F} / \partial \mathbf{a}$, with $\mathbf{a} = \bm{\Pi}^* \mathbf{u}$, giving
\begin{equation}
  \mathbf{A}^{\mathbb{P}}_E = |E| \mathbf{B}_{F}^{\mathrm{T}} \mathbf{C}(\mathbf{F})\mathbf{B}_{F}
\end{equation}
with $\mathbf{C}(\mathbf{F})$ the consistent material tangent evaluated at the deformation gradient 
\begin{equation}
  \mathbf{F} = \mathbf{I} + \nabla (\Pi^\nabla_{k,E}\mathbf{u})
\end{equation}
of the projected displacement, re-evaluated at the current state of each Newton iteration, with $\mathbf{I}$ the identity matrix. At the level of the consistency contraction, the distinction is whether the kinematic map is included in the projection, as in the strain formulation, or in $\mathbf{A}^{\mathbb{P}}_E$, as in the gradient formulation. Both cases then use $(\bm{\Pi}^{c*})^{\mathrm{T}} \mathbf{A}^{\mathbb{P}}_E \bm{\Pi}^{c*}$. Because the projected deformation gradient is affine in the degrees of freedom, the finite-strain consistency tangent contains only this material term and no separate geometric-stiffness contribution, which follows directly from building $\mathbf{F}$ from the projected displacement.

\paragraph{} The split in (\ref{eq:local_general_stiffness_matrix}) enforces the consistency and stability axioms used in the VEM convergence analysis \cite{beirao2013vem,beirao2014hitchhiker}. Let $a_E(\cdot, \cdot)$ denote the exact local bilinear form and $a_{h,E}(\cdot, \cdot)$ its discrete counterpart represented by $\mathbf{K}_E$. The first axiom is $k$-consistency, which requires the discrete form to reproduce the exact form whenever one of its arguments is a polynomial,
\begin{equation}
  \label{eq:axiom_consistency}
  a_{h,E}(p,v) = a_E (p,v), \quad \forall p \in \mathbb{P}_k(E), \quad \forall v \in V_{h,k,E}.
\end{equation}
The second axiom is stability, which requires the discrete form to be bounded above and below by the exact form,
\begin{equation}
  \label{eq:axiom_stability}
  \alpha_* a_E(v,v) \leq a_{h,E}(v,v) \leq \alpha^* a_E (v,v), \quad \forall v \in V_{h,k,E},
\end{equation}
with constants $\alpha_*, \alpha^* > 0$ independent of $h$ and of the element under the admissibility condition of Definition \ref{def:admissible_mesh_family}. The consistency term in (\ref{eq:local_general_stiffness_matrix}) reproduces the polynomial response. The stabilization term adds a positive contribution on the non-polynomial complement, restoring coercivity without changing that response. The same axioms govern the mass operator, with the discrete mass form and the $L^2$ inner product in place of $a_E(\cdot, \cdot)$ and $a_{h,E}(\cdot, \cdot)$.

\paragraph{} Within this structure, the stabilization $\mathbf{S}_E$ must be symmetric positive semidefinite, vanish on the degrees of freedom of polynomials of degree at most $k$, and be positive definite on the kernel of the projection. On that kernel, it must also satisfy uniform upper and lower bounds relative to the exact local form. Polynomial annihilation preserves (\ref{eq:axiom_consistency}), while the kernel bounds supply (\ref{eq:axiom_stability}). The framework provides two interchangeable instances. The linear solvers use the trace-normalized recipe
\begin{equation}
  \mathbf{S}_E = \alpha_E (\mathbf{I} - \mathbf{D}(\mathbf{D}^{\mathrm{T}}\mathbf{D})^{-1}\mathbf{D}^{\mathrm{T}}),
\end{equation}
where the projector $\mathbf{D}(\mathbf{D}^{\mathrm{T}}\mathbf{D})^{-1}\mathbf{D}^{\mathrm{T}}$ maps onto the polynomial degrees of freedom and therefore makes $\mathbf{S}_E$ vanish on $\mathbb{P}_k(E)$ and the scaling
\begin{equation}
  \alpha_E = \frac{\operatorname{tr}(\mathbf{K}^C_E)}{\operatorname{tr}(\mathbf{I} - \mathbf{D}(\mathbf{D}^{\mathrm{T}}\mathbf{D})^{-1}\mathbf{D}^{\mathrm{T}})},
\end{equation}
with $\mathbf{K}^C_E$ the consistency part. The complementary projector $\mathbf{I}-\mathbf{D}(\mathbf{D}^{\mathrm{T}}\mathbf{D})^{-1}\mathbf{D}^{\mathrm{T}}$ has eigenvalue one on the Euclidean complement of the polynomial degree-of-freedom space, so the trace normalization assigns the scale $\alpha_E$ to those complementary directions. The finite-strain element provides the classical surrogate stabilization built on a sub-triangulation of the polygon \cite{vanHuyssteen2020isotropic,vanHuyssteen2021transverselyisotropic} and the decoupled kernel stabilization of \cite{enabe2026stabilization}. The latter has a displacement-independent tangent that is assembled once at element construction. A single construction flag selects between the two instances.

\paragraph{} The same split governs the mass operator of the transient problems. The $L^2$ projection $\bm{\Pi}^{0*}$ of (\ref{eq:l2_projection_enhanced_matrices}) replaces the energy projection, the polynomial operator becomes the monomial mass $\mathbf{H}$, and the local mass is 
\begin{equation}
  \mathbf{M}_E = (\bm{\Pi}^{0*})^{\mathrm{T}} \mathbf{H} \bm{\Pi}^{0*} + \mathbf{S}^M_E.
\end{equation}
The diagonal lumped mass used by the explicit integrator follows from this expression by row summation, which removes the stabilization contribution and leaves one positive entry per degree of freedom.

\subsection{From the abstract method to the software}
\label{subsec:abstract_to_software}

\paragraph{} A formulation is fixed by three quantities: the polynomial order $k$, which determines the dimension (\ref{eq:polynomial_space_dim}) and the degree of freedom count (\ref{eq:number_of_dofs}); the polynomial operator $\mathbf{A}_E^{\mathbb{P}}$, which carries the constitutive law; and the stabilization instance, which supplies $\mathbf{S}_E$. The projections of (\ref{eq:energy_star_projection_matrix}), (\ref{eq:energy_projection_matrix}), and (\ref{eq:l2_projection_enhanced_matrices}), the assembly of (\ref{eq:local_general_stiffness_matrix}), and the solution stage are independent of the formulation. Table \ref{tab:abstract_to_software} maps these abstract objects to their PoliVEM implementations.

\begin{table}[!htbp]
  \centering
  \renewcommand{\arraystretch}{1.3}
  \setlength{\tabcolsep}{6pt}
  \footnotesize
  \caption{Correspondence between the abstract constructs of the unified VEM
  core of Section~\ref{sec:unified_vem_core} and their realization in the
  software, described by function rather than by specific class or routine
  names.}
  \label{tab:abstract_to_software}
  \begin{tabularx}{\textwidth}{@{}L{5.0cm}Y@{}}
  \toprule
  \textbf{Abstract construct} & \textbf{Software realization} \\
  \midrule
  Polytopal partition $\mathcal{T}_h$ &
  A mesh container that stores the node coordinates and a variable-length element
  connectivity, and exposes the local-to-global degree of freedom map for each
  element. \\

  Degrees of freedom, in the vertex, edge, face, interior ordering &
  The ordered local index list of an element, obtained from the mesh at lowest
  order and extended by a per-formulation map that appends the higher-order edge,
  face, and interior entries. \\

  Energy or strain coefficient projection $\bm{\Pi}^{c*}$ &
  A routine that builds the coefficient projection from the element geometry and
  caches it in the element at construction. The lowest-order and finite-strain
  elements use the energy projection, and the higher-order elastic solvers use the
  strain projection. \\

  $L^2$ projection $\bm{\Pi}^{0*}$ &
  A routine that builds the $L^2$ coefficient projection from the enhanced
  moments, cached by the transient solver during system assembly. \\

  Polynomial operator $\mathbf{A}^{\mathbb{P}}_E$ &
  Assembled from the constitutive description supplied by the material model,
  namely the elasticity matrix $\mathbf{C}$ for the linear solvers and the
  material tangent $\mathbf{C}(\mathbf{F})$ of the strain energy for the
  hyperelastic element. \\

  Consistency term &
  The contraction of the cached coefficient projection with the polynomial
  operator, returning the local consistency stiffness or tangent. It is the
  strain-projection contraction for the higher-order elastic solvers and the
  gradient-projection contraction
  $|E|\,(\bm{\Pi}^{*})^{\mathrm{T}}\mathbf{B}_{F}^{\mathrm{T}}\mathbf{C}\mathbf{B}_{F}\,\bm{\Pi}^{*}$,
  re-evaluated at the current state, for the finite-strain element. \\

  Stabilization $\mathbf{S}_E$ &
  An interchangeable routine selected at element construction, namely the
  trace-normalized $\mathbf{D}$-recipe for the linear solvers and the
  constant-kernel form for the finite-strain element. \\

  Global operator and discrete problem &
  An assembler that owns the persistent global sparse structure and the
  per-element scatter map, together with a solution stage that imposes the
  boundary conditions and drives either a sparse factorization or a Newton
  iteration. \\
  \bottomrule
  \end{tabularx}
\end{table}

\paragraph{} The projection matrices are computed once from the element geometry and cached. The operators (\ref{eq:energy_star_projection_matrix}), (\ref{eq:energy_projection_matrix}), and (\ref{eq:l2_projection_enhanced_matrices}) do not depend on the current state, so every subsequent energy, residual, or tangent evaluation reuses them. The implementation also preserves the consistency--stabilization split of (\ref{eq:local_general_stiffness_matrix}) and exposes the two terms separately. This separation allows the finite-strain application of \cite{enabe2026stabilization} to compare stabilization strategies without changing the consistency machinery. A new formulation supplies its order $k$, polynomial operator $\mathbf{A}^{\mathbb{P}}_E$, and stabilization instance; it inherits the projection, assembly, and solution stages. Section \ref{sec:software_arch} describes the architecture that implements this separation.

\section{Software architecture and usage}
\label{sec:software_arch}

\paragraph{} Every formulation shares the projections, the consistency--stabilization split, and the degree of freedom ordering introduced in Section \ref{sec:unified_vem_core}. The software separates this fixed infrastructure from the formulation-specific element and constitutive operators. This section describes the resulting abstractions, the work required to add a formulation, and the interface used to drive a simulation.

\subsection{Layered architecture}

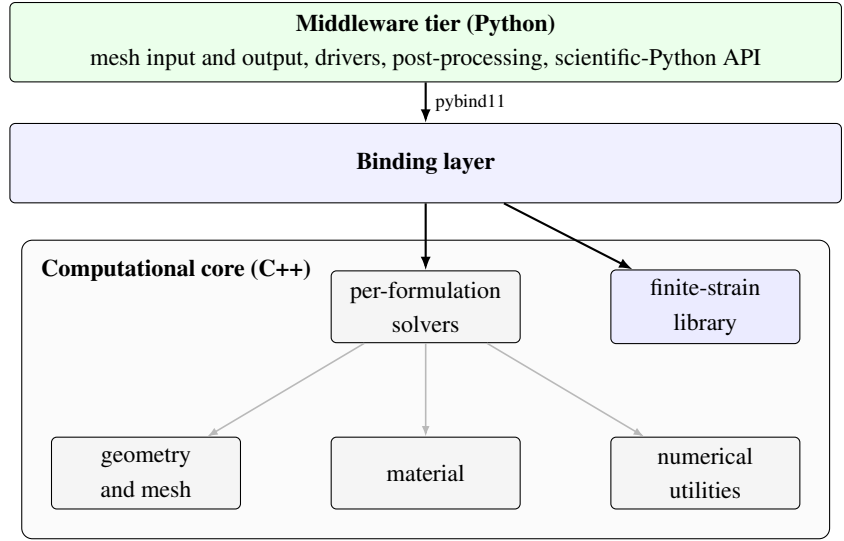
\begin{figure}[htbp]
\centering
\begin{tikzpicture}[
  box/.style={draw, rounded corners=2pt, align=center, font=\footnotesize},
  tier/.style={box, minimum width=11cm, minimum height=1.05cm, font=\small},
  mod/.style={box, minimum width=2.5cm, minimum height=0.95cm, fill=gray!8},
  arrow/.style={-{Latex[length=1.8mm]}, thick},
  dep/.style={-{Latex[length=1.4mm]}, semithick, gray!55}
]
\node[tier, fill=green!8] (mid)  at (0,5.0) {\textbf{Middleware tier (Python)}\\
     mesh input and output, drivers, post-processing, scientific-Python API};
\node[tier, fill=blue!6]  (bind) at (0,3.4) {\textbf{Binding layer}};

\node[mod]              (solv) at (0,1.5)     {per-formulation\\ solvers};
\node[mod, fill=blue!8] (hyp)  at (3.7,1.5)   {finite-strain\\ library};

\node[mod] (mesh) at (-3.7,-0.7) {geometry\\ and mesh};
\node[mod] (mat)  at ( 0,-0.7)   {material};
\node[mod] (util) at ( 3.7,-0.7) {numerical\\ utilities};

\begin{scope}[on background layer]
  \node[draw, rounded corners=4pt, fill=gray!3, inner sep=11pt,
        fit=(solv)(hyp)(mesh)(mat)(util)] (core) {};
\end{scope}
\node[font=\footnotesize\bfseries, anchor=north west]
     at ([shift={(4pt,-4pt)}]core.north west) {Computational core (C++)};

\draw[arrow] (mid)  -- node[right,font=\scriptsize]{pybind11} (bind);
\draw[arrow] (bind) -- (solv);
\draw[arrow] (bind) -- (hyp);

\draw[dep] (solv) -- (mesh);
\draw[dep] (solv) -- (mat);
\draw[dep] (solv) -- (util);
\end{tikzpicture}
\caption{Two-tier architecture. A Python middleware sits on a C++ computational
core through a binding layer. The core is decomposed into role modules with an
acyclic dependency graph, and the per-formulation solvers and the finite-strain
library are the components exposed to the scripting tier.}
\label{fig:architecture_tiers}
\end{figure}

\paragraph{} The framework has two tiers with a one-way dependency. The C++ computational core owns the main data structures and performs mesh handling, projection construction, element evaluation, global assembly, and algebraic solution. The Python middleware exposes this core to scripts, validates and transforms inputs, and provides mesh input, output, and post-processing utilities. The middleware depends on the core; the core has no Python dependency. It can therefore be compiled and benchmarked from a standalone program, while the numerical experiments use Python scripts that import the compiled bindings. Figure \ref{fig:architecture_tiers} shows the two tiers.

\paragraph{} The core itself is decomposed into modules with disjoint responsibilities and an acyclic dependency graph. A geometry module owns the mesh data structures and the mesh generators. A material module provides the constitutive description. A numerical utilities module collects the quadrature rules, the boundary condition handling, the time integration schemes, and the logging facilities. A linear algebra module collects the dense factorizations and the matrix helpers used by the solution stage. A solver module hosts the per-formulation solver components, and a dedicated library hosts the finite-strain hyperelastic infrastructure, which is kept separate because it carries an automatic differentiation dependency and a specialized stabilization machinery that should not propagate to the rest of the core. The binding layer sits above the core and aggregates the modules into a single Python package.

\paragraph{} The project uses CMake and the C++17 standard. A single build description targets Linux, macOS, and Windows and detects the optional libraries available on each system. This removes the need for platform-specific build scripts and lets the same source select its dependency-backed features at configuration time.

\paragraph{} Dense and sparse linear algebra rely on Eigen. When CHOLMOD is available, the solver uses its supernodal Cholesky factorization for large symmetric positive definite systems; otherwise it falls back to Eigen. Shared-memory parallelism uses OpenMP. The automatic differentiation dependency is fetched at configuration time and linked only to the finite-strain library. Isolating the optional dependencies keeps the remaining core build unchanged across the supported platforms.

\subsection{Core abstractions}

\paragraph{} The core is built from five abstractions that map onto the mathematical objects of Section \ref{sec:unified_vem_core}, namely the mesh, the material, the element, the assembler, and the solver. The mesh abstraction stores the node coordinates and a variable-length connectivity that accommodates polygons and polyhedra of arbitrary vertex count without padding, and it exposes the local-to-global degree of freedom map that every later stage consumes. The material abstraction provides the constitutive description independently of the geometric setting, and it builds the elasticity operator of the linear formulations and the strain energy and its derivatives of the finite-strain formulation, the latter obtained either analytically or through automatic differentiation.

\paragraph{} The element abstraction realizes the virtual element construction in software. At construction, it uses the vertex coordinates and material data to compute and cache the projection matrix and the geometric quantities required by the consistency and stabilization terms. It then evaluates the local energy, residual, and tangent from the local degree of freedom vector. The consistency and stabilization contributions remain separate, following Section \ref{subsec:projection_and_stab_split}, so a different stabilization can be selected without changing the consistency evaluation.

\paragraph{} The assembler owns the element objects and persistent sparse data structures. It scatters the local contributions through the degree of freedom map to assemble the global energy, residual, and tangent. The solver hides the algebraic backend behind a small interface. For a linear problem, it imposes the essential boundary conditions and calls the sparse factorization matched to the operator. For a nonlinear problem, it drives Newton iterations through the assembler. For a transient problem, it advances the solution with an explicit or implicit scheme built on the same assembler interface.

\begin{figure}[!htbp]
\centering
\begin{tikzpicture}[
  cls/.style={draw, rounded corners=1pt, rectangle split, rectangle split parts=2,
              rectangle split part fill={blue!10, white},
              align=left, font=\scriptsize, text width=3.5cm},
  own/.style={-{Diamond[length=2.4mm,open]}, thick},
  use/.style={-{Latex[length=1.8mm]}, thick, dashed}
]
\node[cls] (mat)  at (-4.0, 2.9)
  {\textbf{Material}\nodepart{second} constitutive parameters\\ elasticity operator\\ strain energy and derivatives};
\node[cls] (mesh) at ( 1.6, 2.9)
  {\textbf{Mesh}\nodepart{second} node coordinates\\ element connectivity\\ local-to-global DOF map};
\node[cls] (elem) at (-4.0,-0.7)
  {\textbf{Element}\nodepart{second} cached projection\\ consistency term\\ stabilization term\\ energy, residual, tangent};
\node[cls] (asm)  at ( 1.6,-0.7)
  {\textbf{Assembler}\nodepart{second} owns the elements\\ persistent sparse structure\\ scatter map\\ global energy, residual, tangent};
\node[cls] (slv)  at (-1.2,-4.6)
  {\textbf{Solver}\nodepart{second} boundary conditions\\ sparse factorization\\ Newton iteration\\ time integration};

\draw[use] (mat.south)  -- (elem.north);
\draw[use] (mesh.south) -- (asm.north);
\draw[own] (asm.west)   -- (elem.east);
\draw[use] (slv.north east) -- (asm.south west);
\end{tikzpicture}
\caption{Class structure of the core. Open diamonds denote ownership and dashed
arrows denote use. The assembler owns the per-element objects and consumes the
mesh and the material. The element consumes the material and caches the
projection. The solver consumes the assembler and the boundary condition data and
returns the global solution. The abstractions are combined by composition, and no
formulation derives from a common base class.}
\label{fig:core_classes}
\end{figure}
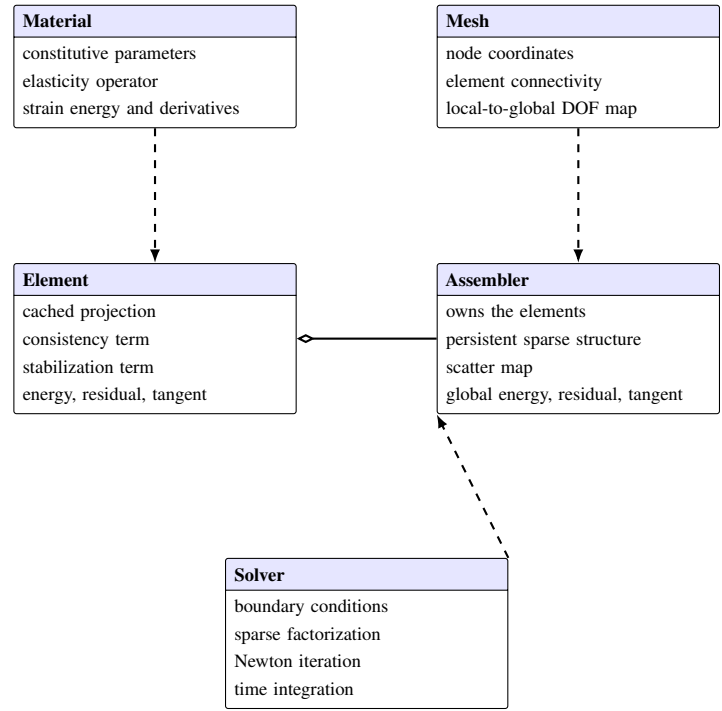

\paragraph{} The five abstractions use composition. Every formulation combines mesh, material, element, assembler, and solver components without deriving from a common element or solver base class. The interfaces differ by formulation: a linear element exposes stiffness and mass matrices, while a finite-strain element exposes an energy, residual, and tangent evaluated at the current state. A common base class would either impose the richer interface on the linear formulations or require an adapter between the two. Figure \ref{fig:core_classes} shows the ownership and use relations.

\subsection{Mesh generation and import}
\label{subsec:mesh_generation_and_import}

\paragraph{} Every solver receives a mesh described by node coordinates and variable-length connectivity. The accompanying metadata records named boundary sets and, when relevant, the geometric parameters and material data of the problem. A mesh can be imported from disk or generated inside the computational core.

\paragraph{} Two file formats are supported for import. The text format uses a fixed JSON schema containing metadata, nodes, elements, boundary sets, and any application-specific problem data. The core reads this format directly, while the middleware reads it into a structured object of NumPy arrays. The binary HDF5 format is intended for larger three-dimensional polyhedral meshes and stores the vertices, element connectivity, and one dataset per named boundary.

\paragraph{} Both formats follow a fixed, documented schema. The JSON schema contains a metadata block, a node list, and an element list. The metadata records the mesh type, spatial dimension, domain extent, and mesh-quality measures such as element-size statistics and the connectivity ratio. It can also include the problem data described above. The node list assigns sequential zero-based identifiers and coordinates, and the element list gives the vertex connectivity of each element. The HDF5 layout stores the corresponding content in a metadata group, a vertex dataset, an element-connectivity dataset, and one dataset per named boundary. A self-contained mesh and problem definition can therefore be archived, exchanged between the core and middleware, or produced by an external mesher without conversion.

\paragraph{} Meshes can also be generated inside the core without an external tool. A structured generator produces regular grids on a rectangular domain. A Voronoi generator tessellates a geometric domain, which may be a rectangle, a circle, or a general polygon, into convex polygonal cells (see \cite{aurenhammer1991voronoi, fortune1987sweepline, talischi2012polymesher}), and a Delaunay triangulator produces simplicial meshes on the same domains (see \cite{lee1980delaunay, watson1981delaunay, ruppert1995delaunay}). A dedicated generator produces the one-dimensional meshes of the beam and frame applications. The generated meshes can be exported to the JSON schema and reused, which makes the built-in generators convenient for the polygonal and polyhedral convergence studies of Section \ref{sec:numerical_experiments}, where families of meshes of decreasing size are required. When a mesh is instead produced by an external mesher, it enters the framework through the same JSON schema, so that no change to the solvers is needed to accommodate a new mesh source.

\subsection{Extensibility and code reuse}

\paragraph{} Adding a formulation follows a regular pattern. A formulation-specific solver encapsulates the projection construction, element evaluation, and global assembly. A formulation that shares the dimensional setting and element interface of an existing one can reuse its element and replace only the constitutive operator or stabilization. A different kinematic setting requires a new projection and element. The line counts reported below quantify this range. The solver connects to the assembler interface and reuses the parallel assembly when its operator has the same algebraic structure as an existing one. The binding recipe then exposes the solver to the middleware, and scripting-tier validation drivers test it against an analytical or high-quality reference solution.

\paragraph{} Table \ref{tab:reuse_boundary} separates the infrastructure reused without modification from the code required for a new VEM or FEM application.

\begin{table}[!htbp]
  \centering
  \renewcommand{\arraystretch}{1.3}
  \setlength{\tabcolsep}{6pt}
  \footnotesize
  \caption{Reuse boundary for a new formulation, listing the components
  inherited unchanged from the fixed infrastructure and those written anew for
  each formulation.}
  \label{tab:reuse_boundary}
  \begin{tabularx}{\textwidth}{@{}L{5.5cm}Y@{}}
  \toprule
  \textbf{Component} & \textbf{Treatment for a new formulation} \\
  \midrule
  Mesh data structures and input or output &
  Reused without modification. \\

  Linear algebra and numerical integration utilities &
  Reused without modification. \\

  Boundary condition machinery &
  Reused without modification. \\

  Sparse linear solver selection &
  Reused, with the solver matched to the algebraic structure of the new
  operator. \\

  Nonlinear driver with line search and regularization &
  Reused without modification for nonlinear problems. \\

  Time integration driver and explicit schemes &
  Reused without modification for transient problems. \\

  Default and optimized parallel assembly &
  Reused, activated by a single construction option. \\

  Binding pattern to the scripting tier &
  Recipe followed for the new formulation. \\

  Material data structures and constitutive interface &
  Reused, optionally extended with new parameters. \\

  Projection construction &
  New, formulation specific. \\

  Discrete bilinear form and element evaluation &
  New, formulation specific. \\

  Stabilization &
  New, or selected from the existing strategies. \\

  Solver orchestration &
  New, but composed from the existing assembler, driver, and projection
  components. \\

  Validation drivers and benchmarks &
  New, application specific. \\
  \bottomrule
  \end{tabularx}
\end{table}

\begin{figure}[!htbp]
\centering
\begin{tikzpicture}[
  box/.style={draw, rounded corners=2pt, minimum width=5.0cm, minimum height=0.7cm,
              align=center, font=\footnotesize},
  reused/.style={box, fill=gray!12},
  newcode/.style={box, fill=blue!12},
  hdr/.style={font=\small\bfseries},
  arrow/.style={-{Latex[length=1.8mm]}, thick, gray}
]
\node[hdr] at (-3.9,4.3) {Reused infrastructure};
\node[hdr] at ( 3.9,4.3) {Formulation specific};
\node[reused] (r1) at (-3.9,3.4) {mesh and input or output};
\node[reused] (r2) at (-3.9,2.5) {material and constitutive interface};
\node[reused] (r3) at (-3.9,1.6) {numerical utilities and integration};
\node[reused] (r4) at (-3.9,0.7) {binding pattern};
\node[reused] (r5) at (-3.9,-0.2) {assembler with parallel scatter};
\node[reused] (r6) at (-3.9,-1.1) {sparse linear solvers};
\node[reused] (r7) at (-3.9,-2.0) {nonlinear driver and time integration};
\node[newcode] (n1) at (3.9,3.4) {projection construction};
\node[newcode] (n2) at (3.9,2.5) {discrete bilinear form and element};
\node[newcode] (n3) at (3.9,1.6) {stabilization};
\node[newcode] (n4) at (3.9,0.7) {solver orchestration};
\node[newcode] (n5) at (3.9,-0.2) {validation driver};
\node[font=\footnotesize\itshape, align=center] at (0,-2.9)
     {connected through the assembler and solver interfaces};
\draw[arrow, dashed] (r5.east) -- (n4.west);
\draw[arrow, dashed] (r7.east) -- (n4.west);
\end{tikzpicture}
\caption{Extension surface. The reused infrastructure on the left covers every
component that does not depend on the physical problem. The formulation-specific
code on the right is restricted to the projection, the discrete bilinear form and
element evaluation, and the stabilization, together with the orchestration and
validation. The two sides communicate only through the assembler and solver
interfaces.}
\label{fig:extension_surface}
\end{figure}
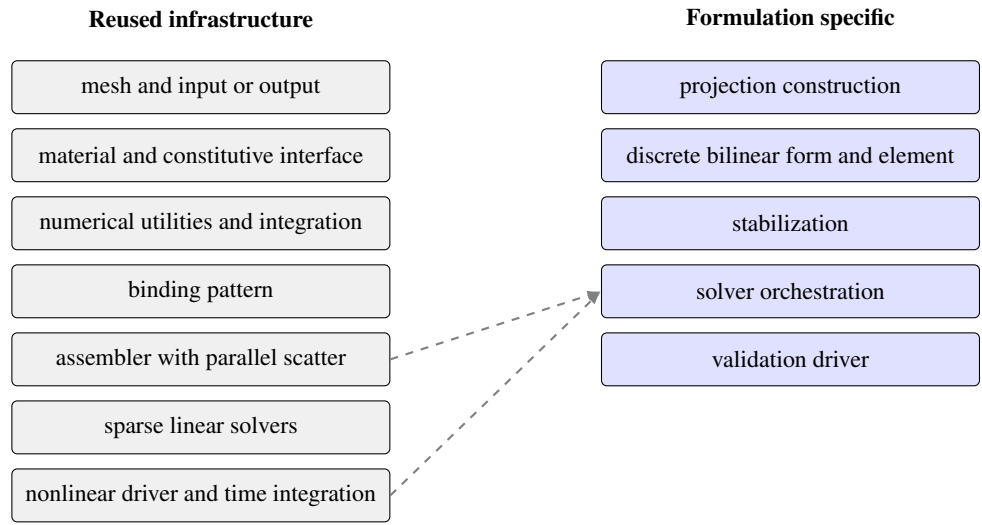

\paragraph{} Formulation-specific code is concentrated in the projection construction, the discrete bilinear form and element evaluation, and the stabilization. Mesh handling, nonlinear iteration, and optimized parallel assembly remain shared infrastructure. Figure \ref{fig:extension_surface} shows this boundary and the assembler and solver interfaces that connect its two sides. A finite element application follows the same division: it reuses the mesh, material, assembler, sparse solvers, and time integration, but supplies its own element evaluation and, when needed, a projection. The two finite element comparison libraries were developed through these interfaces without changes to the shared infrastructure.

\paragraph{} Each formulation exposes its C++ classes through a small binding function with a common signature. The function receives the target scripting module and attaches the formulation's classes to it. A central unit creates the public submodules and calls each binding function inside a protective block, so a failure in one binding does not corrupt the rest of the module. The bindings compile independently, follow the same procedure, and can be tested in isolation from a minimal script.

\paragraph{} Table \ref{tab:code_sizes} reports indicative source sizes, measured as code-only lines with comments and blank lines excluded. The shared count includes only the linked mesh, material, numerical, binding, and mesh-loading components on the assembly and solution path. The formulation-specific count ranges from about two hundred lines for a formulation that reuses an existing dimensional setting to about fourteen hundred lines for one that introduces new machinery.

\begin{table}[!htbp]
  \centering
  \renewcommand{\arraystretch}{1.3}
  \setlength{\tabcolsep}{6pt}
  \footnotesize
  \caption{Indicative source sizes, measured as code-only lines with comments
  and blank lines excluded, for the shared infrastructure and for the
  formulation-specific code of each solver.}
  \label{tab:code_sizes}
  \begin{tabularx}{\textwidth}{@{}YL{6.0cm}@{}}
  \toprule
  \textbf{Code group} & \textbf{Indicative size (code-only lines)} \\
  \midrule
  Shared infrastructure reused by all formulations & $\sim$7700 \\
  One-dimensional Euler--Bernoulli beam            & $\sim$710 \\
  2D linear elasticity, lowest order               & $\sim$360 \\
  2D linear elasticity, higher order               & $\sim$850 \\
  3D linear elasticity, lowest order               & $\sim$300 \\
  3D linear elasticity, higher order               & $\sim$1320 \\
  Axisymmetric elasticity                          & $\sim$230 \\
  Transient diffusion                              & $\sim$1410 \\
  Finite-strain hyperelasticity                    & $\sim$1270 \\
  \bottomrule
  \end{tabularx}
\end{table}

\paragraph{} The axisymmetric solver adds roughly two hundred lines of formulation-specific code to the two-dimensional infrastructure: the radial weight in the bilinear form and the cylindrical correction in the projection. The transient solver is the largest extension because it adds time integration to the assembler, while reusing the mesh, material, projection, and assembly components. Once the projection and local energy of a formulation are specified, its solver can be implemented, exposed to the scripting tier, and validated without rewriting the supporting infrastructure.

\subsection{Driving a simulation from Python}

\paragraph{} A complete simulation passes data between the scripting and computational tiers. The scripting side loads the mesh into a structured object that holds the node coordinates, connectivity, and metadata, using the JSON import of Section \ref{subsec:mesh_generation_and_import}. It then constructs the material and solver and specifies the boundary conditions. The computational core assembles the global operator through the binding layer. The algebraic system can then be solved either by a core solver policy or by a scientific-Python routine, depending on the application. Listing~\ref{lst:driving_python} shows the latter pattern for a two-dimensional linear elastic problem: the mesh is read as NumPy arrays, the solver is constructed for a chosen order, the supports and loads are prescribed, and SciPy solves the assembled operator.

\begin{lstlisting}[style=pythonstyle,
  caption={Driving a two-dimensional linear elastic simulation from the Python
  scripting tier. The mesh enters as NumPy arrays, the solver and the operators
  are built inside the C++ core through the binding layer, and the solution is
  returned to the scripting side for post-processing.},
  label={lst:driving_python}]
import numpy as np
from scipy.sparse import csc_matrix
from scipy.sparse.linalg import spsolve
import polivem_py as pv
from polivem.mesh_io import load_mesh

# 1. Load a polygonal mesh as NumPy arrays
mesh = load_mesh("data/cooks_membrane.json")

# 2. Build the constitutive operator and a virtual element solver of order k
mat = pv.material.Material(); mat.setElasticModule(250.0); mat.setPoissonCoef(0.3)
C = mat.build2DElasticity()
solver = pv.solver.LinearElastic2DSolver(mesh.nodes, mesh.elements_as_lists, 1)  # order k = 1

# 3. Prescribe supports and loaded edges
solver.set_support(support)            # constrained degrees of freedom
solver.set_load(load_edges)            # loaded boundary edges

# 4. Assemble the stiffness and the load vector in the C++ core
K = solver.apply_dirichlet_bc(solver.build_global_stiffness(C))
f = solver.apply_neumann_bc(qx, qy)    # load vector from the loaded edges

# 5. Solve with the SciPy sparse stack; u is a NumPy array of nodal displacements
u = spsolve(csc_matrix(K), f)
\end{lstlisting}

\paragraph{} Every benchmark follows this high-level sequence, but the algebraic backend varies with the problem. Scripting workflows can use SciPy, the beam test uses NumPy's dense solver, and the large three-dimensional systems use the core CHOLMOD policy. The applications also differ in their instantiated formulation, boundary conditions, and post-processing.

\paragraph{} The scripting interface follows the conventions of the scientific Python ecosystem. Inputs are accepted as NumPy arrays and Python lists, while configuration is passed through keyword arguments and structured records. Where practical, sparse matrices and arrays follow the conventions of SciPy and NumPy. The middleware also provides helpers for mesh input and output, relative-error calculations against analytical solutions, distributed loads, and visualizations of deformed configurations and contour fields. An early-stage layer exposes the scripting interface through a REST service. Because this layer has not yet been validated, it is not used in the numerical experiments.

\section{Numerical implementation}
\label{sec:numerical_implementations}

\paragraph{} This section describes the numerical procedures that dominate the cost of a simulation: projection construction, global assembly, linear and nonlinear solution, and time integration.

\subsection{Projection operator construction}
 \paragraph{} The projection is the first numerical object computed for every element. In the lowest-order two-dimensional vector setting on a polygon with $N_v$ vertices, the block coefficient projection has size $6 \times 2N_v$ and maps the local degree-of-freedom vector to six coefficients: two translations and four components of the projected gradient. At higher orders, the number of coefficient rows grows with the dimension of the vector polynomial space, but the construction is structurally identical.

 \paragraph{} Algorithm~\ref{alg:projection_construction} summarizes the construction. It proceeds in three phases. The first phase processes the element geometry to obtain the centroid, the diameter, the signed area, and the outward edge normals and lengths. The second phase builds the scaled monomial basis around the centroid. The third phase assembles the small linear system of Section \ref{subsec:matrix_realization}. Integration by parts expresses its right-hand side as boundary terms, evaluated edge by edge with one-dimensional Gaussian quadrature, and an interior term involving $\Delta q$. The latter vanishes at $k=1$ and is evaluated from the interior moments at higher orders. The constant-fixing row is then imposed. The resulting system has a small fixed size for a fixed polynomial order and is solved by a dense factorization. The projection is computed once during element construction and cached because it depends only on the geometry, not on the current state. Every subsequent evaluation of the energy, residual, or tangent therefore reuses it without recomputation.

\begin{algorithm}[H]
\DontPrintSemicolon
\KwIn{vertex coordinates of the element, order $k$}
\KwOut{coefficient projection matrix $\bm{\Pi}^*$ and geometric cache}
\BlankLine
\textbf{Phase 1. Geometry}\;
\quad compute the centroid, the diameter, and the signed area\;
\quad compute the outward unit normal and the length of every edge\;
\textbf{Phase 2. Monomial basis}\;
\quad build the scaled monomials of degree at most $k$ about the centroid\;
\textbf{Phase 3. Projector assembly}\;
\For{all monomials $a,b$ of degree at most $k$}{
  set $\mathbf{G}_{ab} = \int_E \nabla m_a \cdot \nabla m_b \, dE$\;
}
initialize the right-hand side $\mathbf{B}$ to zero\;
\For{each monomial $q$ of degree at most $k$}{
  \For{each edge $e$ of the element}{
    add the boundary contribution of $q$ on $e$ to $\mathbf{B}$ by 1D Gauss quadrature\;
  }
  \If{$\Delta q \neq 0$}{
    add the interior contribution involving $-\int_E v_h\Delta q\,dE$ from the element moments\;
  }
}
replace the first row of $\mathbf{G}$ and of $\mathbf{B}$ by the constant-fixing condition\;
solve $\mathbf{G}\,\bm{\Pi}^* = \mathbf{B}$ by a dense factorization\;
\Return $\bm{\Pi}^*$ and the geometric cache\;
\caption{Projection construction for one element.}
\label{alg:projection_construction}
\end{algorithm}

\paragraph{} The construction cost is set by the dense algebra on the local system. Let $N_p$ denote the number of polynomial coefficients in the relevant projection: $N_p=N_k$ for a scalar field and $N_p=dN_k$ for its componentwise vector block form. Forming and factorizing the $N_p \times N_p$ matrix costs $\mathcal{O}(N_p^3)$, solving for the $N_E^{\mathrm{dof}}$ columns of the right-hand side costs $\mathcal{O}(N_p^2N_E^{\mathrm{dof}})$, and mapping the result to the degree-of-freedom space costs $\mathcal{O}(N_p(N_E^{\mathrm{dof}})^2)$. Because this work is performed once per element and cached, the setup cost is $\mathcal{O}(\sum_E N_p(N_E^{\mathrm{dof}})^2)$. It is linear in the number of elements when the approximation order and the number of local entities are uniformly bounded. The cached projection adds no reconstruction cost to subsequent energy, residual, and tangent evaluations.

\subsection{Sparse assembly and coloured parallel scatter}
\label{subsec:sparse_assembly}

\paragraph{} Assembly of the global operator is the performance-critical stage of problems that repeatedly assemble the same mesh, such as Newton iterations and transient analyses. The default path collects the local contributions as triplets and rebuilds the global sparse matrix at every call through the standard triplet conversion. This path is simple and portable, but each assembly requires internal sorting and duplicate summation. Let
\begin{equation}
  N_{\mathrm{scat}}=\sum_{E\in\mathcal{T}_h}(N_E^{\mathrm{dof}})^2
\end{equation}
denote the number of local matrix entries scattered before duplicates are combined. Each triplet conversion costs $\mathcal{O}(N_{\mathrm{scat}}\log N_{\mathrm{scat}})$ in the comparison model used here, so $M$ successive assemblies cost $\mathcal{O}(M N_{\mathrm{scat}}\log N_{\mathrm{scat}})$.

\paragraph{} The optimized path removes the repeated symbolic work by separating a symbolic phase from a numerical phase, following \cite{krysl2024parallel}. The local-to-global map and the sparsity pattern of the global operator remain fixed with the mesh, while only the numerical values change between assemblies. The symbolic phase determines which entries are nonzero and where each local contribution is added to the value array. It is executed once during construction and builds the compressed sparse structure from the connectivity. For every element and every pair of local degrees of freedom, a binary search in the corresponding column locates the position of the global entry in the value array. These positions form an element-to-slot map. At every assembly, the numerical phase zeros the value array, evaluates the local matrices, and scatters their entries directly into the recorded positions without sorting or allocation. If $n_c$ is the maximum number of stored entries in a compressed column, the slot searches cost $\mathcal{O}(N_{\mathrm{scat}}\log n_c)$, while each numerical phase costs $\mathcal{O}(N_{\mathrm{scat}}+\mathrm{nnz})$. The $M$ assemblies therefore cost $\mathcal{O}(N_{\mathrm{scat}}\log n_c+M[N_{\mathrm{scat}}+\mathrm{nnz}])$. For fixed order and uniformly bounded local topology, these quantities grow linearly with the number of elements; the optimization removes repeated sorting rather than equating local contributions with unique global nonzeros.

\paragraph{} Parallel scatter requires additional care because adjacent elements contribute to common entries through shared degrees of freedom. The framework uses graph colouring to avoid write conflicts. Two elements are adjacent when they share at least one global degree of freedom, and a greedy algorithm with a heaviest-first ordering assigns colours so that adjacent elements have different colours. Elements of the same colour contribute to disjoint entries of the value array and can therefore be scattered concurrently without locks or atomic operations. The numerical phase processes the colours in sequence and parallelizes the element loop within each colour with OpenMP. Because the symbolic phase has a fixed cost that is amortized over repeated assemblies, the framework activates the optimized path only above a mesh-size threshold determined by benchmarking. Below this threshold, it uses a parallel triplet path with thread-local accumulators to avoid the setup overhead. Algorithm~\ref{alg:parallel_assembly} summarizes the procedure.

\begin{algorithm}[!htb]
\DontPrintSemicolon
\KwIn{mesh with fixed local-to-global map}
\KwOut{global sparse operator in compressed form}
\BlankLine
\textbf{Symbolic phase (once, at construction)}\;
\Indp
build the compressed sparsity pattern from the connectivity\;
\ForEach{element, in parallel}{
  \ForEach{pair of local degrees of freedom}{
    locate the value-array slot by binary search and store it\;
  }
}
build the element adjacency graph on shared degrees of freedom\;
colour the graph greedily with a heaviest-first ordering\;
\Indm
\textbf{Numerical phase (every assembly)}\;
\Indp
zero the value array\;
\ForEach{colour, in sequence}{
  \ForEach{element of the colour, in parallel}{
    evaluate the local matrix\;
    scatter its entries into the stored slots\;
  }
}
\Indm
\caption{Optimized parallel sparse assembly.}
\label{alg:parallel_assembly}
\end{algorithm}

\subsection{Linear solver selection}
\label{subsec:linear_solver_selection}

\paragraph{} Every stationary solve, Newton correction, and implicit time step eventually requires the solution of an algebraic system
\begin{equation}
  \mathbf{A} \mathbf{x} = \mathbf{b}.
\end{equation}
The matrix $\mathbf{A}$ is the stiffness matrix in a linear static problem, the regularized tangent in a Newton iteration, and a combination of the mass and stiffness matrices in an implicit time step. These matrices differ in symmetry, definiteness, sparsity, and size. The solver is therefore selected after assembly, from the known algebraic structure of $\mathbf{A}$, rather than from the element type that produced it.

\paragraph{} Essential boundary conditions are imposed before the algebraic solver is called. Let the subscripts $f$ and $c$ denote free and constrained degrees of freedom, and partition the system as
\begin{equation}
  \left[ 
    \begin{array}{cc}
      \mathbf{A}_{ff} & \mathbf{A}_{fc} \\
      \mathbf{A}_{cf} & \mathbf{A}_{cc}
    \end{array} 
  \right] 
  \left[ 
    \begin{array}{c}
      \mathbf{x}_f \\ 
      \mathbf{x}_c
    \end{array}
  \right] = 
  \left[
    \begin{array}{c}
      \mathbf{b}_f \\
      \mathbf{b}_c
    \end{array}
  \right], \quad \mathbf{x}_c = \overline{\mathbf{x}}_c.
\end{equation}
Elimination gives the reduced problem
\begin{equation}
  \mathbf{A}_{ff} \mathbf{x}_f = \mathbf{b}_f - \mathbf{A}_{fc} \overline{\mathbf{x}}_c.
\end{equation}
The full vector is reconstructed after solving for $\mathbf{x}_f$. This treatment preserves symmetry and positive definiteness when these properties hold on the free subspace. For the homogeneous supports used by the linear benchmarks, the right-hand side correction vanishes. The solvers that expose the assembled operator impose these homogeneous conditions by zeroing the constrained rows and columns, setting each constrained diagonal entry to one, and setting the corresponding right-hand side entries to zero.

\paragraph{} Unlike the row-and-column elimination described above, the integrated uncondensed path of the higher-order three-dimensional solver leaves the off-diagonal entries associated with a constrained degree of freedom unchanged. It uses a large diagonal penalty, which replaces only an existing diagonal entry and therefore leaves the assembled sparse pattern unchanged. For a constrained degree of freedom $j$, the matrix and right-hand side are modified as
\begin{equation}
  A_{jj} \leftarrow \beta, \quad b_j \leftarrow 0, \quad \beta = 10^{10}\max \left\{ 1, \max \limits_i |A_{ii}| \right\}.
\end{equation} 
The scale of $\beta$ is measured against the largest absolute diagonal entry of the assembled matrix. A higher-order edge degree of freedom is constrained when the edge lies on the boundary and both of its end vertices are constrained in that component. A higher-order face-moment degree of freedom is constrained when the face lies on the boundary and all of its vertices are constrained in that component.

\paragraph{} The condensed higher-order path applies these penalties to the complete, uncondensed system before separating the vertex unknowns from the edge, face, and cell moments. This ordering is necessary because constrained edge and face moments belong to the block eliminated during condensation. Penalizing them first modifies that block and hence the Schur complement and the subsequent recovery of the moment unknowns. Applying the conditions only after condensation would account for the constraints on the retained vertex unknowns but not those on the eliminated moments.

\paragraph{} Three direct-solver policies cover the systems produced by the applications. The small systems of the one-dimensional solver use dense factorizations. A general sparse LU factorization is used when the conditions required by Cholesky cannot be assumed, as for a finite-strain tangent along a nonlinear loading path. A sparse Cholesky factorization is used for symmetric positive definite matrices, including the constrained linear-elastic operators. It factors
\begin{equation}
  \mathbf{P} \mathbf{A} \mathbf{P}^{\mathrm{T}} = \mathbf{L} \mathbf{L}^{\mathrm{T}},
\end{equation}
where the permutation $\mathbf{P}$ reduces fill-in. Cholesky stores one triangular factor, uses the symmetry of the operator, and requires no numerical pivoting. LU instead factors a row-permuted general matrix as
\begin{equation}
  \mathbf{P} \mathbf{A} = \mathbf{L} \mathbf{U},
\end{equation}
and introduces pivoting to handle a general nonsingular operator. The factorization algorithms are summarized in Appendix~\ref{ap:direct_solvers}.

\paragraph{} A sparse direct solve has three stages. The symbolic analysis chooses an ordering, builds the elimination tree, predicts the nonzero pattern of the factors, and allocates their storage. The numerical factorization computes the entries of the factors for the current matrix values. Forward and backward substitutions then solve for each right-hand side. For repeated assembly on a fixed mesh with fixed constraints, the matrices retain their sparsity pattern, so the solver can reuse the symbolic analysis even when their values change. The numerical factors can be reused only while both the pattern and the values remain fixed.

\paragraph{} For the largest symmetric positive definite systems, the integrated higher-order three-dimensional solver uses the CHOLMOD supernodal Cholesky implementation described in \cite{chen2008cholmod}. A supernode groups consecutive columns of $\mathbf{L}$ that share their nonzero structure below the diagonal block (see Appendix~\ref{ap:sub_supernodal_method}). The factorization then updates and factors dense blocks using Level-3 Basic Linear Algebra Subprograms (BLAS) kernels, described in Appendix~\ref{ap:blas}, instead of processing isolated sparse columns. The arithmetic count is comparable to that of a column-oriented factorization, but dense kernels obtain better cache reuse and processor utilization. Supernodal execution is therefore most useful once the factors contain blocks large enough to offset the cost of assembling them.

\paragraph{} The status returned by the numerical backend is checked after both factorization and solution. A failed sparse LU factorization indicates a singular or numerically unusable operator. Failure of a Cholesky factorization indicates that the matrix is not numerically positive definite, which can expose an unconstrained rigid mode, an invalid assembly, or an incorrect solver classification. Such a failure is reported rather than silently replacing Cholesky with LU.

\paragraph{} For nested-dissection orderings on regular finite element graphs, the sparse factorization cost scales as $\mathcal{O}(n^{3/2})$ in two dimensions and $\mathcal{O}(n^2)$ in three dimensions. The corresponding factor storage, and hence the cost of a triangular solve, scale as $\mathcal{O}(n\log n)$ and $\mathcal{O}(n^{4/3})$, respectively \cite{george1973nested,davis2016survey}. These estimates depend on the matrix graph rather than on its numerical values. The larger exponents in three dimensions increase the importance of ordering, factor storage, and the supernodal implementation as the system grows.

\paragraph{} The higher-order three-dimensional solve can optionally eliminate its non-vertex degrees of freedom through a global Schur complement. Let $\mathbf{x}_v$ contain the vertex unknowns and let $\mathbf{x}_m$ contain the edge, face, and cell moment unknowns. After the essential conditions have been imposed, the system is partitioned as
\begin{equation}
  \left[
    \begin{array}{cc}
      \mathbf{A}_{vv} & \mathbf{A}_{vm} \\
      \mathbf{A}_{mv} & \mathbf{A}_{mm}
    \end{array}
  \right]
  \left[
    \begin{array}{c}
      \mathbf{x}_v \\
      \mathbf{x}_m
    \end{array}
  \right] = 
  \left[
    \begin{array}{c}
      \mathbf{b}_v \\
      \mathbf{b}_m
    \end{array}
  \right].
\end{equation}
Eliminating the moment block gives
\begin{equation}
  \mathbf{S}_v \mathbf{x}_v = \mathbf{g}_v,
\end{equation}
with
\begin{equation}
  \mathbf{S}_v = \mathbf{A}_{vv} - \mathbf{A}_{vm} \mathbf{A}_{mm}^{-1}\mathbf{A}_{mv}, \quad \mathbf{g}_v = \mathbf{b}_v - \mathbf{A}_{vm} \mathbf{A}_{mm}^{-1}\mathbf{b}_m.
\end{equation}
No inverse is formed explicitly. The implementation factorizes the sparse moment block $\mathbf{A}_{mm}$ and solves against every column of $\mathbf{A}_{mv}$ and against $\mathbf{b}_m$. It then forms $\mathbf{S}_v$ and solves the condensed system with a dense $LDL^{\mathrm{T}}$ factorization. The eliminated unknowns are recovered from
\begin{equation}
  \mathbf{x}_m = \mathbf{A}_{mm}^{-1}(\mathbf{b}_m - \mathbf{A}_{mv} \mathbf{x}_v).
\end{equation}
This condensation is global. Edge and face moments are shared by adjacent polyhedra, so eliminating them independently inside each element would not reproduce the assembled system. At order one the moment block is empty and the standard path is used. At higher order the option reduces the final system to vertex unknowns. With $n_v=\dim(\mathbf{x}_v)$, forming the dense Schur complement requires $\mathcal{O}(n_v^2)$ storage and its dense $LDL^{\mathrm{T}}$ factorization costs $\mathcal{O}(n_v^3)$. The framework therefore exposes condensation as a configuration choice. It is useful when the reduction in unknowns compensates for the multiple solves with the moment block and the dense factorization of $\mathbf{S}_v$.

\subsection{Nonlinear solution}

\paragraph{} The finite-strain problem is written as the stationarity condition of the discrete potential energy. At a load factor $\zeta \in [0,1]$, this potential is
\begin{equation}
  \Phi_h(\mathbf{u};\zeta) = \sum \limits_{E \in \mathcal{T}_h} [U_E^c(\mathbf{u}_E) + U_E^s(\mathbf{u}_E)] - \zeta \mathbf{f}_{\mathrm{ext}}^{\mathrm{T}}\mathbf{u},
\end{equation}
where $U_E^c$ and $U_E^s$ denote the consistency and stabilization energies. Equilibrium requires the gradient of $\Phi_h$ to vanish on the unconstrained degrees of freedom. The resulting out-of-balance residual and tangent are
\begin{equation}
  \mathbf{R}(\mathbf{u}; \zeta) = \mathbf{R}_{\mathrm{int}}(\mathbf{u}) - \zeta \mathbf{f}_{\mathrm{ext}}, \quad \mathbf{K}_{\mathrm{T}}(\mathbf{u}) = \frac{\partial \mathbf{R}}{\partial \mathbf{u}}.
\end{equation}
The internal residual and tangent are obtained by scattering and summing the element contributions $\mathbf{R}_E^c + \mathbf{R}_E^s$ and $\mathbf{K}_E^c + \mathbf{K}_E^s$ according to the local-to-global degree of freedom map.

\paragraph{} For the consistency contribution, the projected deformation gradient is evaluated from the current displacement as
\begin{equation}
  \mathbf{F}_E = \mathbf{I} + \nabla(\Pi^\nabla_{k,E}\mathbf{u}_E).
\end{equation}
Let $W(\mathbf{F})$ denote the strain-energy density of the hyperelastic constitutive model. The element consistency energy is
\begin{equation}
  U_E^c = |E| W(\mathbf{F}_E).
\end{equation}
The first derivative of $W$ with respect to $\mathbf{F}$ supplies the stress used to evaluate the consistency residual,
\begin{equation}
  \mathbf{R}_E^c = \frac{\partial U_E^c}{\partial \mathbf{u}_E},
\end{equation}
while its second derivative is the consistent material tangent $\mathbf{C}(\mathbf{F})$ introduced in Section \ref{sec:unified_vem_core}. The chain rule gives the element tangent in the matrix form
\begin{equation}
  \mathbf{K}_E^c = |E| (\bm{\Pi}^*)^{\mathrm{T}} \mathbf{B}_{F}^{\mathrm{T}} \mathbf{C}(\mathbf{F}_E) \mathbf{B}_{F} \bm{\Pi}^*.
\end{equation}
Because $\mathbf{F}_E$ is affine in the element degrees of freedom, this expression is the complete consistency tangent and no separate geometric-stiffness term is required. The constitutive response is nevertheless re-evaluated at every Newton iteration because both the stress and the material tangent depend on the current deformation.

\paragraph{} The two stabilization strategies described in Section \ref{subsec:projection_and_stab_split} lead to different element-level workloads. The coupled strategy evaluates a nonlinear surrogate energy over a fan triangulation of the polygon. Its stabilization residual and tangent are the first and second derivatives of that energy,
\begin{equation}
  \mathbf{R}_E^s = \frac{\partial U_E^s}{\partial \mathbf{u}_E}, \quad \mathbf{K}_E^s = \frac{\partial^2 U_E^s}{\partial \mathbf{u}_E^2},
\end{equation}
and are computed by automatic differentiation at the current displacement.

\paragraph{} For the decoupled strategy, the stabilization is a quadratic form on the kernel of the VEM projection,
\begin{equation}
  U_E^s = \frac{1}{2} \mathbf{u}_E^{\mathrm{T}} \mathbf{S}_E \mathbf{u}_E, \quad \mathbf{R}_E^s = \mathbf{S}_E \mathbf{u}_E, \quad \mathbf{K}_E^s = \mathbf{S}_E.
\end{equation}
The matrix $\mathbf{S}_E$ depends only on the element geometry, the projection, and the material parameters. It is therefore computed once when the element is constructed and reused throughout the nonlinear analysis. Each subsequent stabilization residual requires only a dense matrix-vector product, while its tangent is inserted directly into the element tangent. The consistency contribution remains nonlinear, so this shortcut removes the repeated differentiation of the stabilization without replacing the Newton iteration itself.

\paragraph{} At Newton iteration $i$, the constrained degrees of freedom are first assigned their prescribed values. Their residual entries are removed from the equilibrium test, and their rows and columns in the tangent system are eliminated. The increment on the remaining degrees of freedom is obtained from
\begin{equation}
  \left[ \mathbf{K}_{\mathrm{T},ff}^{(i)} + \eta_i \mathbf{I} \right] \Delta \mathbf{u}_f^{(i)} = -\mathbf{R}_f^{(i)},
\end{equation}
where the subscript $f$ denotes the free degrees of freedom, $\eta_i \geq 0$ is the diagonal regularization parameter, and
\begin{equation}
  \mathbf{R}^{(i)} = \mathbf{R}(\mathbf{u}^{(i)};\zeta).
\end{equation}
The parameter $\eta_i$ is increased when the sparse factorization fails, when the residual grows markedly, or when successive iterations stagnate. It is reduced after a sufficiently strong decrease in the residual. The shift regularizes nearly singular or indefinite tangent systems that can occur far from equilibrium while recovering the unmodified Newton direction as convergence is approached. The current nonlinear implementation uses a general sparse LU factorization because the tangent is not assumed to remain positive definite along the complete loading path.

\paragraph{} A backtracking line search globalizes the resulting direction. Let $\omega_i \in (0,1]$ denote the step length at iteration $i$, and let $c_{\mathrm{A}} \in (0,1)$ denote the Armijo sufficient-decrease parameter. Beginning with the full Newton step $\omega_i = 1$, the step length is reduced geometrically until
\begin{equation}
  \Phi_h(\mathbf{u}^{(i)} + \omega_i \Delta \mathbf{u}^{(i)};\zeta) \leq \Phi_h(\mathbf{u}^{(i)};\zeta) + c_{\mathrm{A}}\omega_i (\mathbf{R}^{(i)})^{\mathrm{T}}\Delta \mathbf{u}^{(i)}
\end{equation}
is satisfied. Trial configurations with a non-finite potential are rejected. In particular, the logarithmic volumetric energy becomes inadmissible when the determinant of a projected or surrogate deformation gradient is non-positive. The line search therefore also prevents the iteration from accepting element inversions for which the constitutive response is undefined. After accepting the step, the displacement is updated as
\begin{equation}
  \mathbf{u}^{(i+1)} = \mathbf{u}^{(i)} + \omega_i \Delta \mathbf{u}^{(i)},
\end{equation}
and the prescribed values are imposed again to remove any numerical drift.

\paragraph{} The external forces and nonzero prescribed displacements are applied through fixed load increments. At step $s$ of a sequence of $N_s$ increments, the load factor is $\zeta_s = s/N_s$, and the state returned by the preceding step supplies the initial guess. This continuation reduces the distance between successive equilibrium states when the full load lies outside the local convergence region of Newton iteration about the undeformed configuration. Convergence at each load step is monitored through the Euclidean norm of the residual on the free degrees of freedom and the norm of the accepted displacement increment. The implementation records both histories together with the total potential energy, which permits the convergence behaviour and the action of the line search to be inspected after the solve. Algorithm~\ref{alg:incremental_newton} summarizes the procedure.

\begin{algorithm}[!htb]
\DontPrintSemicolon
\KwIn{initial displacement, external load, prescribed values, number of load steps $N_s$}
\KwOut{displacement and convergence history at every load step}
\BlankLine
initialize the displacement with the undeformed state\;
\For{$s \gets 1,\ldots,N_s$}{
  set the load factor $\zeta \gets s/N_s$\;
  scale the external load and prescribed displacement values by $\zeta$\;
  impose the prescribed values on the current displacement\;
  \BlankLine
  \For{each permitted Newton iteration $i$}{
    assemble the internal energy, residual, and tangent\;
    form the out-of-balance residual on the free degrees of freedom\;
    \If{the residual tolerance is satisfied}{
      accept the current state and leave the Newton loop\;
    }
    adapt the diagonal regularization parameter from the residual history\;
    eliminate the constrained rows and columns\;
    solve the regularized tangent system for the Newton increment\;
    \If{the factorization fails}{
      increase the regularization and repeat the iteration\;
    }
    set the trial step length to one\;
    \While{the trial energy is non-finite or violates the Armijo condition}{
      reduce the step length by the backtracking factor\;
    }
    update the displacement with the accepted step\;
    impose the prescribed values again\;
    \If{the increment tolerance is satisfied}{
      accept the current state and leave the Newton loop\;
    }
  }
  store the state and convergence histories\;
  use the returned state as the initial guess for the next load step\;
}
\caption{Incremental Newton solution of the finite-strain problem.}
\label{alg:incremental_newton}
\end{algorithm}

\paragraph{} The cost of one nonlinear iteration consists of the element evaluations, the assembly of the global residual and tangent, and one sparse factorization. With coupled stabilization, the element cost includes an automatic-differentiation Hessian evaluation on every polygon. With decoupled stabilization, this part is replaced by the stored matrix $\mathbf{S}_E$, although the state-dependent consistency tangent and the global factorization remain.

\subsection{Time integration}
\label{subsec:time_integration}

\paragraph{} The time-integration layer starts from the assembled semi-discrete problem
\begin{equation}
  \mathbf{M}\dot{\mathbf{u}}(t) + \mathbf{K}\mathbf{u}(t) = \mathbf{F}(t),
\end{equation}
where $\mathbf{M}$ and $\mathbf{K}$ are the global mass and stiffness matrices, $\mathbf{F}(t)$ is the assembled load vector, and $\mathbf{u}(t)$ contains the global degrees of freedom. The integrator works with these algebraic objects and a load callback. All element calculations finish before the time loop. The construction of the consistent and lumped VEM mass matrices, the positivity of the lumped weights, and the associated stability estimates are given in \cite{enabe2026masslumped}.

\paragraph{} The lumped path replaces $\mathbf{M}$ with the diagonal matrix $\hat{\mathbf{M}}$. After assembly, the active diagonal entries are extracted and their reciprocals are stored, so applying $\hat{\mathbf{M}}^{-1}$ is a componentwise multiplication. Structurally inactive zero-mass degrees of freedom, when present, are added to the constrained set and excluded from this operation. The stiffness matrix remains in sparse compressed form and is assembled once. For a separable load $\mathbf{F}(t)=g(t)\mathbf{F}_0$, the spatial vector $\mathbf{F}_0$ is also assembled once, and only $g(t)$ changes at the Runge--Kutta stages. A general time-dependent load is evaluated through the callback at every stage.

\paragraph{} The transient driver uses the third-order strong-stability-preserving Runge--Kutta scheme (SSP--RK3) described in \cite{gottlieb2001ssp}. With the diagonal lumped mass $\hat{\mathbf{M}}$, one step from $\mathbf{u}^n$ to $\mathbf{u}^{n+1}$ is
\begin{align}
  \mathbf{u}^{(1)}
  &= \mathbf{u}^{n} + \Delta t\,\hat{\mathbf{M}}^{-1}\mathbf{r}_0,
  &\mathbf{r}_0
  &= -\mathbf{K}\mathbf{u}^{n} + \mathbf{F}(t^n), \\
  \mathbf{u}^{(2)}
  &= \frac{3}{4}\mathbf{u}^{n} + \frac{1}{4}\mathbf{u}^{(1)}
     + \frac{1}{4}\Delta t\,\hat{\mathbf{M}}^{-1}\mathbf{r}_1,
  &\mathbf{r}_1
  &= -\mathbf{K}\mathbf{u}^{(1)} + \mathbf{F}(t^n+\Delta t), \\
  \mathbf{u}^{n+1}
  &= \frac{1}{3}\mathbf{u}^{n} + \frac{2}{3}\mathbf{u}^{(2)}
     + \frac{2}{3}\Delta t\,\hat{\mathbf{M}}^{-1}\mathbf{r}_2,
  &\mathbf{r}_2
  &= -\mathbf{K}\mathbf{u}^{(2)} + \mathbf{F}\!\left(t^n+\frac{1}{2}\Delta t\right).
\end{align}
The load is evaluated at $t^n$, $t^n+\Delta t$, and $t^n+\Delta t/2$. The last evaluation therefore occurs at an earlier physical time than the second one, as required by the convex-combination representation of SSP--RK3. The callback uses the supplied stage time rather than an internally incremented time. The transient driver removes homogeneous Dirichlet entries from each residual before applying the inverse mass and reimposes their prescribed values after every stage. This prevents roundoff and intermediate stage updates from changing a constrained degree of freedom.

\paragraph{} The admissible time step depends on both the spatial operator and the Runge--Kutta scheme. The SSP--RK3 stability polynomial is
\begin{equation}
  R_3(z) = 1 + z + \frac{z^2}{2} + \frac{z^3}{6}.
\end{equation}
Let $\{\lambda_j\}_{j=1}^{n_f}$ be the eigenvalues of the active generalized eigenproblem
\begin{equation}
  \mathbf{K}_{ff}\mathbf{v}_j = \lambda_j\hat{\mathbf{M}}_{ff}\mathbf{v}_j,
\end{equation}
where the subscript $f$ denotes the free degrees of freedom and $n_f$ is their number. For the homogeneous linear problem, a fixed step satisfies
\begin{equation}
  \max_{1\leq j\leq n_f}\left|R_3(-\Delta t\,\lambda_j)\right| \leq 1.
\end{equation}
In the work \cite{huang2015stability}, the largest eigenvalue of an explicit parabolic finite element operator is related to the diagonal entries of its stiffness and mass matrices, as well as to the mesh and diffusion tensor. For the mass-lumped VEM operator, the corresponding mesh-independent estimate and its proof are given in \cite{enabe2026masslumped}. The time-integration code uses the resulting diffusion restriction $\Delta t=\mathcal{O}(h^2)$ without repeating that analysis.

\paragraph{} The generic time-integration component accepts a positive $\Delta t$ from the calling driver and does not estimate the spectrum. In the explicit--implicit comparison reported in \cite{enabe2026masslumped}, the driver brackets a stable explicit step through short homogeneous trial integrations and takes a safety fraction of the estimated limit. It then sets
\begin{equation}
  N_t = \left\lceil\frac{T_f-t_0}{\Delta t}\right\rceil,
  \qquad
  \Delta t = \frac{T_f-t_0}{N_t},
\end{equation}
where $t_0$ and $T_f$ are the initial and final times and $N_t$ is the number of steps. This adjustment makes the last step end at $T_f$. Algorithm~\ref{alg:mass_lumped_ssprk3} summarizes the explicit time loop.

\begin{algorithm}[!htb]
\DontPrintSemicolon
\KwIn{lumped mass, stiffness, load callback, initial state, constrained degrees of freedom, $t_0$, $T_f$, selected time step, and output schedule}
\KwOut{state at every requested output time}
\BlankLine
extract and store the reciprocal active diagonal of the lumped mass\;
determine $N_t$ and adjust $\Delta t$ to reach $T_f$ exactly\;
initialize the state with the prescribed initial condition\;
\BlankLine
\For{$n \gets 0,\ldots,N_t-1$}{
  evaluate the first residual at $t^n$ and compute the first stage\;
  impose the prescribed values on the first-stage state\;
  \BlankLine
  evaluate the second residual at $t^n+\Delta t$ and compute the second stage\;
  impose the prescribed values on the second-stage state\;
  \BlankLine
  evaluate the third residual at $t^n+\Delta t/2$ and update the state\;
  impose the prescribed values on the accepted state\;
  \BlankLine
  advance the time and store the state when output is requested\;
}
\caption{Mass-lumped SSP--RK3 time loop.}
\label{alg:mass_lumped_ssprk3}
\end{algorithm}

\paragraph{} The same interface provides backward Euler and Crank--Nicolson for a consistent mass matrix. Their linear systems are, respectively,
\begin{equation}
  \left(\mathbf{M}+\Delta t\,\mathbf{K}\right)\mathbf{u}^{n+1}
  = \mathbf{M}\mathbf{u}^{n} + \Delta t\,\mathbf{F}(t^{n+1}),
\end{equation}
and
\begin{equation}
  \left(\mathbf{M}+\frac{\Delta t}{2}\mathbf{K}\right)\mathbf{u}^{n+1}
  = \left(\mathbf{M}-\frac{\Delta t}{2}\mathbf{K}\right)\mathbf{u}^{n}
  + \frac{\Delta t}{2}\left[\mathbf{F}(t^{n+1})+\mathbf{F}(t^n)\right].
\end{equation}
For a fixed time step and fixed essential constraints, the matrix on the left-hand side is constant. The comparison driver modifies its constrained rows and columns and factors it once before the time loop. Each subsequent backward-Euler step forms a new right-hand side and performs one back-substitution.

\paragraph{} The standalone time-scheme utility provides either a sparse LU factorization or the biconjugate gradient stabilized method (BiCGSTAB) \cite{vanderVorst1992bicgstab}. BiCGSTAB is a Krylov-subspace iterative method for nonsymmetric sparse systems. Each iteration uses sparse matrix-vector products and vector operations instead of forming and storing triangular factors, thereby avoiding factorization fill-in. The implementation uses Eigen's diagonal preconditioner, formed from the diagonal entries of the system matrix, and exposes the convergence tolerance and maximum iteration count; their defaults are $10^{-6}$ and $1000$, respectively. The solve reports failure if the tolerance is not reached within this limit. Its memory requirement is lower than that of sparse LU, but its iteration count depends on the conditioning of the system and the effectiveness of the diagonal preconditioner.

\paragraph{} When a symmetric positive definite transient system is routed through the general solver layer, it can use the Cholesky policy of Section \ref{subsec:linear_solver_selection}. One SSP--RK3 step requires three sparse matrix-vector products with $\mathbf{K}$, three diagonal inverse-mass applications, the stage vector updates, and the load evaluations. With $n$ active degrees of freedom, its algebraic cost is $\mathcal{O}(\mathrm{nnz}(\mathbf{K})+n)$ per step and $\mathcal{O}\!\left(N_t[\mathrm{nnz}(\mathbf{K})+n]\right)$ over the complete run, where $\mathrm{nnz}(\mathbf{K})$ is the number of stored nonzero entries of $\mathbf{K}$. The reciprocal-diagonal setup costs $\mathcal{O}(n)$ and no factorization is required. The implicit path adds one factorization and one sparse back-substitution per step, with the fill-dependent cost given in Section \ref{subsec:linear_solver_selection}. Source assembly adds to these costs when the load cannot be reduced to a preassembled spatial vector.

\section{Numerical experiments}
\label{sec:numerical_experiments}

\paragraph{} The numerical experiments test polynomial reproduction, static condensation, higher-order degree-of-freedom handling, and linear elasticity on polygonal and polyhedral meshes. The three studies cover one-, two-, and three-dimensional discretizations. All numerical data and figures were generated for this paper; no numerical examples from the cited method papers are reused. Those publications supply the formulation details and separate validation of the axisymmetric, transient, and finite-strain solvers. For each test, this section reports the mesh, approximation order, solver settings, and error measures needed to reproduce the result.

\subsection{One-dimensional beam and static condensation}

\paragraph{} The first experiment tests the general-order one-dimensional implementation and its static-condensation path. The element formulation follows the Euler-Bernoulli construction described in \cite{wriggers2022,enabe2025hybrid}. A straight beam of length $L=1$, constant Young modulus $E_Y = 1$, and constant second moment of area $I = 1$ is clamped at both ends and subjected to a uniform transverse load $q_0 = 1$. With upward displacements taken as positive, the boundary-value problem is
\begin{equation}
  \begin{cases}
    E_Y I w^{(4)}(x) = -q_0, \quad x \in (0,L),\\
    w(0) = w'(0) = w(L) = w'(L) = 0.
  \end{cases}
\end{equation}
The exact displacement and the exact midspan displacement are:
\begin{equation}
  w(x) = - \frac{q_0}{24E_Y I}x^2 (L-x)^2, \quad w \left( \frac{L}{2} \right) = - \frac{q_0 L^4}{384 E_Y I} = - \frac{1}{384}.
\end{equation}
The internal bending moments at the supports are both $-q_0 L^2/12$ when sagging moment is positive. The assembled system returns the corresponding generalized rotation reactions $R_{\theta,0} = q_0 L^2 /12$ and $R_{\theta,L}=-q_0L^2/12$. The vertical reactions are $q_0L/2$ at each support.

\paragraph{} The test uses uniform meshes with $n_{\mathrm{el}} = 1,2,4,8$ elements and orders $k=3,4,5$. Here $n_{\mathrm{el}}$ denotes the number of beam elements and is distinct from the edge count used for polygonal and polyhedral meshes in Section \ref{sec:unified_vem_core}. \texttt{PoliVEM} assembles the stiffness matrix and distributed load vector, and the benchmark solves the resulting small system in double precision using NumPy's dense direct solver. No iterative tolerance enters the comparison. On an element of length $l_e$, the local degrees of freedom are the displacement and rotation at both endpoints, followed for $k \geq 4$ by the internal moments
\begin{equation}
  \mathcal{X}^M_{e,j} (w_h) = \frac{1}{l_e^{j+1}} \int \limits^{l_e}_{0} \xi^j w_h(\xi)\,d\xi, \quad j = 0,\ldots,k-4,
\end{equation}
where $\xi$ is the local beam coordinate. The displacement used for error evaluation is reconstructed on each element as the polynomial of degree $k$ determined by these degrees of freedom.

\paragraph{} The complete system contains
\begin{equation}
  n_{\mathrm{full}} = 2(n_{\mathrm{el}}+1) + n_{\mathrm{el}}(k-3)
\end{equation}
unknowns. The moment unknowns are local to individual elements and can be eliminated before solving the global system. Let $\mathbf{d}_N$ collect the nodal displacement and rotation unknowns, and let $\mathbf{d}_M$ collect the internal moments. Partitioning the complete system gives
\begin{equation}
  \left[
    \begin{array}{cc}
      \mathbf{K}_{NN} & \mathbf{K}_{NM} \\
      \mathbf{K}_{MN} & \mathbf{K}_{MM}
    \end{array}
  \right] 
  \left[
    \begin{array}{c}
      \mathbf{d}_N \\
      \mathbf{d}_M
    \end{array}
  \right] = 
  \left[
    \begin{array}{c}
      \mathbf{f}_N \\
      \mathbf{f}_M
    \end{array}
  \right].
\end{equation}
The condensed nodal problem is 
\begin{equation}
  \left( \mathbf{K}_{NN} - \mathbf{K}_{NM} \mathbf{K}_{MM}^{-1}\mathbf{K}_{MN} \right) \mathbf{d}_N = \mathbf{f}_N - \mathbf{K}_{NM} \mathbf{K}_{MM}^{-1}\mathbf{f}_M.
\end{equation}
After the nodal solution has been computed, the eliminated variables are recovered from
\begin{equation}
  \mathbf{d}_M = \mathbf{K}^{-1}_{MM} \left(\mathbf{f}_M - \mathbf{K}_{MN}\mathbf{d}_N\right).
\end{equation}
The condensed system therefore has
\begin{equation}
  n_{\mathrm{cond}} = 2(n_{\mathrm{el}}+1)
\end{equation}
unknowns, independent of $k$. The four clamped degrees of freedom are included in the systems reported below. The corresponding numbers of free unknowns are $n_{\mathrm{full}}-4$ and $n_{\mathrm{cond}}-4$.

The comparison uses the normalized midspan displacement 
\begin{equation}
  \overline{w}_{1/2} = \frac{384 E_Y I}{q_0 L^4} \left| w_h\left( \frac{L}{2} \right) \right|,
\end{equation}
the normalized magnitude of the left support moment
\begin{equation}
  \overline{M}_0 = \frac{12}{q_0 L^2}|R_{\theta, 0}|,
\end{equation}
and the sampled maximum displacement error. Let
\begin{equation}
  x_m = \frac{mL}{400}, \qquad m=0,\ldots,400,
\end{equation}
denote the 401 equally spaced sampling points, including both endpoints. The error is
\begin{equation}
  e^{(401)}_{\infty}
  = \max \limits_{m=0,\ldots,400}
  \left|w_h(x_m)-w(x_m)\right|.
\end{equation}
Table~\ref{tab:beam_field_error} reports this error for each mesh and approximation order.

\begin{table}[!htbp]
  \centering
  \caption{Sampled maximum displacement error for the fixed--fixed beam.}
  \label{tab:beam_field_error}
  \begin{tabular}{@{}cccc@{}}
    \toprule
    $n_{\mathrm{el}}$
      & $e_{\infty}^{(401)}$, $k=3$
      & $e_{\infty}^{(401)}$, $k=4$
      & $e_{\infty}^{(401)}$, $k=5$ \\
    \midrule
    1 & $2.604 \times 10^{-3}$ & $1.389 \times 10^{-17}$ & $2.732 \times 10^{-17}$ \\
    2 & $1.628 \times 10^{-4}$ & $2.168 \times 10^{-18}$ & $1.214 \times 10^{-17}$ \\
    4 & $1.017 \times 10^{-5}$ & $7.373 \times 10^{-18}$ & $4.012 \times 10^{-16}$ \\
    8 & $6.358 \times 10^{-7}$ & $1.010 \times 10^{-16}$ & $9.073 \times 10^{-15}$ \\
    \bottomrule
  \end{tabular}
\end{table}

Except for the one-element $k=3$ solution, the computed $\overline{w}_{1/2}$ differs from one by at most $3.5 \times 10^{-12}$. For the one-element case, $\overline{w}_{1/2} = 0$. The normalized magnitudes of both end moments differ from one by at most $3.1 \times 10^{-12}$, and the two vertical reactions differ from their exact values of $1/2$ by at most $1.2 \times 10^{-12}$.

\begin{figure}[!htbp]
  \centering
  \begin{subfigure}[t]{0.49\textwidth}
    \centering
    \includegraphics[width=\linewidth]{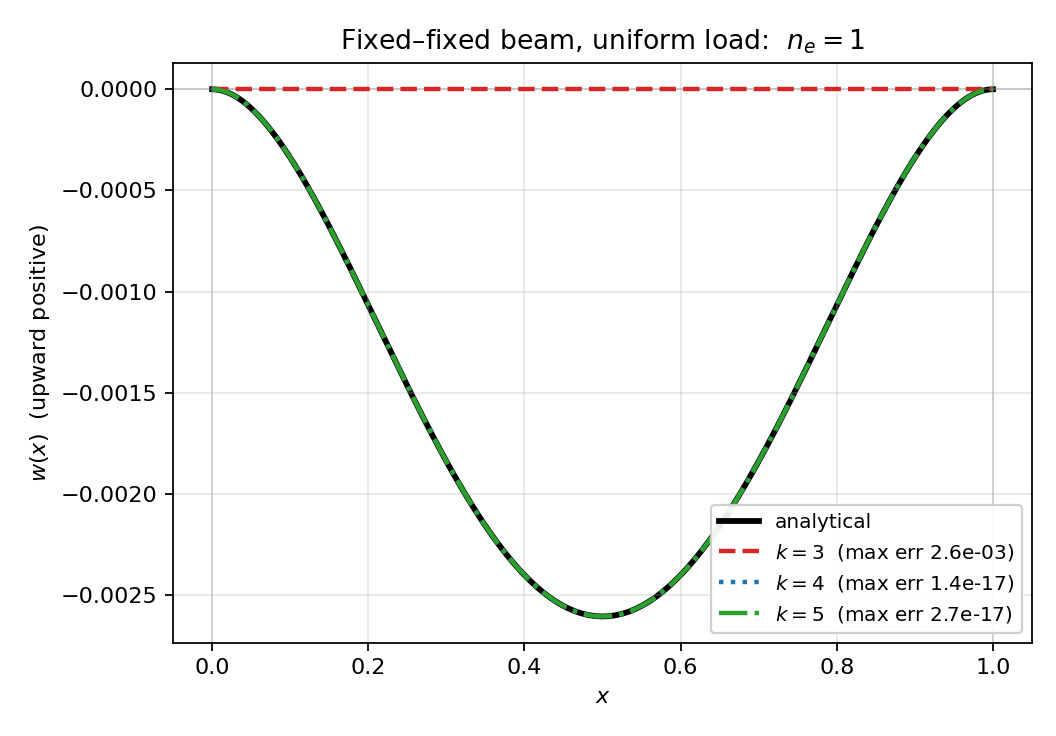}
    \caption{$n_{\mathrm{el}}=1$}
    \label{fig:beam_deflection_ne1}
  \end{subfigure}\hfill
  \begin{subfigure}[t]{0.49\textwidth}
    \centering
    \includegraphics[width=\linewidth]{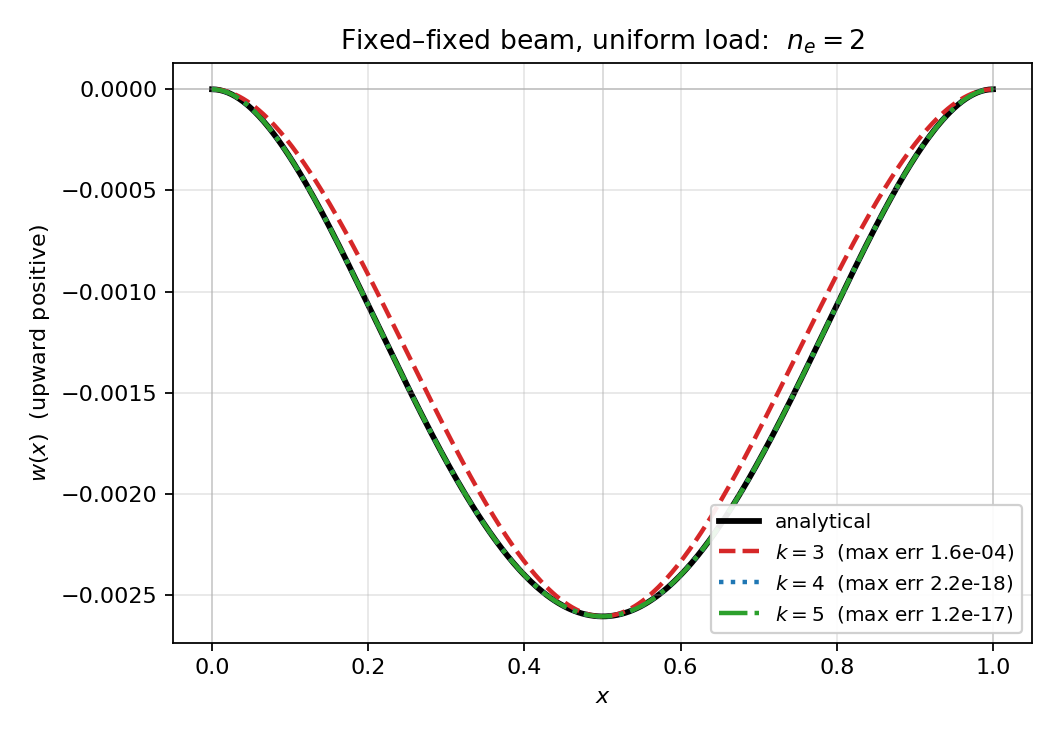}
    \caption{$n_{\mathrm{el}}=2$}
    \label{fig:beam_deflection_ne2}
  \end{subfigure}

  \begin{subfigure}[t]{0.49\textwidth}
    \centering
    \includegraphics[width=\linewidth]{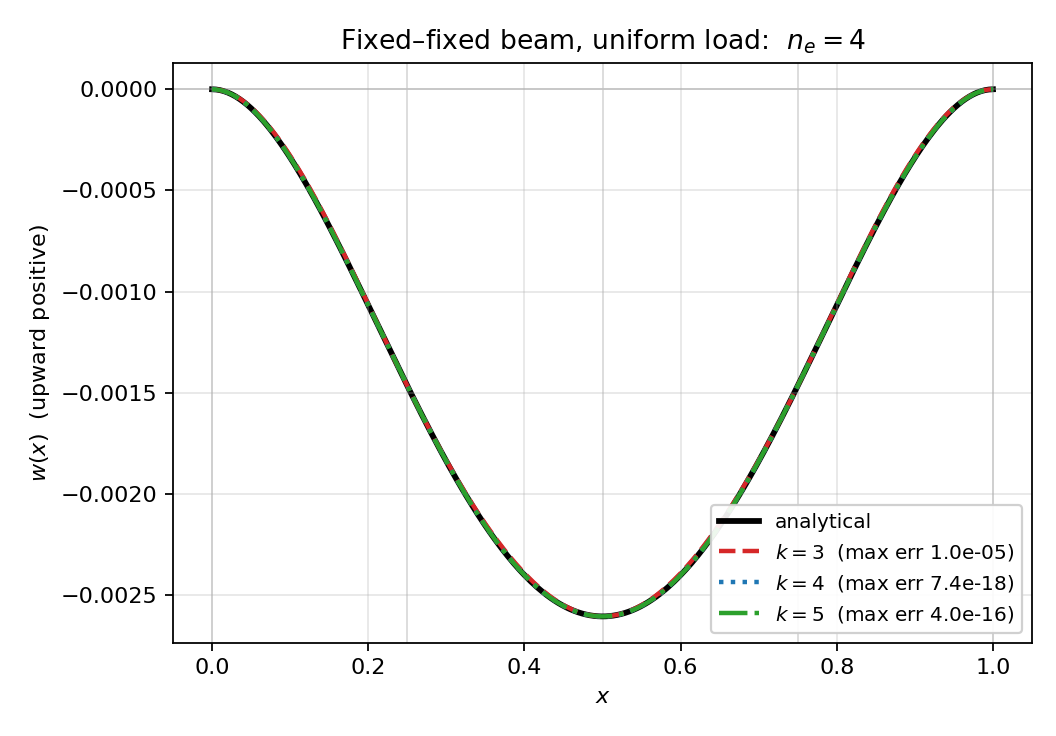}
    \caption{$n_{\mathrm{el}}=4$}
    \label{fig:beam_deflection_ne4}
  \end{subfigure}\hfill
  \begin{subfigure}[t]{0.49\textwidth}
    \centering
    \includegraphics[width=\linewidth]{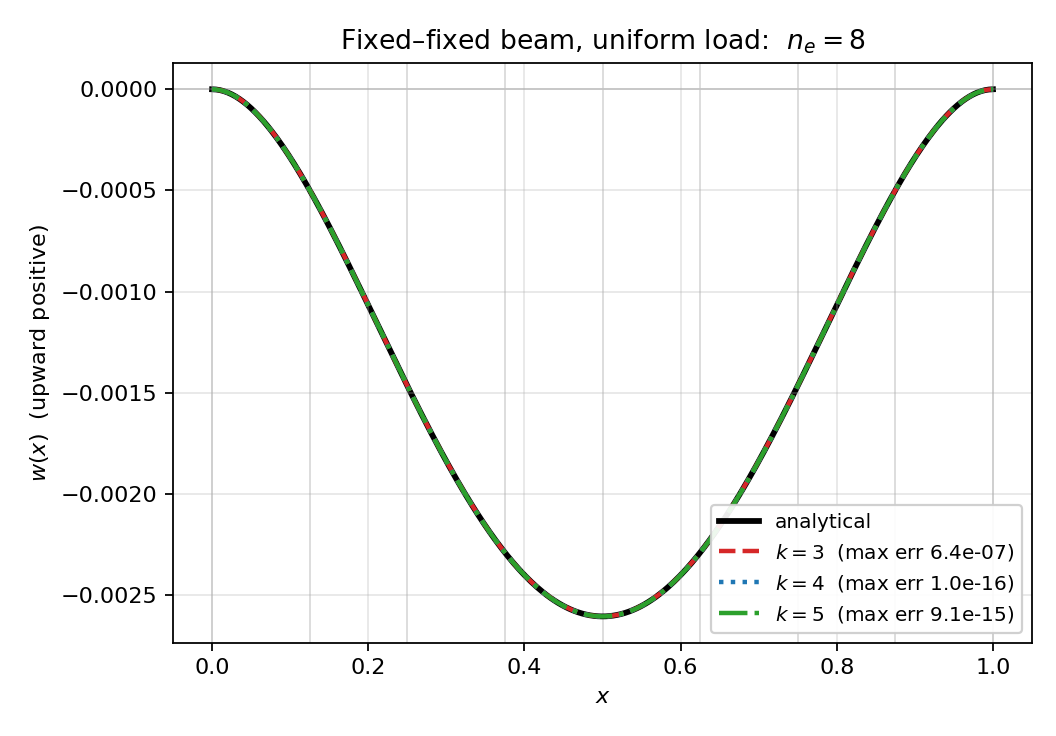}
    \caption{$n_{\mathrm{el}}=8$}
    \label{fig:beam_deflection_ne8}
  \end{subfigure}
  \caption{Reconstructed displacement and analytical solution for the fixed--fixed beam with approximation orders $k=3,4,5$.}
  \label{fig:beam_deflection}
\end{figure}

\paragraph{} The exact displacement is a polynomial of degree four. It belongs to the $k=4$ and $k=5$ approximation spaces, and both orders reproduce the complete displacement field to roundoff on every mesh in Figure~\ref{fig:beam_deflection}. For $k=3$, the one-element problem in Figure~\ref{fig:beam_deflection_ne1} has no free degrees of freedom. On the remaining meshes in Figures~\ref{fig:beam_deflection_ne2}--\ref{fig:beam_deflection_ne8}, the midpoint is a node and its displacement agrees with the exact value to roundoff, while the sampled full-field error in Table~\ref{tab:beam_field_error} decreases by a factor of sixteen whenever the element length is halved. The observed field error is therefore fourth order in the element length.

\paragraph{} Static condensation is checked against a direct solution of the complete system through
\begin{equation}
  \varepsilon_{\mathrm{cond}} = \frac{\| \mathbf{d}_{\mathrm{full}} - \mathbf{d}_{\mathrm{rec}}\|_2}{\| \mathbf{d}_{\mathrm{full}}\|_2},
\end{equation}
where $\mathbf{d}_{\text{rec}}$ contains the condensed nodal solution and the recovered moment variables. Table~\ref{tab:beam_static_condensation} gives the system sizes and the discrepancy for $k=4$ and $k=5$.

\begin{table}[!htbp]
  \centering
  \caption{Complete and condensed numbers of unknowns and the relative discrepancy after recovery of the internal moments. The counts include the four constrained endpoint degrees of freedom.}
  \label{tab:beam_static_condensation}
  \begin{tabular}{@{}ccccc@{}}
    \toprule
    $n_{\mathrm{el}}$
      & $n_{\mathrm{full}}/n_{\mathrm{cond}}$, $k=4$
      & $\varepsilon_{\mathrm{cond}}$, $k=4$
      & $n_{\mathrm{full}}/n_{\mathrm{cond}}$, $k=5$
      & $\varepsilon_{\mathrm{cond}}$, $k=5$ \\
    \midrule
    1 & 5/4  & $0$                      & 6/4   & $0$                      \\
    2 & 8/6  & $2.671 \times 10^{-15}$ & 10/6  & $4.292 \times 10^{-13}$ \\
    4 & 14/10 & $6.499 \times 10^{-16}$ & 18/10 & $3.059 \times 10^{-12}$ \\
    8 & 26/18 & $5.290 \times 10^{-14}$ & 34/18 & $3.950 \times 10^{-11}$ \\
    \bottomrule
  \end{tabular}
\end{table}

For $k=3$, there are no internal moment unknowns, so static condensation leaves the system unchanged. For meshes with 1, 2, 4, and 8 elements, the complete and condensed system sizes are, respectively, 4/4, 6/6, 10/10, and 18/18, with $\varepsilon_{\mathrm{cond}}=0$. For the higher orders, the largest relative discrepancy is $3.950\times10^{-11}$, obtained for \(k=5\) with eight elements. In this case, static condensation reduces the total number of unknowns from 34 to 18. After prescribing the four clamped endpoint degrees of freedom, the corresponding numbers of free unknowns are 30 and 14.

\subsection{Two-dimensional higher-order elasticity}
\begin{figure}[!htbp]
  \centering
  \begin{tikzpicture}[
    x=0.62cm,
    y=0.62cm,
    font=\small,
    boundary/.style={black, very thick},
    symmetry/.style={blue!65!black, very thick},
    traction/.style={
      red!70!black,
      thick,
      -{Latex[length=2.2mm]}
    },
    dimension/.style={
      thin,
      {Latex[length=1.8mm]}-{Latex[length=1.8mm]}
    }
  ]

    \def\plateW{10}
    \def\plateH{10}
    \def\holeR{1}

    \path[
      fill=blue!7
    ]
      (\holeR,0)
      -- (\plateW,0)
      -- (\plateW,\plateH)
      -- (0,\plateH)
      -- (0,\holeR)
      arc[start angle=90,end angle=0,radius=\holeR]
      -- cycle;

    \draw[boundary]
      (\holeR,0)
      -- (\plateW,0)
      -- (\plateW,\plateH)
      -- (0,\plateH)
      -- (0,\holeR)
      arc[start angle=90,end angle=0,radius=\holeR];

    \draw[symmetry]
      (\holeR,0) -- (\plateW,0);

    \draw[symmetry]
      (0,\holeR) -- (0,\plateH);

    \foreach \xx in {1.7,2.7,...,9.7}{
      \draw[blue!65!black, thin]
        (\xx,0) -- ++(-0.20,-0.28);
    }

    \foreach \yy in {1.7,2.7,...,9.7}{
      \draw[blue!65!black, thin]
        (0,\yy) -- ++(-0.28,-0.20);
    }

    \foreach \yy in {0.8,2.2,3.6,5.0,6.4,7.8,9.2}{
      \draw[traction]
        (\plateW,\yy) -- ++(0.85,0);
    }

    \node[
      red!70!black,
      anchor=west,
      align=left
    ] at (11.0,5.0)
      {$\overline{\mathbf{t}}
        =\sigma_0\mathbf{e}_x$};

    \node[
      blue!65!black,
      fill=blue!7,
      inner sep=2pt
    ] at (5.4,0.42)
      {$u_y=0$};

    \node[
      blue!65!black,
      fill=blue!7,
      inner sep=2pt,
      rotate=90
    ] at (0.42,5.4)
      {$u_x=0$};

    \node[
      anchor=south
    ] at (5.2,10.15)
      {$\overline{\mathbf{t}}=\mathbf{0}$};

    \node[
      anchor=west,
      align=left
    ] at (2.05,1.35)
      {traction-free hole};

    \draw[
      thin,
      -{Latex[length=1.8mm]}
    ]
      (1.95,1.25)
      -- ({\holeR*cos(45)},{\holeR*sin(45)});

    \draw[dimension]
      (0,0)
      --
      ({\holeR*cos(35)},{\holeR*sin(35)})
      node[midway, below right] {$a$};

    \node at (5.2,5.2) {$\Omega$};

    \draw[thin]
      (0,-1.10) -- (0,-1.70);

    \draw[thin]
      (\plateW,-1.10) -- (\plateW,-1.70);

    \draw[dimension]
      (0,-1.45)
      --
      node[
        midway,
        fill=white,
        inner sep=2pt
      ] {$W=10a$}
      (\plateW,-1.45);

    \draw[thin]
      (-1.10,0) -- (-1.70,0);

    \draw[thin]
      (-1.10,\plateH) -- (-1.70,\plateH);

    \draw[dimension]
      (-1.45,0)
      --
      node[
        midway,
        fill=white,
        inner sep=2pt,
        rotate=90
      ] {$H=10a$}
      (-1.45,\plateH);

    \draw[-{Latex[length=1.8mm]}]
      (8.1,8.3) -- ++(1.0,0)
      node[right] {$x$};

    \draw[-{Latex[length=1.8mm]}]
      (8.1,8.3) -- ++(0,1.0)
      node[above] {$y$};

  \end{tikzpicture}

  \caption{
    Quarter model of the perforated plate. Symmetry conditions are
    imposed on \(x=0\) and \(y=0\), a uniform horizontal traction acts
    on \(x=W\), and the upper and circular boundaries are traction free.
    Here \(a=1\) and \(W=H=10a\).
  }
  \label{fig:perforated_plate_geometry}
\end{figure}
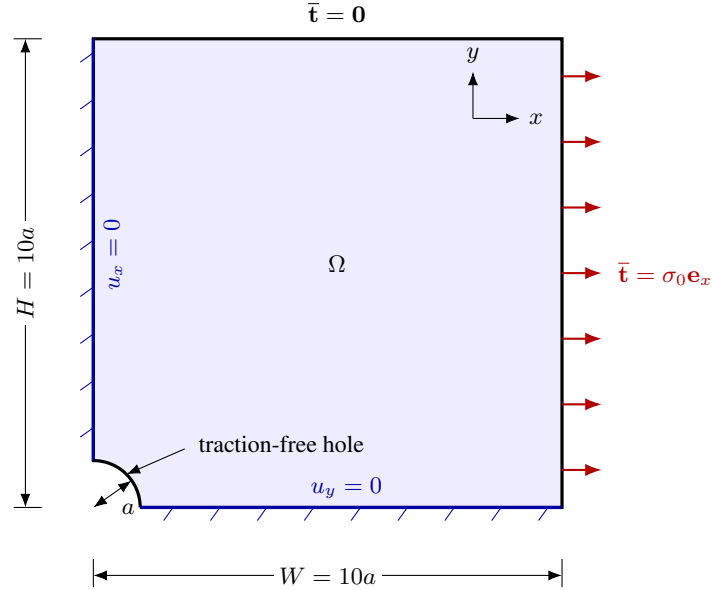

\paragraph{} The two-dimensional experiment examines the higher-order elasticity formulation on quadrilateral and general polygonal meshes. A square plate with a central circular hole is subjected to uniaxial tension. Symmetry reduces the computation to the quarter domain shown in Figure~\ref{fig:perforated_plate_geometry}. The parameters are $a=1$, $W=H=10$, $E_Y=1$, $\nu=0.3$, $t_p=1$, and $\sigma_0=1$, where $t_p$ is the out-of-plane thickness. The material is isotropic, and the calculation assumes plane stress. No body force is applied. Symmetry is enforced by prescribing $u_x=0$ on $x=0$ and $u_y=0$ on $y=0$. A uniform traction $(\sigma_0,0)^{\mathrm{T}}$ acts on $x=W$, while the boundary $y=H$ and the hole are traction free. For $k\geq2$, the edge moments of the constrained displacement component on each symmetry boundary are prescribed together with the corresponding vertex values. The displacement is computed with the strain-projection formulation described in \cite{artioli2017} for $k=1,2,3$.

\begin{figure}[htbp]
  \centering
  \includegraphics[width=\textwidth]{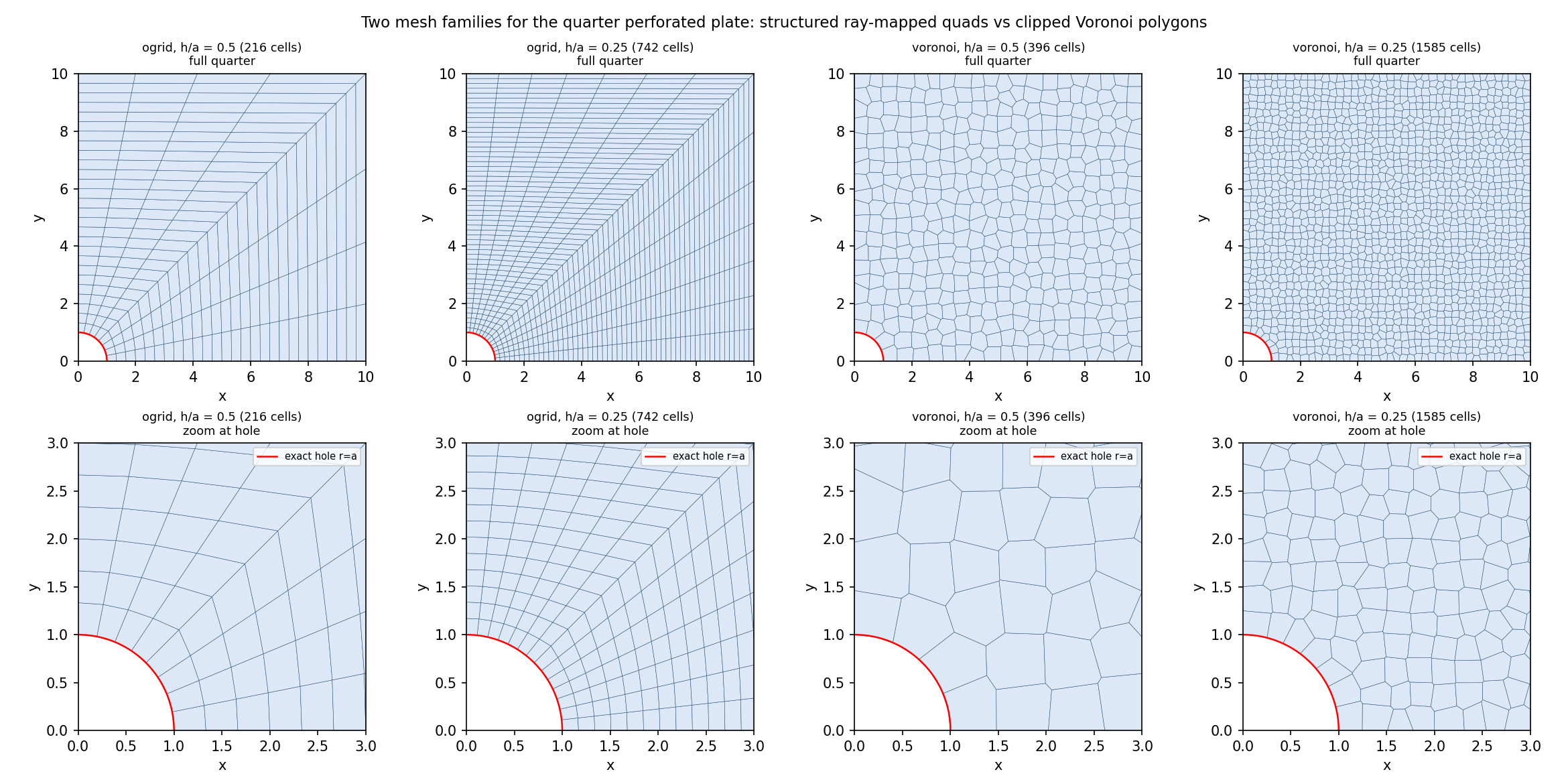}
  \caption{Structured O-grid quadrilateral and clipped Voronoi meshes at the first two resolution levels. The enlargements show the polygonal representation of the circular hole.}
  \label{fig:perforated_plate_meshes}
\end{figure}

\paragraph{} Two mesh families are used at the nominal resolutions $h/a=0.5,\;0.25,\;0.125,\;0.0625$. The first is a deterministic, structured O-grid of quadrilateral cells generated by connecting corresponding points along rays extending from the circular hole to the outer square boundary. Its angular divisions are chosen so that one ray reaches the outer corner, and a common number of radial divisions keeps the mesh conforming. The second family consists of clipped Voronoi cells with four to nine vertices. A fixed random seed makes each mesh reproducible, but the four levels are separately generated, non-nested realizations rather than successive refinements of one grid. Both families replace the circular boundary by polygonal segments, so the geometric approximation can limit the convergence of quantities evaluated at the hole. Figure~\ref{fig:perforated_plate_meshes} shows the two families and the resolution of the hole boundary. The O-grid construction is described in Appendix~\ref{ap:mesh_o_grid}; background on Voronoi tessellations and polygonal mesh generation is given in \cite{aurenhammer1991voronoi,fortune1987sweepline,talischi2012polymesher}.

\paragraph{} Table~\ref{tab:perforated_plate_mesh_sizes} gives the number of cells and the number of vector degrees of freedom for $k=2$. The counts follow the degree-of-freedom hierarchy of Section~\ref{sec:unified_vem_core} and include the constrained degrees of freedom.

\begin{table}[htbp]
  \centering
  \caption{Mesh sizes and vector degrees of freedom for the $k=2$ perforated-plate calculations.}
  \label{tab:perforated_plate_mesh_sizes}
  \begin{tabular}{@{}crrrr@{}}
    \toprule
    $h/a$ & \textbf{O-grid cells} & \textbf{O-grid $n_{\mathrm{dof}}$}
      & \textbf{Voronoi cells} & \textbf{Voronoi $n_{\mathrm{dof}}$} \\
    \midrule
    0.5    & 216    & 1,870  & 396    & 4,854   \\
    0.25   & 742    & 6,206  & 1,585  & 19,210  \\
    0.125  & 2,756  & 22,578 & 6,347  & 76,594  \\
    0.0625 & 10,972 & 88,830 & 25,393 & 305,514 \\
    \bottomrule
  \end{tabular}
\end{table}

\paragraph{} The principal local quantity is the hoop stress on the hole. Physically, this is the normal stress acting in the circumferential direction tangent to the hole boundary. The associated hoop strain is the relative stretching or contraction of a small material line in that direction: tension lengthens the local perimeter, whereas compression shortens it. Let $\mathbf{e}_\theta=(-\sin\theta,\cos\theta)^{\mathrm{T}}$, where $\theta$ is measured counterclockwise from the positive $x$-axis, and let $\bm{\sigma}_h$ denote the stress tensor reconstructed from the polynomial strain projection. The discrete hoop stress is
\begin{equation}
  \sigma_{\theta\theta,h}
  =
  \mathbf{e}_\theta^{\mathrm{T}}\bm{\sigma}_h\mathbf{e}_\theta.
\end{equation}
The subscript $h$ identifies a quantity obtained from the discrete VEM solution and reconstructed on a mesh of nominal size $h$. By contrast, $\sigma_{\theta\theta}$ without the subscript $h$ denotes a continuum hoop stress. In the Kirsch comparison below, it denotes the exact continuum stress for an infinite plate rather than the stress reconstructed from the numerical solution.

\paragraph{} Two dimensionless stress-concentration measures are reported. Both compare the discrete hoop stress with the nominal applied stress $\sigma_0$. A value of one means that the local stress equals the applied stress, whereas a value greater than one measures the amplification caused by the load path passing around the hole. The first is sampled directly on the polygonal hole boundary,
\begin{equation}
  K_{t,h}^{\mathrm{mid}}
  =
  \max_{\mathbf{x}_m\in\mathcal{M}_h^\circ}
  \frac{\sigma_{\theta\theta,h}(\mathbf{x}_m)}{\sigma_0},
\end{equation}
where $\mathcal{M}_h^\circ$ is the set of hole-edge midpoints. The second is evaluated at the crown of the hole,
\begin{equation}
  K_{t,h}^{\mathrm{crown}}
  =
  \frac{\sigma_{\theta\theta,h}(0,a)}{\sigma_0}.
\end{equation}
Here, $K_{t,h}^{\mathrm{mid}}$ is the largest amplification sampled on the polygonal approximation of the hole boundary, whereas $K_{t,h}^{\mathrm{crown}}$ measures the amplification at the fixed physical point $(0,a)$, the top of the hole. This point is shared by all meshes and is the location of the maximum tensile hoop stress in the classical infinite-plate solution under horizontal tension.

\paragraph{} The crown value is used for the convergence comparison. No hole-edge midpoint lies exactly at $\theta=\pi/2$, and the distance between the closest midpoint and the crown changes with the angular mesh spacing. Consequently, $K_{t,h}^{\mathrm{mid}}$ contains a sampling error in addition to the stress-approximation error.

\begin{remark}
For an infinite plate, the exact continuum Kirsch stress satisfies
\begin{equation}
  \frac{\sigma_{\theta\theta}(a,\theta)}{\sigma_0}
  =
  1-2\cos(2\theta).
\end{equation}
At the crown, where $\theta=\pi/2$, this expression yields
\begin{equation}
  \frac{\sigma_{\theta\theta}(a,\pi/2)}{\sigma_0}
  =
  3.
\end{equation}
The infinite-plate crown concentration factor is therefore $3$. This value provides physical context but is not the convergence target for the present domain. The outer boundaries of the computational domain are located at $10a$, so the continuum solution of the finite-domain problem need not equal the Kirsch solution. Consequently, $K_{t,h}^{\mathrm{crown}}$ need not converge to $3$ as $h\to0$.
\end{remark}

\paragraph{} Two global quantities accompany the local stress. The discrete strain energy is
\begin{equation}
  U_h = \frac{1}{2} \mathbf{u}_h^{\mathrm{T}} \mathbf{K} \mathbf{u}_h,
\end{equation}
where $\mathbf{u}_h$ is the discrete displacement field and $\mathbf{K}$ is the stiffness matrix before the boundary conditions are applied. The mean horizontal displacement of the loaded boundary is approximated with the composite trapezoidal rule,
\begin{equation}
  \overline{u}_{x,h}
  =
  \frac{1}{H}
  \sum_{e=(p,q)\subset\{x=W\}}
  \frac{|e|}{2}\left[u_{x,h}(p)+u_{x,h}(q)\right].
\end{equation}
Here $p$ and $q$ are the endpoints of boundary edge $e$, and $|e|$ is its length. The edge-length weighting approximates the same boundary mean on both mesh families, unlike an unweighted average of the boundary vertices.

\paragraph{} An independent reference was computed with biquadratic Q9 finite elements and a curved quadratic representation of the hole. Four successively refined meshes were used, and the crown stress was recovered both by direct nodal evaluation and by extrapolation from the element interior. The resulting reference estimates are included in Table~\ref{tab:perforated_plate_extrapolation}. The Q9 crown concentration is $3.0861$, about $2.9\%$ above the infinite-plate Kirsch value of $3$.

\paragraph{} Only the deterministic O-grid family is used for Richardson extrapolation. If $Q_1$, $Q_2$, and $Q_3$ denote a quantity on three consecutive levels, ordered from coarse to fine with a nominal refinement factor of two, the observed rate and extrapolated value are
\begin{equation}
  p = \log_2 \left| \frac{Q_2 - Q_1}{Q_3 - Q_2} \right|, \quad Q_{\text{ext}} = Q_3 + \frac{Q_3 - Q_2}{2^p - 1}.
\end{equation}
A sequence is eligible for extrapolation only when it is monotone, its successive differences decrease and remain above the measured numerical uncertainty, and the corresponding energy sequence also converges. The Voronoi results are not extrapolated because their levels are separately generated rather than systematically refined. Instead, they test whether the result persists when the element shapes and connectivity are changed. Table~\ref{tab:perforated_plate_extrapolation} reports the accepted $k=2$ extrapolates, their observed rates, and the independent Q9 reference values.

\begin{table}[htbp]
  \centering
  \small
  \caption{Extrapolated $k=2$ VEM quantities and the independent Q9 finite element reference. The uncertainty attached to a VEM extrapolate is the larger of its distance from the finest computed value and the measured numerical uncertainty. The midpoint stress is not compared directly with the crown stress of the Q9 reference.}
  \label{tab:perforated_plate_extrapolation}
  \begin{tabular}{@{}lccc@{}}
    \toprule
    \textbf{Quantity} & \textbf{O-grid $k=2$ extrapolate} & \textbf{Observed $p$} & \textbf{Independent Q9 reference} \\
    \midrule
    $K_{t,h}^{\mathrm{crown}}$ & $3.086\pm0.004$ & $1.780$ & $3.0861\pm0.0001$ \\
    $K_{t,h}^{\mathrm{mid}}$ & $3.088\pm0.008$ & $1.521$ & -- \\
    $U_h$ & $51.2121\pm0.0002$ & $2.341$ & $51.2121769\pm4.8\times10^{-8}$ \\
    $\overline{u}_{x,h}$ & $10.24244\pm0.00009$ & $2.053$ & $10.2424354\pm9.7\times10^{-9}$ \\
    \bottomrule
  \end{tabular}
\end{table}

\begin{figure}[htbp]
  \centering
  \includegraphics[width=\textwidth]{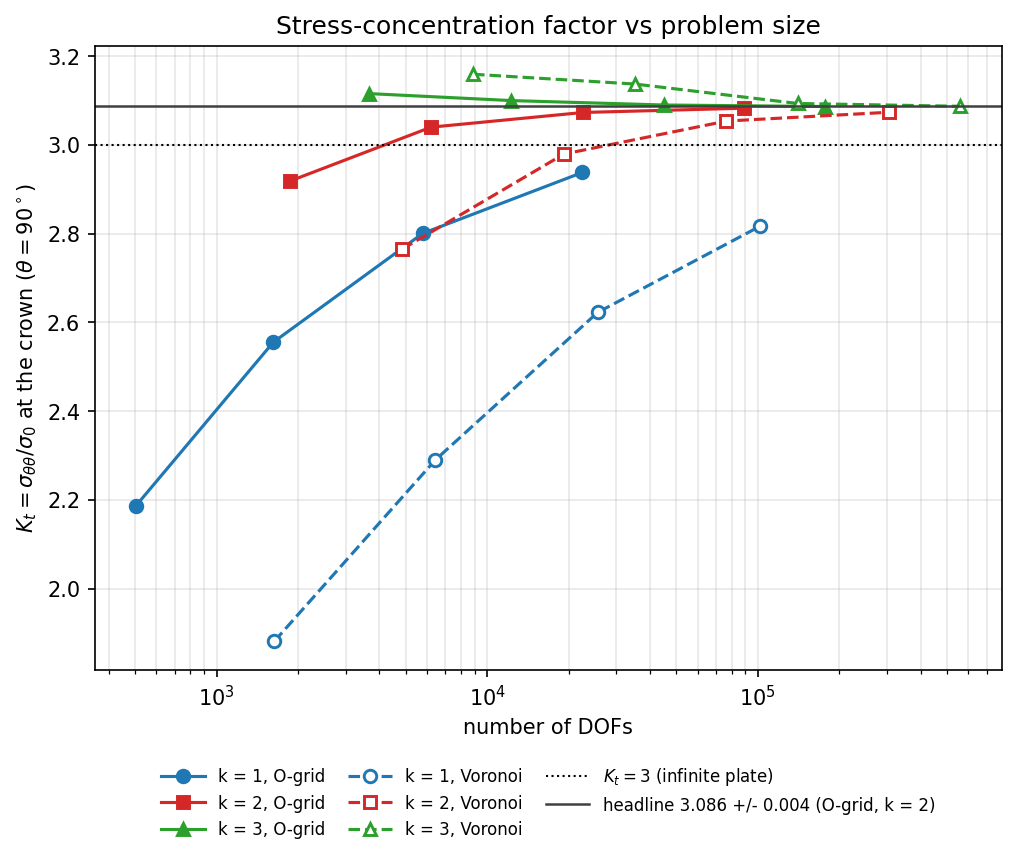}
  \caption{Convergence of the crown stress-concentration factor with respect to the number of vector degrees of freedom for the O-grid and Voronoi mesh families. The dotted line indicates the infinite-plate Kirsch value, while the solid grey line shows the accepted $k=2$ O-grid extrapolate.}
  \label{fig:perforated_plate_kt_convergence}
\end{figure}

\paragraph{} Figure~\ref{fig:perforated_plate_kt_convergence} shows the convergence of the crown stress-concentration factor. On the finest $k=2$ meshes, the O-grid and Voronoi values are $3.0823679$ and $3.0732601$, respectively. Their difference is $9.108 \times 10^{-3}$, or $2.95 \times 10^{-3}$ relative to the O-grid value. At the same resolution, the relative differences between the two families are $9.68 \times 10^{-8}$ for the strain energy and $4.12 \times 10^{-6}$ for the loaded-boundary displacement. This agreement indicates that the $k=2$ result is insensitive to the mesh topology. Because both families use straight boundary segments, however, their difference does not capture the shared geometric-error component and is not an overall error estimate. The cross-family spread in the crown stress is $2.3$ times the O-grid extrapolation uncertainty, so that uncertainty is not an overall error bound either. The Q9 calculation supplies the external comparison.

\paragraph{} The $k=3$ O-grid sequence requires qualification. Its strain energies are $51.2003$, $51.2086$, $51.2082$, and $51.1996$ as the mesh is refined, so the differences between the last three levels increase rather than decrease even though the crown stress approaches the Q9 value. The componentwise backward error of the finest linear solve is $2.34\times 10^{-16}$, whereas the stiffness-matrix norm grows markedly for the scaled-monomial basis on the anisotropic O-grid cells. The loss of convergence is therefore consistent with basis-scaling and conditioning effects rather than a failure of the sparse solver. No $k=3$ Richardson values are reported. These results motivate better scaling or orthogonalization of the high-order polynomial basis.

\paragraph{} All 24 combinations of mesh family, resolution, and approximation order completed the same validation checks. The relative matrix-symmetry defects remained below $10^{-15}$, the assembled horizontal traction resultant was $\sigma_0 Ht_p=10\sigma_0t_p$, and the implemented degree-of-freedom counts agreed with the hierarchy in Section~\ref{sec:unified_vem_core}. Each order also passed a constant-stress patch test before the perforated-plate calculations were run.

\subsection{Three-dimensional polyhedral cantilever}

\begin{figure}[htbp]
  \centering
  \begin{tikzpicture}[
    font=\small,
    boundary/.style={black, very thick},
    hidden/.style={black!55, dashed, thick},
    clamp/.style={black!65, thin},
    traction/.style={red!75!black, thick, -{Latex[length=2.4mm]}},
    dimension/.style={thin, {Latex[length=1.8mm]}-{Latex[length=1.8mm]}},
    axis/.style={thick, -{Latex[length=2.0mm]}}
  ]

  \coordinate (A) at (0,0);
  \coordinate (B) at (0,3.0);
  \coordinate (C) at (1.35,3.82);
  \coordinate (D) at (1.35,0.82);
  \coordinate (E) at (8.0,0);
  \coordinate (F) at (8.0,3.0);
  \coordinate (G) at (9.35,3.82);
  \coordinate (H) at (9.35,0.82);

  \fill[blue!7] (A)--(E)--(F)--(B)--cycle;
  \fill[blue!11] (B)--(F)--(G)--(C)--cycle;
  \fill[blue!5] (A)--(D)--(H)--(E)--cycle;
  \fill[red!6] (E)--(H)--(G)--(F)--cycle;
  \fill[black!10] (A)--(B)--(C)--(D)--cycle;

  \draw[boundary] (A)--(B)--(C)--(D)--cycle;
  \draw[boundary] (E)--(F)--(G)--(H)--cycle;
  \draw[boundary] (A)--(E);
  \draw[boundary] (B)--(F);
  \draw[boundary] (C)--(G);
  \draw[hidden] (D)--(H);

  \foreach \yy in {0.15,0.55,...,2.95}{
    \draw[clamp] (0,\yy) -- ++(-0.34,-0.22);
  }
  \node[anchor=east, align=right] at (-0.48,1.62)
    {$\Gamma_0$\\$\mathbf{u}=\mathbf{0}$};

  \foreach \xx/\yy in {8.15/2.62,8.50/2.82,8.85/3.02,9.18/3.22}{
    \draw[traction] (\xx,\yy) -- ++(0,-0.82);
  }
  \node[red!75!black, anchor=east] at (7.65,1.05)
    {$\overline{\mathbf{t}}=-t_0\mathbf{e}_y$};
  \node[red!75!black] at (8.72,1.18) {$\Gamma_L$};

  \node[anchor=south] at (4.75,3.96)
    {$\bm{\sigma}\mathbf{n}=\mathbf{0}$};

  \draw[thin] (A) -- ++(0,-0.82);
  \draw[thin] (E) -- ++(0,-0.82);
  \draw[dimension]
    (0,-0.60) -- node[midway, fill=white, inner sep=2pt] {$L=6$} (8.0,-0.60);

  \draw[thin] (G) -- ++(0.62,0);
  \draw[thin] (H) -- ++(0.62,0);
  \draw[dimension]
    (9.78,0.82) -- node[midway, fill=white, inner sep=1.5pt, rotate=90] {$b=1$}
    (9.78,3.82);

  \draw[thin] (F) -- ++(-0.08,0.50);
  \draw[thin] (G) -- ++(-0.08,0.50);
  \draw[dimension]
    (7.98,3.98) -- node[midway, above left, inner sep=1pt] {$c=1$} (9.30,4.78);

  \coordinate (O) at (3.15,1.15);
  \draw[axis] (O) -- ++(1.10,0) node[right] {$x$};
  \draw[axis] (O) -- ++(0,1.05) node[above] {$y$};
  \draw[axis] (O) -- ++(0.72,0.44) node[above right] {$z$};

  \node at (5.05,1.55) {$\Omega$};

\end{tikzpicture}

  \caption{
    Three-dimensional cantilever. The root face $\Gamma_0$ is fully clamped, the end face $\Gamma_L$ carries the uniform traction $\overline{\mathbf{t}}=-t_0 \mathbf{e}_y$, and the lateral faces are traction free. The drawing is not to scale. 
  }
  \label{fig:polyhedral_cantilever}
\end{figure}
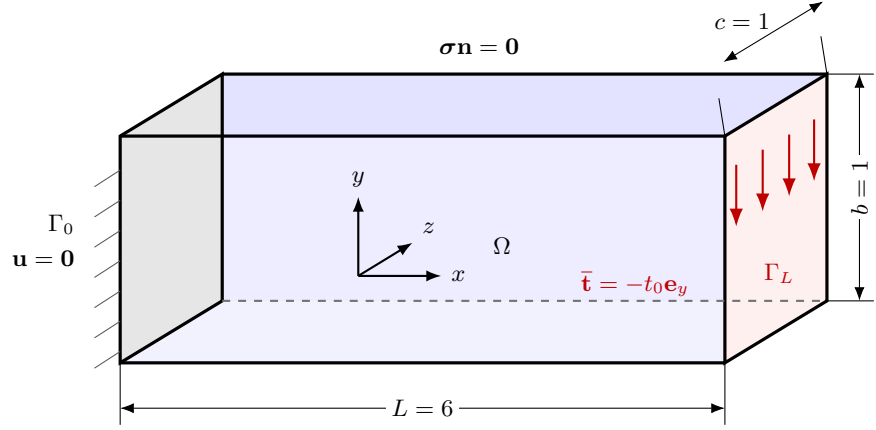

\paragraph{} The three-dimensional experiment considers a cantilever with a square cross-section. Its domain is $\Omega = [0,L] \times [-b/2, b/2] \times [-c/2,c/2]$, with $L=6$, $b=c=1$, $E_Y=1000$, and $\nu = 0.3$. Here, $b$ and $c$ are cross-sectional dimensions in the $y$- and $z$-directions, respectively. The material is homogeneous, isotropic, and linearly elastic. No body force is applied. Let
\begin{equation}
  \Gamma_0 = \{ (x,y,z) \in \partial \Omega : \; x = 0 \}, \quad \Gamma_L = \{ (x,y,z) \in \partial \Omega: \; x = L \}.
\end{equation}
The root face is fully clamped, so all three displacement components vanish on $\Gamma_0$. A uniform traction acts in the negative $y$-direction on $\Gamma_L$ while the four lateral faces are traction free. The boundary conditions are therefore $\mathbf{u} = \mathbf{0}$ on $\Gamma_0$, $\bm{\sigma} \mathbf{n} = \overline{\mathbf{t}}$ on $\Gamma_L$, and $\bm{\sigma} \mathbf{n} = \mathbf{0}$ on $\partial \Omega \setminus (\Gamma_0 \cup \Gamma_L)$, where $\bm{\sigma}$ is the Cauchy stress tensor and $\mathbf{n}$ is the outward unit normal vector. The prescribed traction is
\begin{equation}
  \overline{\mathbf{t}} = -t_0 \mathbf{e}_y = (0,-1,0)^{\mathrm{T}}, \quad t_0 = 1.
\end{equation}
Here $t_0$ is the traction magnitude and $\mathbf{e}_y$ is the unit vector in the positive $y$-direction. Since the loaded face has area $bc=1$, the traction produces the resultant force $\mathbf{P}=(0,-1,0)^{\mathrm{T}}$. Figure~\ref{fig:polyhedral_cantilever} summarizes the geometry, dimensions, and boundary conditions.

\paragraph{} The discretization uses structured hexahedral and clipped Voronoi mesh families at the nominal spacings $h_{\mathrm{nom}}=0.5$, $0.25$, and $0.125$. Here $h_{\mathrm{nom}}$ is the prescribed spacing used by the mesh generator, not the maximum cell diameter. The structured meshes contain $12\times2\times2$, $24\times4\times4$, and $48\times8\times8$ cubes. The Voronoi meshes contain the same numbers of cells but have different vertex, edge, and face connectivities. Table~\ref{tab:polyhedral_cantilever_meshes} summarizes the cell and vertex counts and the range of polyhedral face sizes for both mesh families.

\begin{table}[htbp]
  \centering
  \caption{Mesh-family data for the three-dimensional cantilever. The hexahedral cells have six quadrilateral faces. At the three levels, 47 of 48, 384 of 384, and 3,071 of 3,072 Voronoi cells, respectively, are not hexahedra.}
  \label{tab:polyhedral_cantilever_meshes}
  \small
  \begin{tabular}{@{}crrrrr@{}}
    \toprule
    $h_{\mathrm{nom}}$ & \textbf{Cells} & \shortstack{\textbf{Hexahedral}\\\textbf{vertices}} & \shortstack{\textbf{Voronoi}\\\textbf{vertices}} & \shortstack{\textbf{Voronoi faces}\\\textbf{per cell}} & \shortstack{\textbf{Vertices per}\\\textbf{Voronoi face}} \\
    \midrule
    0.5   & 48    & 117   & 241    & 6--13 & 3--8  \\
    0.25  & 384   & 625   & 2,157  & 7--19 & 3--12 \\
    0.125 & 3,072 & 3,969 & 18,933 & 6--23 & 3--12 \\
    \bottomrule
  \end{tabular}
\end{table}

\begin{figure}[!htbp]
  \centering
  \begin{tikzpicture}[
    x=1cm,
    y=1cm,
    font=\scriptsize,
    cellface/.style={fill=blue!5},
    topface/.style={fill=blue!11},
    sideface/.style={fill=blue!8},
    celledge/.style={black!80, thick},
    hiddenedge/.style={black!45, dashed},
    tetedge/.style={orange!80!black, densely dashed, thick},
    fanedge/.style={blue!70!black, thick},
    point/.style={circle, fill=black, inner sep=1.15pt},
    facepoint/.style={circle, fill=blue!70!black, inner sep=1.15pt},
    firstvertex/.style={circle, fill=red!75!black, inner sep=1.25pt}
  ]

  \path[use as bounding box] (-0.45,-1.82) rectangle (14.75,2.55);

  \begin{scope}
    \coordinate (aA)  at (0,0);
    \coordinate (aB)  at (2.35,0);
    \coordinate (aC)  at (2.35,1.70);
    \coordinate (aD)  at (0,1.70);
    \coordinate (aAp) at (0.55,0.45);
    \coordinate (aBp) at (2.90,0.45);
    \coordinate (aCp) at (2.90,2.15);
    \coordinate (aDp) at (0.55,2.15);
    \coordinate (aE)  at (1.34,0.91);

    \fill[cellface] (aA)--(aB)--(aC)--(aD)--cycle;
    \fill[sideface] (aB)--(aBp)--(aCp)--(aC)--cycle;
    \fill[topface] (aD)--(aC)--(aCp)--(aDp)--cycle;
    \fill[orange!30, opacity=0.72] (aE)--(aD)--(aC)--(aCp)--cycle;

    \draw[celledge] (aA)--(aB)--(aC)--(aD)--cycle;
    \draw[celledge] (aD)--(aDp)--(aCp)--(aC);
    \draw[celledge] (aB)--(aBp)--(aCp);
    \draw[hiddenedge] (aA)--(aAp)--(aBp);
    \draw[hiddenedge] (aAp)--(aDp);
    \draw[fanedge] (aD)--(aCp);
    \draw[tetedge] (aE)--(aD) (aE)--(aC) (aE)--(aCp);

    \node[firstvertex] at (aD) {};
    \node[red!75!black, above left=1pt] at (aD) {$v_1$};
    \node[point] at (aE) {};
    \node[below=1pt] at (aE) {$c_E$};
    \node[orange!80!black] at (1.80,1.43) {$T$};

    \node[align=center] at (1.45,-0.62)
      {$V_E=\displaystyle\sum_{T\subset E}|T|,
        \qquad V_h=\displaystyle\sum_E V_E$};
    \node[align=center, text width=4.25cm] at (1.45,-1.48)
      {\textbf{(a)} Face-vertex fan; cellwise sum.};
  \end{scope}

  \begin{scope}[xshift=5.05cm]
    \coordinate (bA)  at (0,0);
    \coordinate (bB)  at (2.35,0);
    \coordinate (bC)  at (2.35,1.70);
    \coordinate (bD)  at (0,1.70);
    \coordinate (bAp) at (0.55,0.45);
    \coordinate (bBp) at (2.90,0.45);
    \coordinate (bCp) at (2.90,2.15);
    \coordinate (bDp) at (0.55,2.15);
    \coordinate (bE)  at (1.34,0.91);
    \coordinate (bF)  at (1.45,1.925);

    \fill[cellface] (bA)--(bB)--(bC)--(bD)--cycle;
    \fill[sideface] (bB)--(bBp)--(bCp)--(bC)--cycle;
    \fill[topface] (bD)--(bC)--(bCp)--(bDp)--cycle;
    \fill[green!28, opacity=0.72] (bE)--(bF)--(bC)--(bCp)--cycle;

    \draw[celledge] (bA)--(bB)--(bC)--(bD)--cycle;
    \draw[celledge] (bD)--(bDp)--(bCp)--(bC);
    \draw[celledge] (bB)--(bBp)--(bCp);
    \draw[hiddenedge] (bA)--(bAp)--(bBp);
    \draw[hiddenedge] (bAp)--(bDp);
    \draw[fanedge] (bF)--(bD) (bF)--(bC) (bF)--(bCp) (bF)--(bDp);
    \draw[green!50!black, densely dashed, thick]
      (bE)--(bF) (bE)--(bC) (bE)--(bCp);

    \node[point] at (bE) {};
    \node[below=1pt] at (bE) {$c_E$};
    \node[facepoint] at (bF) {};
    \node[above=1pt, blue!70!black] at (bF) {$c_F$};
    \node[green!45!black] at (1.95,1.48) {$\widehat T$};

    \node[align=center] at (1.45,-0.62)
      {$V_E=\displaystyle\sum_{\widehat T\subset E}|\widehat T|,
        \qquad V_h=\displaystyle\sum_E V_E$};
    \node[align=center, text width=4.25cm] at (1.45,-1.48)
      {\textbf{(b)} Face-centroid fan; cellwise sum.};
  \end{scope}

  \begin{scope}[xshift=10.15cm]
    \coordinate (cA)   at (0,0);
    \coordinate (cM)   at (1.20,0);
    \coordinate (cB)   at (2.40,0);
    \coordinate (cD)   at (0,1.55);
    \coordinate (cN)   at (1.20,1.55);
    \coordinate (cC)   at (2.40,1.55);
    \coordinate (cAp)  at (0.40,0.34);
    \coordinate (cMp)  at (1.60,0.34);
    \coordinate (cBp)  at (2.80,0.34);
    \coordinate (cDp)  at (0.40,1.89);
    \coordinate (cNp)  at (1.60,1.89);
    \coordinate (cCp)  at (2.80,1.89);
    \coordinate (cEone) at (0.70,0.82);
    \coordinate (cEtwo) at (1.90,0.82);

    \fill[cellface] (cA)--(cB)--(cC)--(cD)--cycle;
    \fill[sideface] (cB)--(cBp)--(cCp)--(cC)--cycle;
    \fill[topface] (cD)--(cC)--(cCp)--(cDp)--cycle;
    \fill[orange!30, opacity=0.72]
      (cEone)--(cD)--(cN)--(cNp)--cycle;
    \fill[violet!24, opacity=0.72]
      (cEtwo)--(cN)--(cC)--(cCp)--cycle;

    \draw[celledge] (cA)--(cB)--(cC)--(cD)--cycle;
    \draw[celledge] (cD)--(cDp)--(cCp)--(cC);
    \draw[celledge] (cB)--(cBp)--(cCp);
    \draw[celledge] (cM)--(cN)--(cNp)--(cMp)--cycle;
    \draw[hiddenedge] (cA)--(cAp)--(cBp);
    \draw[hiddenedge] (cAp)--(cDp);
    \draw[fanedge] (cD)--(cNp) (cN)--(cCp);
    \draw[tetedge]
      (cEone)--(cD) (cEone)--(cN) (cEone)--(cNp)
      (cEtwo)--(cN) (cEtwo)--(cC) (cEtwo)--(cCp);

    \node[point] at (cEone) {};
    \node[point] at (cEtwo) {};
    \node[below=1pt] at (cEone) {$c_{E_1}$};
    \node[below=1pt] at (cEtwo) {$c_{E_2}$};
    \node[orange!80!black] at (0.77,1.30) {$T_1$};
    \node[violet!70!black] at (1.98,1.30) {$T_m$};
    \node[above, align=center] at (1.40,2.08)
      {same tetrahedra as in (a)};

    \node at (1.40,-0.28)
      {$T_1\rightarrow T_2\rightarrow\cdots\rightarrow T_{N_T}$};
    \node[align=center] at (1.40,-0.78)
      {$V_h=\displaystyle\sum_{m=1}^{N_T}|T_m|$};
    \node[align=center, text width=4.25cm] at (1.40,-1.48)
      {\textbf{(c)} Face-vertex fan; flat sum.};
  \end{scope}

\end{tikzpicture}
  \caption{Evaluation of the total mesh volume. (a) Each face is triangulated from its first stored vertex, and the tetrahedral volumes are accumulated first within each cell. (b) Each face is triangulated from its vertex-average point, while the same cellwise accumulation is retained. (c) The tetrahedra from (a) are accumulated in one global sequence without cellwise partial sums.}
  \label{fig:volume_evaluations}
\end{figure}
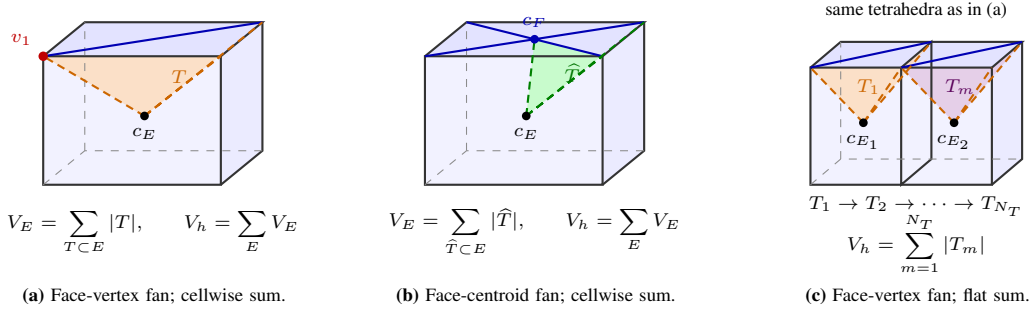

\paragraph{} The mesh checks verify that every interior face is shared by two cells, every boundary face by one cell, and every stored face is simple and planar. The largest absolute out-of-plane displacement of a face vertex is $8.9\times10^{-15}$. The total volume is evaluated in three ways. In the first, each face is divided into triangles by a fan from its first stored vertex, and each triangle forms a tetrahedron with the vertex-average point of the cell. The tetrahedral volumes are first accumulated within each cell and the cell volumes are then summed over the mesh. The second evaluation replaces the face-vertex fan with triangles formed by the vertex-average point of the face and each of its edges. It therefore uses a different tetrahedral decomposition of the same polyhedron. The third evaluation uses the tetrahedra from the first decomposition but adds all their volumes directly in a single sequence, without forming cellwise partial sums. Figure~\ref{fig:volume_evaluations} illustrates these three procedures. 

\paragraph{} In exact arithmetic, all three procedures give the same total volume, $Lbc=6$. Their small differences isolate the effects of the chosen tetrahedral decomposition and of floating-point summation order. Across both mesh families, the largest relative residual from the exact volume is $3.6\times10^{-13}$. Figure~\ref{fig:polyhedral_mesh_face_distribution} shows the resulting distributions of faces per cell.

\begin{figure}[!htbp]
  \centering
  \input{figures/ex3_mesh_face_distributions.tex}
  \caption{Distributions of the number of faces per cell. The columns show structured hexahedra and clipped Voronoi cells, respectively; the rows correspond, from top to bottom, to the nominal mesh spacings $h_{\mathrm{nom}}=0.5$, $0.25$, and $0.125$.}
  \label{fig:polyhedral_mesh_face_distribution}
\end{figure}

\paragraph{} Orders $k=1,\;2,\;3$ are run on both families. The complete systems use the sparse assembly path with one OpenMP thread and the guarded CHOLMOD Cholesky solver described in Section~\ref{subsec:linear_solver_selection}. Essential conditions are imposed with the diagonal penalty $\beta=10^{10}\max_i |A_{ii}|$. Seventeen of eighteen combinations completed. The finest Voronoi problem at $k=3$ has $518,853$ unknowns, but its predicted Cholesky factor requires approximately $51.2$~GB on a machine with $38.7$~GB of memory; the solver therefore rejects that factorization before allocation. All degree-of-freedom counts include the constrained unknowns.

\paragraph{} Two downward displacement measures are retained. On the hexahedral meshes, the centre of the loaded face is a vertex, and 
\begin{equation}
  \delta_h^c = -u_{y,h}(L,0,0).
\end{equation}
The measure used to compare both mesh families is the loaded-face area average,
\begin{equation}
  \delta_h^A = -\frac{1}{bc} \int \limits_{\Gamma_L} u^{\text{lin}}_{y,h}dS,
\end{equation}
where $u^{\text{lin}}_{y,h}$ is the piecewise-linear interpolant of the vertex values on a fan triangulation of each polygonal face. Table~\ref{tab:polyhedral_cantilever_displacements} reports the complete-system sizes and the area-averaged displacement for every completed run.

\begin{table}[htbp]
  \centering
  \small
  \setlength{\tabcolsep}{5pt}
  \caption{Complete system sizes and downward area-averaged end-face displacements. The finest Voronoi $k=3$ solution is omitted because the estimated sparse-factor memory exceeds the available memory.}
  \label{tab:polyhedral_cantilever_displacements}
  \begin{tabular}{@{}ccrrrr@{}}
    \toprule
    $h_{\mathrm{nom}}$ & $k$
    & \textbf{Hexahedral} $n_{\mathrm{dof}}$
    & \textbf{Hexahedral} $\delta_h^A$
    & \textbf{Voronoi} $n_{\mathrm{dof}}$
    & \textbf{Voronoi} $\delta_h^A$ \\
    \midrule
    0.5   & 1 & 351     & 0.741212 & 723     & 0.954007 \\
    0.5   & 2 & 1,875   & 0.858581 & 3,159   & 0.864726 \\
    0.5   & 3 & 4,275   & 0.868113 & 6,741   & 0.863746 \\
    0.25  & 1 & 1,875   & 0.832109 & 6,471   & 0.889947 \\
    0.25  & 2 & 11,907  & 0.869905 & 28,167  & 0.871608 \\
    0.25  & 3 & 28,323  & 0.870932 & 59,781  & 0.870190 \\
    0.125 & 1 & 11,907  & 0.861396 & 56,799  & 0.876942 \\
    0.125 & 2 & 84,099  & 0.872245 & 245,607 & 0.872665 \\
    0.125 & 3 & 204,867 & 0.872189 & 518,853 & \textemdash \\
    \bottomrule
  \end{tabular}
\end{table}

\begin{figure}[!htbp]
  \centering
  \begin{subfigure}[t]{\textwidth}
    \centering
    \includegraphics[width=\linewidth]{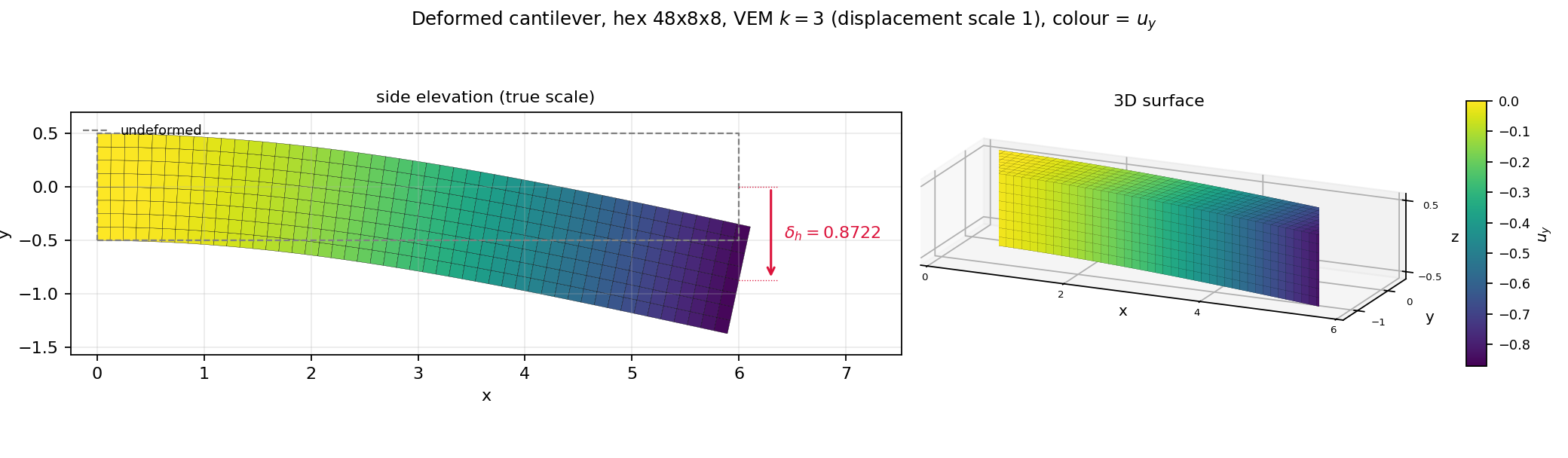}
    \caption{Structured $48\times8\times8$ hexahedral mesh.}
    \label{fig:polyhedral_cantilever_deformed_hex}
  \end{subfigure}

  \vspace{0.6em}

  \begin{subfigure}[t]{\textwidth}
    \centering
    \includegraphics[width=\linewidth]{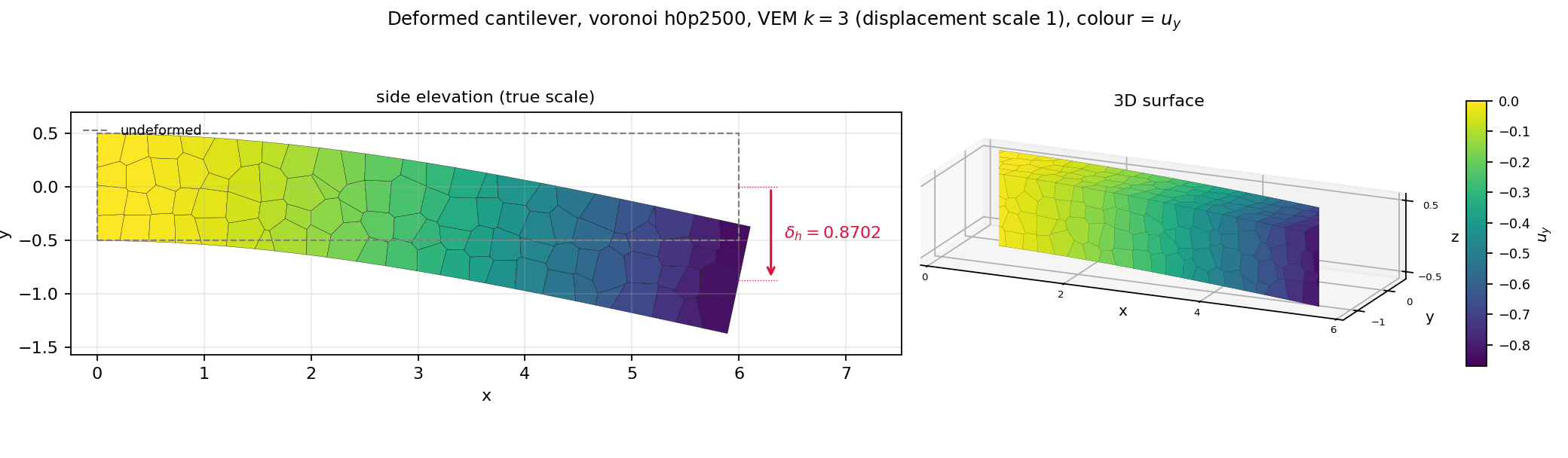}
    \caption{Clipped Voronoi mesh with 384 cells.}
    \label{fig:polyhedral_cantilever_deformed_voronoi}
  \end{subfigure}
  \caption{Deformed cantilever at $k=3$: (a) the $48\times8\times8$ hexahedral mesh with 204,867 unknowns and (b) the 384-cell Voronoi mesh with 59,781 unknowns.}
  \label{fig:polyhedral_cantilever_deformed}
\end{figure}

\paragraph{} Across the three resolutions, the $k=1$ hexahedral values increase from $0.741212$ to $0.861396$, whereas the Voronoi values decrease from $0.954007$ to $0.876942$. The higher-order results occupy a much narrower interval. At $k=2$, the finest completed area averages are $0.872245$ and $0.872665$ for the hexahedral and Voronoi families, respectively. Their absolute difference is $4.20\times10^{-4}$, or $0.048\%$ relative to the hexahedral value. The corresponding strain energies are $0.4361226$ and $0.4363314$. The close agreement persists despite the different mesh geometries. For each family and available resolution, the change from $k=2$ to $k=3$ is smaller than the preceding change from $k=1$ to $k=2$. Its sign is not consistent across families or levels, so the data do not support a claim that $k=3$ is uniformly more efficient. Figure~\ref{fig:polyhedral_cantilever_deformed} shows the displacement fields on the finest completed $k=3$ mesh of each family. The displacement scale is one, and the colour field represents $u_{y,h}$.

\paragraph{} The same cantilever is also solved with a separate finite element implementation to compare the VEM result with a conventional discretization. The finite element implementation uses eight-node trilinear hexahedra (H8) and 27-node triquadratic hexahedra (H27). Its shape functions, quadrature rules, assembly procedure, and treatment of the boundary conditions do not call the VEM code. Before the cantilever calculation, both elements reproduce a linear displacement field with relative errors below $5.24 \times 10^{-16}$. H27 also reproduces a quadratic field with its compatible constant body force to $5.8 \times 10^{-16}$, whereas H8 has the expected interior field error of $3.34 \times 10^{-3}$.

\paragraph{} On the three successively refined meshes, H27 gives centre displacements of $0.869115$, $0.871539$, and $0.872344$. Richardson extrapolation of this sequence gives $0.872744$. The corresponding hexahedral VEM extrapolates for $k=1,2,3$ are $0.872854$, $0.872713$, and $0.873064$, respectively. The H27 estimate therefore lies between the smallest and the largest VEM estimates. The independently assembled finite element and virtual element calculations predict the centre displacement within the same narrow interval. None of the extrapolated values is treated as the exact solution of the three-dimensional elasticity problem.

\paragraph{} Observed rates are not reported because the fully clamped root meets the traction-free lateral faces along mixed-boundary-condition edges, where the stress field is singular. The same slow approach appears in the independent H27 sequence under uniform refinement; it is therefore a feature of the boundary-value problem rather than of the virtual element discretization.

\paragraph{} For physical comparison, the standard closed-form Euler--Bernoulli and Timoshenko cantilever expressions are evaluated using the geometry, load, and material parameters of this problem. No Euler--Bernoulli or Timoshenko beam solver is implemented. The resulting downward tip-displacement estimates are $0.864000$ and $0.882720$. The Timoshenko value uses the shear correction factor $5/6$. These are engineering references, not exact solutions of the three-dimensional boundary-value problem. In particular, the full-face clamp suppresses transverse contraction and warping at the root. Replacing it in the independent H27 model by pointwise $u_x=0$, zero mean values of $u_y$ and $u_z$, and zero mean twist gives $0.88187$, close to the Timoshenko estimate. The full-face clamp is therefore retained when comparing the VEM and finite element solutions.

\paragraph{} A patch test was performed because convergence of the cantilever displacement does not by itself verify that the implementation reproduces the polynomial fields contained in the virtual element space. The test provides a direct check of element consistency on both hexahedral and Voronoi meshes. It prescribes the linear displacement field associated with uniform uniaxial tension in the $x$-direction. Let $\mathbf{e}_x$ denote the unit vector along the beam axis and set $\bm{\sigma} = \sigma_0 \mathbf{e}_x \otimes \mathbf{e}_x$, where $\otimes$ denotes the tensor product and $\sigma_0 = 1$. For the material parameters used in the cantilever problem, the exact displacement field, up to a rigid-body motion, is
\begin{equation}
  \mathbf{u}_{\mathbf{e}_x}(x,y,z) = \frac{\sigma_0}{E_Y} \left[ \begin{array}{ccc}
    x & -\nu y & -\nu z
  \end{array} \right]^{\mathrm{T}}.
\end{equation}

\begin{figure}[!htbp]
  \centering
  \begin{tikzpicture}[
    x=1cm,
    y=1cm,
    font=\scriptsize,
    boundary/.style={black, very thick},
    meshline/.style={black!42, thin},
    hidden/.style={black!50, dashed, thick},
    traction/.style={red!75!black, thick, -{Latex[length=2.2mm]}},
    axis/.style={thick, -{Latex[length=1.8mm]}},
    prescribed/.style={circle, fill=blue!70!black, inner sep=1.35pt},
    freevertex/.style={circle, draw=orange!85!black, fill=white,
      line width=0.8pt, inner sep=1.25pt},
    freedof/.style={rectangle, draw=orange!85!black, fill=white,
      line width=0.8pt, inner sep=1.15pt}
  ]

  \path[use as bounding box] (-0.70,-1.90) rectangle (14.90,4.15);

  \begin{scope}
    \coordinate (aA) at (0,0);
    \coordinate (aB) at (0,2.15);
    \coordinate (aC) at (0.95,2.72);
    \coordinate (aD) at (0.95,0.57);
    \coordinate (aE) at (5.05,0);
    \coordinate (aF) at (5.05,2.15);
    \coordinate (aG) at (6.00,2.72);
    \coordinate (aH) at (6.00,0.57);

    \fill[blue!7] (aA)--(aE)--(aF)--(aB)--cycle;
    \fill[blue!11] (aB)--(aF)--(aG)--(aC)--cycle;
    \fill[blue!5] (aA)--(aD)--(aH)--(aE)--cycle;
    \fill[red!5] (aA)--(aB)--(aC)--(aD)--cycle;
    \fill[red!5] (aE)--(aF)--(aG)--(aH)--cycle;

    \draw[boundary] (aA)--(aB)--(aC)--(aD)--cycle;
    \draw[boundary] (aE)--(aF)--(aG)--(aH)--cycle;
    \draw[boundary] (aA)--(aE) (aB)--(aF) (aC)--(aG);
    \draw[hidden] (aD)--(aH);

    \foreach \yy in {0.48,1.08,1.68}{
      \draw[traction] (0.10,\yy) -- ++(-0.72,0);
      \draw[traction] (5.48,\yy) -- ++(0.72,0);
    }
    \node[red!75!black, anchor=east] at (-0.18,2.47)
      {$-\sigma_0\mathbf{e}_x$};
    \node[red!75!black, anchor=west] at (5.42,2.47)
      {$+\sigma_0\mathbf{e}_x$};

    \node[anchor=south] at (3.05,2.82)
      {$\bm{\sigma}\mathbf{n}=\mathbf{0}$ on the lateral faces};
    \node at (3.05,1.72)
      {$\bm{\sigma}=\sigma_0\mathbf{e}_x\otimes\mathbf{e}_x$};
    \node at (3.05,1.28) {$\mathbf{b}=\mathbf{0}$};

    \coordinate (aO) at (2.15,0.54);
    \draw[axis] (aO) -- ++(0.92,0) node[right] {$x$};
    \draw[axis] (aO) -- ++(0,0.75) node[above] {$y$};
    \draw[axis] (aO) -- ++(0.55,0.34) node[above right] {$z$};

    \node[align=center, text width=6.4cm] at (2.75,-1.58)
      {\textbf{(a)} Uniform-tension state and boundary tractions.};
  \end{scope}

  \begin{scope}[xshift=8.25cm]
    \coordinate (bA) at (0,0);
    \coordinate (bB) at (0,2.15);
    \coordinate (bC) at (0.90,2.69);
    \coordinate (bD) at (0.90,0.54);
    \coordinate (bE) at (4.55,0);
    \coordinate (bF) at (4.55,2.15);
    \coordinate (bG) at (5.45,2.69);
    \coordinate (bH) at (5.45,0.54);
    \coordinate (bI) at (2.275,0);
    \coordinate (bJ) at (2.275,2.15);
    \coordinate (bK) at (3.175,2.69);
    \coordinate (bL) at (3.175,0.54);
    \coordinate (bQ) at (2.725,1.345);

    \fill[blue!7] (bA)--(bE)--(bF)--(bB)--cycle;
    \fill[blue!11] (bB)--(bF)--(bG)--(bC)--cycle;
    \fill[blue!5] (bA)--(bD)--(bH)--(bE)--cycle;

    \draw[boundary] (bA)--(bB)--(bC)--(bD)--cycle;
    \draw[boundary] (bE)--(bF)--(bG)--(bH)--cycle;
    \draw[boundary] (bA)--(bE) (bB)--(bF) (bC)--(bG);
    \draw[hidden] (bD)--(bH);
    \draw[meshline] (bI)--(bJ)--(bK)--(bL)--cycle;

    \foreach \p in {bA,bB,bC,bD,bE,bF,bG,bH,bI,bJ,bK,bL}{
      \node[prescribed] at (\p) {};
    }

    \node[freedof] at ($(bA)!0.50!(bI)$) {};
    \node[freedof] at ($(bI)!0.50!(bE)$) {};
    \node[freedof] at ($(bB)!0.50!(bJ)$) {};
    \node[freedof] at ($(bJ)!0.50!(bF)$) {};
    \node[freedof] at ($(bB)!0.50!(bK)$) {};
    \node[freedof] at ($(bK)!0.50!(bG)$) {};
    \node[freedof] at ($(bA)!0.50!(bJ)$) {};
    \node[freedof] at ($(bI)!0.50!(bF)$) {};
    \node[freedof] at (1.42,1.23) {};
    \node[freedof] at (3.62,1.23) {};
    \node[freevertex] at (bQ) {};

    \foreach \yy in {0.48,1.08,1.68}{
      \draw[traction] (4.98,\yy) -- ++(0.62,0);
    }
    \node[red!75!black, anchor=west] at (4.90,2.42)
      {consistent end-face load};

    \node[blue!70!black, anchor=south] at (2.72,2.86)
      {$\mathbf{u}_v=\mathbf{u}_{\mathbf{e}_x}(\mathbf{x}_v)$};

    \node[prescribed] at (0.12,-0.42) {};
    \node[anchor=west] at (0.28,-0.42)
      {prescribed boundary vertex};
    \node[freevertex] at (0.12,-0.78) {};
    \node[anchor=west] at (0.28,-0.78) {free interior vertex};
    \node[freedof] at (0.12,-1.14) {};
    \node[anchor=west] at (0.28,-1.14)
      {free higher-order DOF (schematic)};

    \node[align=center, text width=6.4cm] at (2.72,-1.70)
      {\textbf{(b)} Exact boundary-vertex values and free remaining degrees of freedom.};
  \end{scope}

\end{tikzpicture}
  \caption{Three-dimensional uniform-tension patch test. (a) Exact constant-stress state and boundary tractions. (b) Discrete treatment: the exact linear displacement is prescribed at boundary vertices, while the remaining degrees of freedom are free and the end-face tractions are assembled consistently. The degree-of-freedom symbols in (b) are schematic.}
  \label{fig:polyhedral_patch_test}
\end{figure}
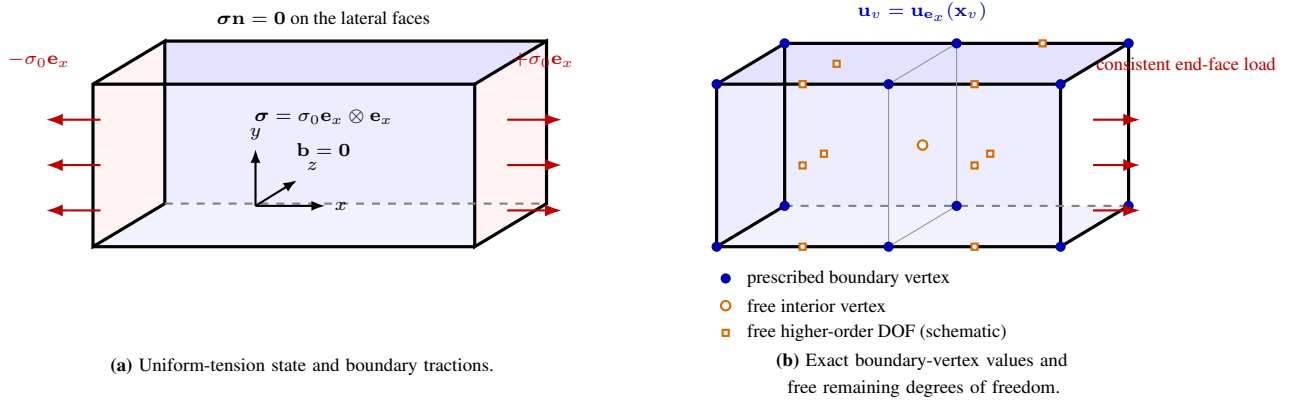

\paragraph{} The body force and the tractions on the lateral faces vanish, while the end faces carry $-\sigma_0 \mathbf{e}_x$ at $x=0$ and $+\sigma_0 \mathbf{e}_x$ at $x=L$. At every boundary vertex, the three displacement components are prescribed from $\mathbf{u}_{\mathbf{e}_x}$. The remaining degrees of freedom are left free, and the nonzero end-face tractions are assembled consistently so that the higher-order edge and face degrees of freedom receive the corresponding loads. The prescribed vertex values are imposed by exact elimination. The error is the maximum componentwise difference between $\mathbf{u}_h$ and $\mathbf{u}_{\mathbf{e}_x}$ at the interior vertices, normalized by the largest absolute component of $\mathbf{u}_{\mathbf{e}_x}$ over all vertices. Figure~\ref{fig:polyhedral_patch_test} summarizes the stress state, boundary tractions, and discrete treatment.

\paragraph{} On the coarsest hexahedral and Voronoi meshes, this test is performed for $k=1,2,3$. All six systems reproduce the linear field to roundoff. The largest relative interior-vertex error is $2.42 \times 10^{-12}$. Spectral checks on the corresponding unconstrained stiffness matrices identify six near-zero eigenvalues followed by a positive seventh eigenvalue, as required for the six rigid-body modes of three-dimensional elasticity and the absence of additional zero-energy modes. The largest magnitude in the six-dimensional cluster is $7.17 \times 10^{-16}$ relative to the largest eigenvalue.

\paragraph{} The same uniform-tension problem is then repeated through the solver workflow used for the cantilever calculations, including the penalty conditions and Neumann-load assembly. All twelve tested combinations pass the conditioning-aware check. In particular, the $k=1$ errors on the coarse and medium Voronoi meshes are $3.10\times 10^{-12}$ and $1.84 \times 10^{-12}$. The traction is integrated against the face projector at every order, so the irregular polygonal faces no longer require equal lumping of the face resultant among their vertices. The largest raw error in the full-workflow test is $2.30 \times 10^{-10}$, obtained on the $59,781$-unknown Voronoi problem at $k=3$. To estimate the roundoff sensitivity, the same system is solved again after permuting the node labels. The relative difference between the two solutions is $4.24\times 10^{-10}$, which is used as the measured roundoff floor.

\paragraph{} These uniform-tension calculations complete the patch-test verification. The discussion now returns to the cantilever simulations. Additional assembled-matrix checks are performed for every completed cantilever case whose complete system contains at most 20,000 unknowns. This criterion selects ten cases: all three orders on the $12\times2\times2$ hexahedral mesh; $k=1,2$ on the $24\times4\times4$ hexahedral mesh; $k=1$ on the $48\times8\times8$ hexahedral mesh; all three orders on the Voronoi mesh with $h_{\mathrm{nom}}=0.5$; and $k=1$ on the Voronoi mesh with $h_{\mathrm{nom}}=0.25$. For these cases, the relative symmetry defects are below $2.27\times10^{-16}$, the free-degree-of-freedom residuals are below $1.78\times10^{-9}$ relative to the load norm, and the relative differences between the strain energy and half the external work are below $1.27\times10^{-10}$. The other completed cases are excluded because the diagnostic accessor converts the sparse matrix to dense storage. Across all 17 completed cases, the maximum displacement magnitude on a constrained degree of freedom is $8.10\times10^{-13}$.

\paragraph{} Static condensation is checked independently of these displacement comparisons. The recovered full vector $\mathbf{u}_{\text{rec}}$ is compared with the solution $\mathbf{u}_{\text{full}}$ of the complete system through
\begin{equation}
  \varepsilon_{\mathrm{cond}} = \frac{\| \mathbf{u}_{\mathrm{full}} - \mathbf{u}_{\mathrm{rec}}\|_2}{\| \mathbf{u}_{\mathrm{full}} \|_2}.
\end{equation}

\paragraph{} Table~\ref{tab:polyhedral_cantilever_condensation} reports the eight nontrivial comparisons for $k=2,3$. Unlike the assembled-matrix checks, the condensation comparison is restricted to retained vertex blocks with at most 7,000 unknowns because the present implementation forms the corresponding Schur complement in dense storage. Four additional $k=1$ runs satisfy this restriction, but at $k=1$ every unknown is a vertex value, so no condensation or recovery occurs; these identity cases are omitted from the table.

\begin{table}[!htbp]
  \centering
  \caption{Complete and retained system sizes and relative discrepancies after recovery of the edge, face, and cell unknowns. The counts include constrained vertex unknowns.}
  \label{tab:polyhedral_cantilever_condensation}
  \begin{tabular}{@{}lrrr r@{}}
    \toprule
    \textbf{Family} & $h_{\mathrm{nom}}$ & $k$ &
    \textbf{Complete/retained unknowns} & $\varepsilon_{\mathrm{cond}}$ \\
    \midrule
    Hexahedral & 0.5  & 2 & 1,875/351   & $1.197\times10^{-11}$ \\
    Hexahedral & 0.5  & 3 & 4,275/351   & $5.975\times10^{-10}$ \\
    Hexahedral & 0.25 & 2 & 11,907/1,875 & $1.749\times10^{-11}$ \\
    Hexahedral & 0.25 & 3 & 28,323/1,875 & $6.925\times10^{-10}$ \\
    Voronoi    & 0.5  & 2 & 3,159/723   & $8.021\times10^{-12}$ \\
    Voronoi    & 0.5  & 3 & 6,741/723   & $6.474\times10^{-10}$ \\
    Voronoi    & 0.25 & 2 & 28,167/6,471 & $6.025\times10^{-11}$ \\
    Voronoi    & 0.25 & 3 & 59,781/6,471 & $4.083\times10^{-9}$ \\
    \bottomrule
  \end{tabular}
\end{table}

\paragraph{} The retained block contains only the three displacement components at each vertex, so its size is $3N_v^{\mathrm{mesh}}$, where $N_v^{\mathrm{mesh}}$ is the total number of mesh vertices. The largest measured discrepancy among the nontrivial comparisons is $4.083\times10^{-9}$. These runs test the algebraic equivalence of the complete and condensed solution paths; the dense condensed solve is not used as a performance optimization.

\section{Conclusion}
\label{sec:conclusion}

\paragraph{} \texttt{PoliVEM} places one-dimensional beams, two- and three-dimensional elasticity, axisymmetric elasticity, transient diffusion, and finite-strain hyperelasticity in a common implementation. The C++17 core stores the vertex, edge, face, and cell degrees of freedom in the same hierarchy and constructs the energy, strain, and $L^2$ projections from the corresponding polynomial data. It also retains the separation between polynomial consistency and stabilization at the element level. The Python interface uses this core for model definition, execution, and post-processing. A new formulation supplies its projection, discrete form, and stabilization while reusing the mesh representation, degree-of-freedom numbering, sparse assembly, boundary-condition treatment, algebraic solvers, and binding pattern.

\paragraph{} The fixed--fixed beam test checks polynomial reproduction and static condensation independently. The $k=4$ and $k=5$ spaces reproduce the fourth-degree displacement field to roundoff on every mesh, while the sampled $k=3$ field error decreases by a factor of sixteen under each halving of the element length. For the eight-element $k=5$ problem, condensation reduces the complete system from 34 to 18 unknowns, and the recovered solution differs from the complete solution by $3.950\times10^{-11}$. These results verify the higher-order beam degrees of freedom, the load construction, and the recovery of eliminated internal moments.

\paragraph{} The perforated-plate experiment tests the two-dimensional formulation on structured quadrilateral and clipped Voronoi meshes. For $k=2$, Richardson extrapolation of the deterministic O-grid sequence gives a crown stress-concentration factor of $3.086\pm0.004$, compared with $3.0861\pm0.0001$ from the independent Q9 calculation. The extrapolated strain energy and loaded-boundary displacement agree with their Q9 references within the reported uncertainties. The finest O-grid and Voronoi meshes also give relative differences of $9.68\times10^{-8}$ in strain energy and $4.12\times10^{-6}$ in loaded-boundary displacement. The nonmonotone $k=3$ O-grid energy sequence prevents the same extrapolation at that order. The small backward error of the linear solve, together with the growth of the stiffness-matrix norm, points to the scaling of the high-order polynomial basis on anisotropic cells as the limiting factor.

\paragraph{} The three-dimensional cantilever extends the test to hexahedral and general polyhedral cells. On the finest completed $k=2$ meshes, the two families give area-averaged end displacements of $0.872245$ and $0.872665$, a relative difference of $0.048\%$. An independently assembled H27 sequence extrapolates to a centre displacement of $0.872744$, within the interval formed by the three hexahedral VEM extrapolates. The uniform-tension patch test reproduces the linear displacement field to roundoff, and the stiffness spectra contain the expected six rigid-body modes without additional zero-energy modes. Static condensation recovers the complete solutions with relative discrepancies no larger than $4.083\times10^{-9}$ in the tested cases.

\paragraph{} The tests also expose current computational limits. The finest Voronoi $k=3$ cantilever system has 518,853 unknowns, but the predicted Cholesky factor requires $51.2$~GB on a machine with $38.7$~GB of memory, so the guarded solver rejects the factorization before allocation. The current three-dimensional condensation and matrix-diagnostic paths form dense objects and are therefore restricted to smaller retained systems. Better scaling or orthogonalization of the high-order polynomial basis, sparse Schur-complement operations, and memory-conscious iterative solvers with suitable preconditioners are the next numerical developments indicated by these tests. The axisymmetric, transient, and finite-strain solvers share the same infrastructure, but their numerical assessment remains in the corresponding method papers \cite{enabe2025axisymmetric,enabe2026masslumped,enabe2026stabilization}. The results reported here verify the common core for the three test classes; conditioning and scalable solution of the largest higher-order three-dimensional systems remain open implementation problems.

\bibliographystyle{unsrt}  


\newpage

\appendix

\section{Mesh}
\label{ap:mesh}

\paragraph{} \texttt{PoliVEM} accepts meshes produced by its internal generators or by external programs through the same coordinate, connectivity, and boundary-data representation. This appendix documents the two mesh components needed to reproduce and exchange the numerical examples. Appendix~\ref{ap:mesh_o_grid} specifies the deterministic O-grid used for the perforated plate, including the corner-alignment rule and the radial indexing that define the mesh family. Appendix~\ref{ap:mesh_schema} specifies the JSON and HDF5 layouts used to transfer meshes and metadata between the Python and C++ tiers. The clipped Voronoi meshes follow the standard constructions in \cite{aurenhammer1991voronoi,fortune1987sweepline,talischi2012polymesher} and are not described separately.

\subsection{Structured O-grid quadrilateral mesh}
\label{ap:mesh_o_grid}

\paragraph{} Let $a$ be the radius of the circular hole and $h$ be the prescribed nominal mesh size. The structured mesh is indexed by an angular coordinate $i = 0,\ldots,N_{\theta}$ and a radial coordinate $j = 0,\ldots,N_r$. The initial number of angular intervals is
\begin{equation}
  N_{\theta}^{(0)} = \left\lceil \frac{\pi a}{h} \right\rceil. 
\end{equation}
This choice makes the hole-boundary chords approximately $h/2$ long. The ray that reaches the outer corner has angle
\begin{equation}
  \theta_c = \arctan \left( \frac{H}{W} \right).
\end{equation}
Starting from $N_\theta^{(0)}$, the implementation increases $N_\theta$ until
\begin{equation}
  \frac{N_\theta \theta_c}{\pi/2} \in \mathbb{N}.
\end{equation}
The condition places one of the rays at the corner $(W,H)$. For the square domain, $\theta_c = \pi/4$, and the condition requires $N_\theta$ to be even. Without this adjustment, an odd number of angular intervals would replace the outer corner by a chord and would therefore change the computational domain.

\begin{remark}
  The notation $\lceil x \rceil$ denotes the ceiling of $x$, namely the smallest integer greater than or equal to $x$. The symbols $\lceil \cdot \rceil$ are used with this meaning throughout this text.
\end{remark}

\paragraph{} The angular coordinates are 
\begin{equation}
  \theta_i = \frac{i \pi}{2N_\theta}, \quad i = 0,\ldots,N_\theta.
\end{equation}
The inner endpoint of ray $i$ lies on the circular boundary 
\begin{equation}
  \mathbf{p}^{\text{in}}_i = a (\cos \theta_i, \sin \theta_i)^{\mathrm{T}}.
\end{equation}
Its outer endpoint is the intersection of the same ray with the square boundary
\begin{equation}
  \mathbf{p}^{\text{out}}_i = \begin{cases}
    (W, W \tan \theta_i)^{\mathrm{T}}, \quad &\tan \theta_i \leq H/W, \\
    (H\cot \theta_i, H)^{\mathrm{T}}, \quad &\tan \theta_i > H/W.
  \end{cases}
\end{equation}
The endpoint at $\theta_i = \theta_c$ is assigned directly as $(W,H)^{\mathrm{T}}$, which avoids a coordinate mismatch caused by floating-point evaluation of the trigonometric functions.

\begin{figure}[htbp]
\centering
\begin{tikzpicture}[
  font=\small,
  line cap=round,
  line join=round,
  >=Latex,
  gridline/.style={draw=gridgray, line width=0.45pt},
  boundary/.style={draw=black, line width=0.9pt},
  point/.style={circle, fill=black, inner sep=1.35pt},
  raypoint/.style={circle, fill=rayblue, inner sep=1.55pt},
  cornerpoint/.style={circle, fill=cornergreen, inner sep=1.55pt}
]

\begin{scope}
  \coordinate (O)     at (0,0);
  \coordinate (in0)   at (1.000,0.000);
  \coordinate (in15)  at (0.966,0.259);
  \coordinate (in30)  at (0.866,0.500);
  \coordinate (in45)  at (0.707,0.707);
  \coordinate (in60)  at (0.500,0.866);
  \coordinate (in75)  at (0.259,0.966);
  \coordinate (in90)  at (0.000,1.000);

  \coordinate (out0)  at (4.500,0.000);
  \coordinate (out15) at (4.500,1.206);
  \coordinate (out30) at (4.500,2.598);
  \coordinate (out45) at (4.500,4.500);
  \coordinate (out60) at (2.598,4.500);
  \coordinate (out75) at (1.206,4.500);
  \coordinate (out90) at (0.000,4.500);

  \path[fill=domainblue]
    (in0) -- (out0) -- (out15) -- (out30) -- (out45)
    -- (out60) -- (out75) -- (out90) -- (in90)
    arc[start angle=90,end angle=0,radius=1] -- cycle;

  \foreach \ang in {0,15,30,45,60,75,90}{
    \draw[gridline] (in\ang) -- (out\ang);
  }
  \foreach \t in {0.25,0.50,0.75}{
    \draw[gridline]
      ($(in0)!\t!(out0)$) --
      ($(in15)!\t!(out15)$) --
      ($(in30)!\t!(out30)$) --
      ($(in45)!\t!(out45)$) --
      ($(in60)!\t!(out60)$) --
      ($(in75)!\t!(out75)$) --
      ($(in90)!\t!(out90)$);
  }

  \coordinate (xij)     at ($(in30)!0.25!(out30)$);
  \coordinate (xipj)    at ($(in45)!0.25!(out45)$);
  \coordinate (xipjp)   at ($(in45)!0.50!(out45)$);
  \coordinate (xijp)    at ($(in30)!0.50!(out30)$);
  \path[fill=cellorange!48, draw=cellorange!80!black, line width=1.0pt]
    (xij) -- (xipj) -- (xipjp) -- (xijp) -- cycle;

  \draw[rayblue, line width=1.25pt] (in30) -- (out30);
  \draw[cornergreen, dashed, line width=1.15pt] (O) -- (out45);

  \draw[boundary]
    (in0) -- (out0) -- (out15) -- (out30) -- (out45)
    -- (out60) -- (out75) -- (out90) -- (in90)
    arc[start angle=90,end angle=0,radius=1];

  \node[raypoint] at (in30) {};
  \node[raypoint] at (out30) {};
  \node[cornerpoint] at (out45) {};
  \node[point] at (xij) {};
  \node[point] at (xipj) {};
  \node[point] at (xipjp) {};
  \node[point] at (xijp) {};

  \node[rayblue, anchor=north west] at ($(in30)+(0.05,-0.03)$)
    {$\bm p_i^{\mathrm{in}}$};
  \node[rayblue, anchor=west] at ($(out30)+(0.08,0)$)
    {$\bm p_i^{\mathrm{out}}$};
  \node[cornergreen, anchor=south west] at ($(out45)+(0.04,0.04)$)
    {$(W,H)$};

  \node[anchor=south east, inner sep=1pt] at ($(xij)+(-0.03,0.03)$)
    {$\bm x_{ij}$};
  \node[anchor=south east, inner sep=1pt] at ($(xipj)+(-0.03,0.03)$)
    {$\bm x_{i+1,j}$};
  \node[anchor=south west, inner sep=1pt] at ($(xipjp)+(0.03,0.03)$)
    {$\bm x_{i+1,j+1}$};
  \node[anchor=north west, inner sep=1pt] at ($(xijp)+(0.03,-0.03)$)
    {$\bm x_{i,j+1}$};
  \node[font=\small\bfseries] at ($(xij)!0.50!(xipjp)$) {$E_{ij}$};

  \draw[rayblue, -{Latex[length=1.6mm]}, line width=0.8pt]
    (0.72,0) arc[start angle=0,end angle=30,radius=0.72];
  \node[rayblue] at (0.83,0.24) {$\theta_i$};

  \draw[-{Latex[length=1.7mm]}] (O) -- (1.45,0)
    node[below right] {$x$};
  \draw[-{Latex[length=1.7mm]}] (O) -- (0,1.45)
    node[above left] {$y$};
  \draw[<->, line width=0.6pt] (O) -- (in75)
    node[midway, left] {$a$};

  \node[anchor=north, font=\bfseries] at (2.25,-1.05)
    {(a) Ray and cell indexing};
\end{scope}

\begin{scope}[xshift=7.0cm]
  \coordinate (Ob) at (0,0);
  \coordinate (Rb) at (4.5,0);
  \coordinate (C)  at (4.5,4.5);
  \coordinate (Tb) at (0,4.5);
  \coordinate (qm) at (4.5,3.22);
  \coordinate (qp) at (3.22,4.5);

  \path[fill=domainblue] (Ob) -- (Rb) -- (C) -- (Tb) -- cycle;
  \path[fill=cutred!22] (qm) -- (C) -- (qp) -- cycle;

  \draw[boundary] (Rb) -- (C) -- (Tb);
  \draw[gridgray, line width=0.55pt] (Ob) -- (Rb);
  \draw[gridgray, line width=0.55pt] (Ob) -- (Tb);
  \draw[cutred, line width=1.1pt] (Ob) -- (qm);
  \draw[cutred, line width=1.1pt] (Ob) -- (qp);
  \draw[cutred, dashed, line width=1.2pt] (qm) -- (qp);
  \draw[cornergreen, line width=1.3pt] (Ob) -- (C);

  \node[point] at (qm) {};
  \node[point] at (qp) {};
  \node[cornerpoint] at (C) {};

  \draw[cornergreen, -{Latex[length=1.6mm]}, line width=0.8pt]
    (0.78,0) arc[start angle=0,end angle=45,radius=0.78];
  \node[cornergreen] at (0.93,0.34) {$\theta_c$};

  \node[cornergreen, anchor=south west] at (4.56,4.56) {$(W,H)$};

  \draw[cornergreen, line width=1.2pt] (0.24,4.17) -- (0.76,4.17);
  \node[anchor=west, cornergreen] at (0.86,4.17)
    {ray at $\theta_c$};
  \draw[cutred, dashed, line width=1.2pt] (0.24,3.80) -- (0.76,3.80);
  \node[anchor=west, cutred] at (0.86,3.80)
    {chord if omitted};

  \node[align=center] at (2.25,-0.55)
    {$N_\theta\theta_c/(\pi/2)\in\mathbb N$};
  \node[anchor=north, font=\bfseries] at (2.25,-1.05)
    {(b) Corner-alignment condition};
\end{scope}

\end{tikzpicture}
\caption{Construction of the structured O-grid. (a) Inner and outer ray endpoints, linearly interpolated nodes, and the four vertices of $E_{ij}$. (b) The ray at $\theta_c$ places a node at $(W,H)$. If this ray is omitted, the segment joining the neighbouring outer endpoints cuts off the corner and changes the computational domain.}
\label{fig:ogrid_construction}
\end{figure}
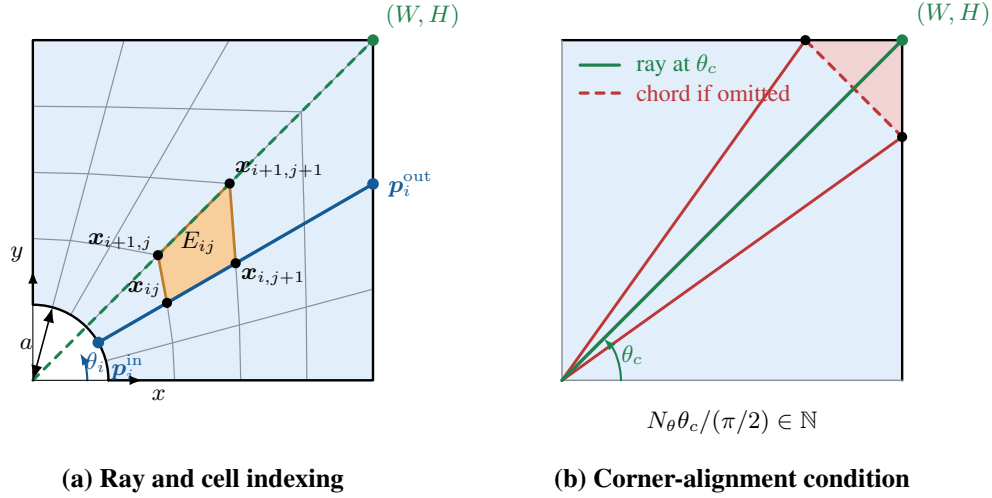

\paragraph{} Let 
\begin{equation}
  l_i = \| \mathbf{p}^{\text{out}}_i - \mathbf{p}^{\text{in}}_i \|_2
\end{equation}
be the length of ray $i$ inside the domain. A single radial count is used for the complete mesh
\begin{equation}
  N_r = \left\lceil \frac{\max_i l_i}{h} \right\rceil.
\end{equation}
With $s_j = j /N_r$, the mesh nodes are obtained by linear interpolation along each ray,
\begin{equation}
  \mathbf{x}_{ij} = (1-s_j)\mathbf{p}^{\mathrm{in}}_i + s_j \mathbf{p}^{\mathrm{out}}_i.
\end{equation}
The quadrilateral in angular interval $i$ and radial interval $j$ has the ordered vertices
\begin{equation}
  E_{ij} = (\mathbf{x}_{ij}, \mathbf{x}_{i+1,j}, \mathbf{x}_{i+1,j+1}, \mathbf{x}_{i,j+1}).
\end{equation}
The vertex order is checked and reversed when necessary so that every stored element is counterclockwise. Figure \ref{fig:ogrid_construction} summarizes the ray construction, the indexing of a representative cell, and the corner-alignment condition.

\paragraph{} Using the same $N_r$ on every ray ensures that adjacent rays contain corresponding nodes and that the resulting mesh is conforming. It also ensures that every radial increment is no larger than $h$. The longest ray determines $N_r$, however, so the cells become anisotropic away from that ray. The construction is deterministic, i.e. $N_\theta$, $N_r$, every node coordinate, and every connectivity entry are functions of $h$. Rebuilding a mesh with the same value of $h$ therefore produces the same arrays. Meshes at successive values of $h$ are not necessarily node-nested, because $N_\theta$ and $N_r$ change, but they form a systematic parametric family.

\subsection{Mesh file schema}
\label{ap:mesh_schema}

\paragraph{} The framework reads and writes meshes through a documented schema with a JSON text form and an HDF5 binary form. A mesh, or a complete problem definition when the problem data are included, can be archived, reproduced, and exchanged between the two tiers and external tools. A JSON mesh file has three top-level members: a metadata object, a node list, and an element list, as shown in Listing~\ref{lst:mesh_schema}.

\begin{lstlisting}[basicstyle=\ttfamily\footnotesize, backgroundcolor=\color{codebg},
  frame=single, rulecolor=\color{codegray}, columns=fullflexible,
  keepspaces=true, showstringspaces=false, captionpos=b,
  caption={Top-level structure of a JSON mesh file, with the metadata object,
  the node list, and the element list.}, label={lst:mesh_schema}]
{ "metadata": { ... }, "nodes": [ ... ], "elements": [ ... ] }
\end{lstlisting}

\paragraph{} The metadata object records the mesh type, the spatial dimension, the node and element counts, and the domain extent, together with a set of mesh-quality measures computed from the elements. Table~\ref{tab:mesh_metadata} lists the common fields. For the structured and distorted quadrilateral meshes the metadata additionally records the element type and the grid resolution, and for the polygonal Voronoi meshes it records the per-element vertex-count statistics and the edge and connectivity counts. For the applications whose problem data travel with the mesh, such as the axisymmetric benchmarks, the metadata additionally carries the geometric parameters, the material data, the named boundary sets, and the boundary conditions, which turns the file into a self-contained problem specification.

\begin{table}[!htbp]
  \centering
  \renewcommand{\arraystretch}{1.3}
  \setlength{\tabcolsep}{6pt}
  \footnotesize
  \caption{Common fields of the metadata object in a JSON mesh file.}
  \label{tab:mesh_metadata}
  \begin{tabularx}{\textwidth}{@{}L{3.7cm}L{1.3cm}L{1.9cm}Y@{}}
  \toprule
  \textbf{Field} & \textbf{Type} & \textbf{Required} & \textbf{Description} \\
  \midrule
  \texttt{meshType} & string & yes &
  Mesh-type identifier, for example \texttt{voronoi}, \texttt{serendipity\_quad},
  or \texttt{distorted\_quad\_grid} \\

  \texttt{dimension} & integer & recommended & Spatial dimension \\

  \texttt{nodeCount} & integer & yes & Number of nodes \\

  \texttt{elementCount} & integer & yes & Number of elements \\

  \texttt{domain} & object & optional &
  Domain extent, with \texttt{xmin}, \texttt{xmax}, \texttt{ymin}, \texttt{ymax} \\

  \texttt{actualHMax}, \texttt{actualHMin}, \texttt{actualHAvg} & number & optional &
  Element-size statistics \\

  \texttt{minVerticesPerElement}, \texttt{maxVerticesPerElement},
  \texttt{avgVerticesPerElement} & number & optional &
  Per-element vertex-count statistics \\

  \texttt{interiorEdges}, \texttt{boundaryEdges}, \texttt{connectivityRatio} &
  number & optional & Connectivity statistics \\

  \texttt{totalArea} & number & optional &
  Total area covered by the elements \\

  \texttt{geometry}, \texttt{material}, \texttt{boundaries},
  \texttt{boundaryConditions} & object & optional &
  Self-contained problem definition, when present \\
  \bottomrule
  \end{tabularx}
\end{table}

\paragraph{} The node list gives one entry per node, each with a sequential zero-based identifier and its coordinates, including a third coordinate for three-dimensional meshes. The identifiers run from zero to the node count minus one without gaps. The element list gives one entry per element, each with a sequential zero-based identifier and an ordered list of node identifiers. Counterclockwise vertex ordering produces a positive signed area. Variable-length connectivity allows arbitrary polygons and fixed four- and eight-node quadrilaterals to use the same schema. Listing~\ref{lst:mesh_example} shows a single hexagonal element on the unit square.

\begin{lstlisting}[basicstyle=\ttfamily\footnotesize, backgroundcolor=\color{codebg},
  frame=single, rulecolor=\color{codegray}, columns=fullflexible,
  keepspaces=true, showstringspaces=false, captionpos=b,
  caption={A complete JSON mesh file for a single hexagonal element on the unit
  square, with its metadata, node list, and element list.},
  label={lst:mesh_example}]
{
  "metadata": { "meshType": "voronoi", "dimension": 2,
                "nodeCount": 7, "elementCount": 1,
                "domain": {"xmin":0.0,"xmax":1.0,"ymin":0.0,"ymax":1.0},
                "actualHMax": 0.5, "totalArea": 1.0 },
  "nodes": [ {"id":0,"x":0.5,"y":0.0}, {"id":1,"x":1.0,"y":0.25},
             {"id":2,"x":1.0,"y":0.75}, {"id":3,"x":0.5,"y":1.0},
             {"id":4,"x":0.0,"y":0.75}, {"id":5,"x":0.0,"y":0.25} ],
  "elements": [ {"id":0,"vertices":[0,1,2,3,4,5]} ]
}
\end{lstlisting}

\paragraph{} A valid file satisfies the following conditions: all three sections are present; the node and element counts match the array lengths; the node and element identifiers are zero-based and sequential; every vertex identifier references an existing node; and every element has positive area under counterclockwise ordering. A JSON Schema description distributed with the source supports automated validation. The file-naming convention encodes the mesh type and characteristic size so that each family of refined meshes is self-describing.

\paragraph{} Larger three-dimensional polyhedral meshes store the same content in binary form. An HDF5 mesh file has a metadata group that records the dimension and mesh statistics, a double-precision vertex dataset of shape $n_{nodes} \times 3$, an integer dataset of element connectivity, and one integer dataset per named boundary listing the nodes on that boundary. The binary form carries the same information as the text form but uses less storage and loads faster for meshes with large node counts.

\paragraph{} A mesh can be archived with its results and reloaded to reproduce them. When the metadata includes the geometry, material, and boundary definitions, the file is a self-contained problem specification and is archived with the released benchmarks. Meshes produced by external generators use the same schema and require no changes to the solvers.

\section{Direct solvers for the discrete systems}
\label{ap:direct_solvers}

\paragraph{} The discrete problems considered here reduce to linear systems of the form
\begin{equation}
  \label{eq:generic_discrete_problem}
  \mathbf{A}\mathbf{x} = \mathbf{b},
\end{equation}
in which $\mathbf{A}=[a_{ij}]$ is the assembled global operator. The framework uses direct factorization because it does not require an iterative convergence criterion whose behaviour depends on conditioning. The factors can also be reused when the operator remains fixed, as in a fixed-step implicit transient run. Following \cite{burden2011numerical}, this appendix summarizes the three factorizations used by the linear solver: general LU factorization, Cholesky factorization for symmetric positive definite operators, and the supernodal sparse Cholesky variant.

\subsection{LU factorization}
\paragraph{} Gaussian elimination applied to (\ref{eq:generic_discrete_problem}) can be recorded as a factorization of the coefficient matrix. When the elimination proceeds without row interchanges, the multipliers used to eliminate the entries below each pivot form a unit lower triangular matrix $\mathbf{L} = [l_{ij}]$ ($l_{ij} = 0$ for $i<j$), and the reduced coefficient matrix at the end of the process is an upper triangular matrix $\mathbf{U} = [u_{ij}]$ ($u_{ij} = 0$ for $i>j$), so that
\begin{equation}
  \label{eq:lu_decomposition}
  \mathbf{A} = \mathbf{L} \mathbf{U}.
\end{equation}
The normalization is not unique, and fixing the diagonal of $\mathbf{L}$ to unity gives the variant usually called Doolittle form, while fixing the diagonal of $\mathbf{U}$ instead gives the variant called Crout form. In the Doolittle form the entries follow from equating the two sides of (\ref{eq:lu_decomposition}) one row and one column at a time
\begin{equation}
  a_{ij} = \sum \limits^{\min \{i,j \}}_{k=1} l_{ik} u_{kj},
\end{equation}
which gives
\begin{equation}
  \label{eq:algebraic_form_lu_decomposition}
  u_{jk} = a_{jk} - \sum \limits^{j-1}_{m=1}l_{jm}u_{mk}, \quad k \geq j, \quad l_{ij} = \frac{1}{u_{jj}} \left( a_{ij} - \sum \limits^{j-1}_{m=1} l_{im}u_{mj} \right), \quad i > j.
\end{equation}
Each entry depends only on the entries already computed at earlier steps, so the two expressions in (\ref{eq:algebraic_form_lu_decomposition}) are evaluated for $j=1,\ldots,n$ in order, and the factors overwrite the coefficient matrix in place because the entry $a_{ij}$ is not needed once $l_{ij}$ or $u_{ij}$ has been formed.

\paragraph{} Once the factors are available, the system (\ref{eq:generic_discrete_problem}) is solved in two triangular sweeps, namely a forward substitution $\mathbf{L} \mathbf{y} = \mathbf{b}$ followed by a backward substitution $\mathbf{U} \mathbf{x} = \mathbf{y}$. Each sweep costs $\mathcal{O}(n^2)$ operations for a dense matrix of order $n$, whereas building the factors beforehand costs $\mathcal{O} (n^3)$ operations, approximately $n^3/3$ multiplications and divisions. The two stages therefore differ by a full power of $n$, which is the practical argument for factorizing once rather than eliminating repeatedly, because every additional right-hand side costs only the two triangular sweeps once the factors are known.

\paragraph{} A factorization of the form (\ref{eq:lu_decomposition}) does not exist for every nonsingular matrix, and it fails as soon as a leading principal submatrix is singular, the simplest instance being a zero pivot in the first position. Row interchanges repair this, and the elimination with partial pivoting produces
\begin{equation}
  \mathbf{P} \mathbf{A} = \mathbf{L} \mathbf{U},
\end{equation}
in which $\mathbf{P}=[p_{ij}]$ is the permutation matrix that records the interchanges. The choice of the interchange is made column by column and reuses the quantities already present in (\ref{eq:algebraic_form_lu_decomposition}). At step $j$ the updated entries of the $j$-th column are formed for every candidate row,
\begin{equation}
  \label{eq:candidate_row}
  s_{ij} = a_{ij} - \sum \limits^{j-1}_{m=1}l_{im}u_{mj}, \quad i = j,\ldots,n,
\end{equation}
and the pivot row is the one that maximizes their magnitude,
\begin{equation}
  \label{eq:pivoting_index}
  p = \arg \; \max_{j \leq i \leq n} |s_{ij}|.
\end{equation}
Rows $j$ and $p$ are interchanged in the coefficient matrix, in the already computed columns of $\mathbf{L}$, and in the record of the permutation, after which the step is completed by
\begin{equation}
  u_{jj} = s_{jj}, \quad l_{ij} = \frac{s_{ij}}{u_{jj}}, \quad i > j,
\end{equation}
with the indices understood after the interchange. The search in (\ref{eq:pivoting_index}) uses the updated values (\ref{eq:candidate_row}) rather than the original coefficients because the updated column is divided by the pivot. The interchange also applies to the previously computed part of $\mathbf{L}$: permuting the rows of $\mathbf{A}$ permutes the corresponding rows of $\mathbf{L}$. The rule (\ref{eq:pivoting_index}) guarantees $|l_{ij}| \leq 1$ for every multiplier, which bounds entry growth during elimination and controls the propagation of rounding errors. In practice, the permutation is stored as an index vector rather than a matrix, and the triangular sweeps solve $\mathbf{L} \mathbf{y} = \mathbf{P} \mathbf{b}$ followed by $\mathbf{U} \mathbf{x} = \mathbf{y}$. Partial pivoting is the default safeguard for a matrix whose structure is not known in advance.

\subsection{Cholesky factorization}

\paragraph{} A symmetric positive definite operator admits the unique factorization
\begin{equation}
  \label{eq:cholesky_factorization}
  \mathbf{A} = \mathbf{L} \mathbf{L}^{\mathrm{T}},
\end{equation}
in which $\mathbf{L}$ is lower triangular with strictly positive diagonal entries. The entries follow directly from equating the two sides of (\ref{eq:cholesky_factorization}) column by column, which gives
\begin{equation}
  \label{eq:algebraic_form_cholesky_decomposition}
  l_{jj} = \left( a_{jj} - \sum \limits^{j-1}_{k=1} l_{jk}^2 \right)^{1/2}, \quad l_{ij} = \frac{1}{l_{jj}} \left( a_{ij} - \sum \limits^{j-1}_{k=1} l_{ik}l_{jk} \right), \quad i>j.
\end{equation}
Positive definiteness guarantees that the quantity under the square root in (\ref{eq:algebraic_form_cholesky_decomposition}) remains positive, so the process cannot break down and requires no pivoting for stability. A related variant factors
\begin{equation}
  \mathbf{A} = \mathbf{L} \mathbf{D} \mathbf{L}^{\mathrm{T}}
\end{equation}
with a unit lower triangular $\mathbf{L}$ and a diagonal $\mathbf{D}$, which avoids the square roots of (\ref{eq:algebraic_form_cholesky_decomposition}) at the cost of storing the diagonal separately.

\paragraph{} Two properties make (\ref{eq:cholesky_factorization}) preferable whenever it applies. Only the triangular factor is formed and stored, which halves the memory of the factorization, and the arithmetic is likewise about half that of the general elimination, of the order of $n^3/6$ multiplications and divisions. Since the elastic operators of this work are SPD once the essential boundary conditions are imposed, and the mass operators are SPD by construction, the Cholesky factorization is the natural default for them.

\subsection{Sparse Cholesky and the supernodal method}
\label{ap:sub_supernodal_method}

\paragraph{} The operators assembled from a mesh are sparse, and a dense factorization is unusable at the sizes considered here. The sparse case introduces a difficulty that has no dense counterpart, namely that the factor $\mathbf{L}$ generally contains nonzero entries in positions where $\mathbf{A}$ is zero. These entries are called fill-in, and their number depends strongly on the ordering of the unknowns. A sparse Cholesky factorization is therefore computed for a permuted matrix
\begin{equation}
  \mathbf{P} \mathbf{A} \mathbf{P}^{\mathrm{T}} = \mathbf{L} \mathbf{L}^{\mathrm{T}},
\end{equation}
in which the permutation $\mathbf{P}$ is chosen to reduce fill rather than to control pivot growth, since the latter is unnecessary for a positive definite operator. The orderings used in practice are heuristics, the most common being the minimum-degree family and nested dissection, and they are applied to the graph of the operator before any arithmetic is performed.

\paragraph{} The computation separates into symbolic and numeric stages, as does the assembly in Section \ref{subsec:sparse_assembly}. The symbolic stage uses only the sparsity structure to determine the nonzero pattern of $\mathbf{L}$, fixing the storage and order of later operations. The elimination tree \cite{liu1990etree,liu1993supernodes} describes the dependencies between columns of the factor; the parent of a column is the row index of its first subdiagonal nonzero. The tree determines the contributing columns, the pattern of the factor, and a valid evaluation order. The numeric stage then computes the values in this fixed pattern.

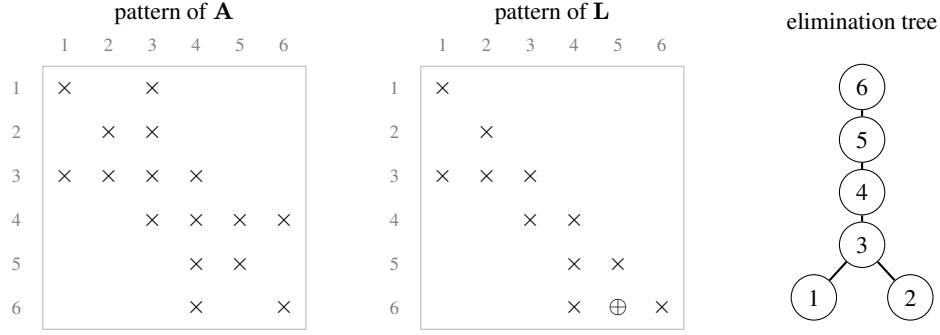
\begin{figure}[!htbp]
\centering
\begin{tikzpicture}[
  x=0.58cm, y=0.58cm,
  ent/.style={font=\footnotesize},
  idx/.style={font=\scriptsize, gray},
  tn/.style={draw, circle, inner sep=0pt, minimum size=6mm, font=\footnotesize}
]
\begin{scope}
  \node[font=\small] at (3.5,1.7) {pattern of $\mathbf{A}$};
  \draw[gray!55] (0.5,0.5) rectangle (6.5,-5.5);
  \foreach \i in {1,...,6}{
    \node[idx] at (\i,1.0) {\i};
    \node[idx] at (-0.1,{1-\i}) {\i};
  }
  \foreach \r/\c in {1/1,2/2,3/3,4/4,5/5,6/6,
                     3/1,1/3,3/2,2/3,4/3,3/4,5/4,4/5,6/4,4/6}
    \node[ent] at (\c,{1-\r}) {$\times$};
\end{scope}
\begin{scope}[xshift=5.0cm]
  \node[font=\small] at (3.5,1.7) {pattern of $\mathbf{L}$};
  \draw[gray!55] (0.5,0.5) rectangle (6.5,-5.5);
  \foreach \i in {1,...,6}{
    \node[idx] at (\i,1.0) {\i};
    \node[idx] at (-0.1,{1-\i}) {\i};
  }
  \foreach \r/\c in {1/1,2/2,3/3,4/4,5/5,6/6,3/1,3/2,4/3,5/4,6/4}
    \node[ent] at (\c,{1-\r}) {$\times$};
  \node[ent] at (5,-5) {$\oplus$};
\end{scope}
\begin{scope}[xshift=10.2cm, yshift=-0.1cm]
  \node[font=\small] at (1.6,1.7) {elimination tree};
  \node[tn] (n1) at (0.5,-4.6) {1};
  \node[tn] (n2) at (2.7,-4.6) {2};
  \node[tn] (n3) at (1.6,-3.4) {3};
  \node[tn] (n4) at (1.6,-2.2) {4};
  \node[tn] (n5) at (1.6,-1.0) {5};
  \node[tn] (n6) at (1.6, 0.2) {6};
  \draw[thick] (n1) -- (n3);
  \draw[thick] (n2) -- (n3);
  \draw[thick] (n3) -- (n4);
  \draw[thick] (n4) -- (n5);
  \draw[thick] (n5) -- (n6);
\end{scope}
\end{tikzpicture}
\caption{Sparsity patterns and elimination tree of a symmetric positive definite
matrix of order six. The symbol $\times$ marks an entry that is nonzero in the
matrix, and the symbol $\oplus$ marks the fill-in entry that the factorization
creates at the position $(6,5)$. The parent of each column of $\mathbf{L}$ is the row
index of its first subdiagonal nonzero, which gives the tree on the right. Columns
one and two lie in disjoint subtrees and can be eliminated independently.}
\label{fig:etree}
\end{figure}

\paragraph{} A small example makes the construction concrete. Figure \ref{fig:etree} shows the pattern of a symmetric positive definite matrix of order six, the pattern of its Cholesky factor, and the associated elimination tree. Eliminating the fourth column couples the fifth and the sixth unknowns, which were not coupled in the original matrix, so the factor acquires a nonzero at position $(6,5)$ that the matrix does not have. This single entry is the fill-in of the example. Reading the first subdiagonal nonzero of each column of the factor gives the parent, namely $3$ for the first and second columns, $4$ for the third, $5$ for the fourth, and $6$ for the fifth, while the sixth column has no subdiagonal entry and is therefore the root. The resulting tree records two facts that the numeric stage uses. A column receives contributions only from its descendants, which is what allows the pattern of the factor to be predicted before any arithmetic is performed, and two columns that lie in disjoint subtrees, such as the first and the second in the figure, depend on no common ancestor below them and can therefore be eliminated independently, which is the property that parallel and supernodal implementations exploit.

\paragraph{} In factors arising from practical problems, consecutive columns of $\mathbf{L}$ frequently share the same nonzero pattern below their diagonal block. Such a group forms a supernode and can be stored and manipulated as a dense rectangular block instead of separate sparse columns. The factorization advances one supernode at a time through four dense-block operations: a symmetric update of the diagonal block by descendant contributions, a dense Cholesky factorization of the updated block, a product forming the off-diagonal contributions, and a dense triangular solve that scales them. Optimized Level-3 BLAS kernels execute these operations with cache reuse, performing the same arithmetic faster than a column-by-column sparse factorization.

\paragraph{} For very sparse operators, the supernodes are small and the overhead of assembling dense blocks can exceed their benefit. A column-oriented sparse factorization is then faster. Established sparse Cholesky libraries select between the two at run time from the predicted density of the factor, and the framework uses the policy of its linked library.

\section{Basic Linear Algebra Subprograms (BLAS)}
\label{ap:blas}

\paragraph{} The Basic Linear Algebra Subprograms (BLAS) are a standardized collection of low-level kernels for common vector and matrix operations. The standard specifies the operation performed by each routine, together with its name, arguments, and numerical requirements, but it does not prescribe a unique implementation. A higher-level algorithm can therefore call the same interface while the linked numerical library supplies an implementation optimized for the target architecture. Besides improving portability and performance, this separation gives linear algebra software a small set of common building blocks with well-defined behaviour \cite{dongarra1990level3,blackford2002blas}.

\paragraph{} The classical BLAS hierarchy distinguishes the routines by the type of operands on which they act. For square operands of order $n$, the representative costs also increase with the level:

\begin{enumerate}
  \item \textbf{Level 1} contains scalar and vector operations with a representative cost of $\mathcal{O}(n)$. Examples include the scaled vector update $\mathbf{y} \leftarrow \alpha \mathbf{x}+\mathbf{y}$ (\texttt{AXPY}), dot products, and vector norms.
  \item \textbf{Level 2} contains matrix-vector operations with a representative cost of $\mathcal{O}(n^2)$. Examples include the general matrix-vector product $\mathbf{y} \leftarrow \alpha\mathbf{A}\mathbf{x}+\beta\mathbf{y}$ (\texttt{GEMV}), rank-one updates (\texttt{GER}), and triangular solves with one right-hand side (\texttt{TRSV}).
  \item \textbf{Level 3} contains matrix-matrix operations with a representative cost of $\mathcal{O}(n^3)$. The set includes the general matrix product $\mathbf{C} \leftarrow \alpha\mathbf{A}\mathbf{B}+\beta\mathbf{C}$ (\texttt{GEMM}), symmetric or Hermitian rank-$k$ updates (\texttt{SYRK} and \texttt{HERK}), triangular matrix products (\texttt{TRMM}), and triangular solves with multiple right-hand sides (\texttt{TRSM}) \cite{dongarra1990level3,blackford2002blas}.
\end{enumerate}

\paragraph{} Level-3 operations provide more data reuse than Level-2 operations. A Level-2 matrix-vector product performs $\mathcal{O}(n^2)$ arithmetic on a matrix containing $\mathcal{O}(n^2)$ entries, so each matrix entry offers limited reuse within one call. A Level-3 matrix-matrix operation performs $\mathcal{O}(n^3)$ arithmetic on $\mathcal{O}(n^2)$ data. When the matrices are partitioned into blocks, a block loaded into cache or local memory can participate in many operations before it must be moved again. The same organization exposes parallelism both between independent blocks and within each dense-block kernel. These properties motivated the Level-3 specification for hierarchical-memory and parallel computers \cite{dongarra1990level3}.

\paragraph{} Level-3 BLAS routines are deliberately kernels rather than complete factorization algorithms. A blocked factorization organizes its work so that most arithmetic is delegated to these kernels. In blocked Cholesky factorization, for example, the diagonal blocks are factorized in sequence, triangular solves form the off-diagonal blocks, and symmetric rank-$k$ updates modify the remaining matrix. The Level-3 paper uses this factorization to illustrate how the block size controls data movement and the amount of work passed to the kernels \cite{dongarra1990level3}. The same principle underlies the supernodal sparse Cholesky method of Appendix \ref{ap:sub_supernodal_method}: grouping structurally similar sparse columns exposes dense blocks that can be updated by Level-3 BLAS routines.

\paragraph{} The BLAS interface is portable, but performance depends on the implementation and hardware. The calling algorithm remains unchanged, while an architecture-specific BLAS implementation chooses blocking, vectorization, and parallel execution appropriate to the machine. The benefit is greatest when the dense blocks are large enough to amortize the call and packing overhead; for small supernodes, a column-oriented sparse method may remain preferable. The updated BLAS standard retains this separation between standardized functionality and optimized implementations while extending the available dense, banded, sparse, and mixed-precision operations \cite{blackford2002blas}.

\subsection{Unblocked and blocked algorithms}
\label{ap:unblocked_blocked_algorithms}

\paragraph{} Blocked algorithms reorganize a factorization so that most of its
arithmetic is performed by Level-3 BLAS routines. The mathematical factorization
and its leading operation count remain unchanged; blocking changes how the work is
grouped and how often data can be reused after they have been loaded into cache
\cite{dongarra1990level3}.

\paragraph{} An unblocked LU factorization processes the matrix one column at a
time. At each step, it computes the multipliers and applies a rank-one update to the
remaining submatrix, commonly through the Level-2 routine \texttt{GER}. Because each
matrix entry participates in few arithmetic operations before the next data transfer,
memory traffic can limit the performance of this column-oriented formulation. A
blocked factorization instead groups $b$ consecutive columns, with $b\ll n$, into a
panel.

\paragraph{} The panel is factorized by an unblocked kernel, which also selects
and applies the row interchanges when partial pivoting is used. After these
interchanges have been applied to the remaining columns, the factorization uses
\texttt{TRSM} to compute the corresponding block row and \texttt{GEMM} to update the
trailing submatrix. With the partition understood after the row interchanges, these
two operations have the form

\begin{equation}
  \mathbf{U}_{12}
  =
  \mathbf{L}_{11}^{-1}\mathbf{A}_{12},
  \qquad
  \mathbf{A}_{22}
  \leftarrow
  \mathbf{A}_{22}-\mathbf{L}_{21}\mathbf{U}_{12}.
\end{equation}

\paragraph{} The trailing-submatrix update contains most of the arithmetic for a
large matrix and reuses the entries of its operands within the Level-3 kernel. The
panel factorization remains less efficient, but it accounts for a smaller fraction of
the total work when $b\ll n$. Setting $b=1$ recovers the unblocked algorithm. The
chosen block size must be large enough for the Level-3 kernels to run efficiently but
small enough to limit the panel work and associated overhead. It therefore depends on
the matrix size and the memory hierarchy of the target architecture
\cite{dongarra1990level3}.

\end{document}